\documentclass[preprint,12pt]{elsarticle}

\usepackage{graphicx}
\usepackage{dcolumn}
\usepackage{bm}
\usepackage{epsfig}
\usepackage{subfig}
\usepackage{epstopdf}

\usepackage{placeins}

\usepackage{amssymb,amsmath}
\usepackage{hyperref}

\newcommand{\calP}{\mathcal{P}}
\newcommand{\re}{\mathrm{e}}

\journal{Physica D}

\begin{document}

\begin{frontmatter}



\title{Gaussian noise and the two-network frustrated Kuramoto model}


\author[label1]{Andrew B. Holder}
\ead{andrew.holder@dsto.defence.gov.au}
\author[label1]{Mathew L. Zuparic}
\ead{mathew.zuparic@dsto.defence.gov.au}
\author[label1]{Alexander C. Kalloniatis}
\ead{alexander.kalloniatis@dsto.defence.gov.au}
\address[label1]{Defence Science and Technology Group, Canberra,
ACT 2600, Australia}

\begin{abstract}
We examine analytically and numerically a variant of the stochastic Kuramoto model for phase oscillators coupled on a general network. Two populations of phased oscillators are considered, labelled `Blue' and `Red', each with their respective networks, internal and external couplings, natural frequencies, and frustration parameters in the dynamical interactions of the phases. 
We disentagle the different ways that additive Gaussian noise may influence the dynamics by applying it
separately on zero modes or normal modes corresponding to a Laplacian decomposition for the sub-graphs for Blue and Red.
Under the linearisation \textit{ansatz} that the oscillators of each respective network remain relatively phase-sychronised centroids or clusters, we are able to obtain simple closed-form expressions using the Fokker-Planck approach for the dynamics of the average angle of the two centroids. 
In some cases, this leads to subtle effects of metastability that we may analytically describe using the theory of ratchet potentials.
These considerations are extended to a regime where one of the populations has fragmented in two.
The analytic expressions we derive largely predict the dynamics of the non-linear system
seen in numerical simulation. 
In particular, we find that noise acting on a more tightly coupled population allows for improved synchronisation of the other population 
where deterministically
it is fragmented. 

\end{abstract}


\end{frontmatter}



\section{Introduction}
Many real world complex systems display both cooperative and competitive phenomena with tendencies for both order and disorder in tension.
Where such systems involve connectivity between component entities and elementary limit cycles, a generalisation of the Kuramoto model \cite{Kura1984}
for synchronising oscillators to multi-networks \cite{Bocc2014,MKB2004,BHOS2008,KNAKK2010,SkarRes2012} provides a useful paradigm for modelling. Add to this certain `lags', otherwise known as {\it frustrations} \cite{SakKur1986,Cool2003,NVCDGL2013,KirkSev2015} giving
what is often called the Kuramoto-Sakaguchi model, and
noise \cite{ABS2005,Bag2007,Khosh2008,Toen2010,Esfah2012,Deville2012,Traxl2014} and the key elements for modelling such systems are in place. 
Previously we studied such a model \cite{KallZup2015} for its equilibria both analytically and numerically. In this paper we add Gaussian noise to the system,
a generalisation to finite arbitrary networks of approaches such as \cite{KNAKK2010b}.

The starting point is the system of equations describing the dynamics of Kuramoto oscillators on a network \cite{DGM2008,ADKMZ2008,DorBull2014}
\begin{eqnarray}
\dot{\theta}_i =\omega_i -\sigma \sum^N_{j=1}A_{ij} \sin (\theta_i - \theta_j), \;\; i = \{1,\dots,N\},
\label{originalKura}
\end{eqnarray}
where $\theta_i$ is the phase angle for node $i$, $\omega_i$ is the intrinsic frequency of the aforementioned node, and $\sigma$ is a global coupling constant. The adjacency matrix, $A_{ij}$, encodes the network topology of the system, where $A_{ij} = 1$ or $0$ if nodes $i$ and $j$ are connected or disconnected, respectively.  
For non-identical native frequencies, at strong coupling all phases eventually frequency synchronise and approach phase locking: $\theta_i\approx \theta_j$ $\forall $  $i,j$.

In this work we focus on a two-network generalisation of Eq.(\ref{originalKura}), with double the defining equations, and additional coupling constants for internal and external network interactions. Also we add frustration parameters to the cross-network interaction interactions
take the form $\sin(\theta_i - \theta_j + \phi)$ if $i$ and $j$ belong to different populations. 
With no frustration within a population and, possibly asymmetric, frustrations between
populations there is competition between the two networks: members of the same population seek to phase synchronise while competing agents seek to frequency synchronise by a phase $\phi$. As in \cite{KallZup2015}, we refer to the populations as \textit{Blue} and \textit{Red} and individual phase oscillator at network nodes as agents. 
The asymmetry means that if Blue agents seek to be $\phi$ ahead of Red agents, the latter may seek a phase shift of $\psi$ with respect to Blue,
where $\phi\neq\psi$. 

In \cite{KallZup2015} we employed approximations in a fixed point analysis to derive thresholds for sharp changes in the dynamics.
Two fixed points were considered, one at which one population achieves near-phase synchronisation but places the other population at the point of dislocation from their competitors where there are two clusters in frequency synchrony. The other gives the threshold for internal fragmentation where there are three clusters in frequency synchrony.  
We now test these same thresholds for robustness against increasing noise strength when noise is applied to nodes in different sub-structures in the network by comparing numerical solution of the full non-linear 'Langevin' equation with analytic solutions in the linearised system using the Fokker-Planck equation.
An important tool is the Laplacian matrix for the various networks, leading to its eigenvalues, both zero and non-zero - or `normal' - modes, and eigenvectors. As in \cite{Zup2013} for the ordinary Kuramoto model, applying noise to these then triggers incoherence. We show that the approximations largely predict the behaviours in the numerical solution for the non-linear stochastic system. In particular, we show that noise applied separately to zero or normal modes leads to quite different dynamics elegantly understandable within the Fokker-Planck approach.

We explore a surprising behaviour, where noise on a tightly coupled population, whose deterministic interaction with the other population
triggers fragmentation of the latter, {\it enhances} synchronisation. Thus noise is not 
always a `nuisance', but may improve order, much like gentle rocking of the hand helps  balancing a vertical rod. 
Such phenomena has been observed elsewhere \cite{Kost2002,Aceb2007,Kawamura2014} but to our knowledge this is the first example of this 
arising from noise in a multi-network Kuramoto based system. Our `Blue-vs-Red' formalism is a special case of that of \cite{Kori2009,Kawamura2014b} though we additionally apply noise, and provide numerical illustrations of larger more complex systems.

First we summarise the Blue-vs-Red model and the linearisation around the two cluster fixed point. We then set up the stochastic model and its linearisation.
We present an illustrative case with Blue a tree network and Red a random graph used in \cite{KallZup2015}. We then separately examine the impact of noise on zero modes and non-zero, or normal, modes, 
deriving analytical results from the
Fokker-Planck approach near the two cluster fixed point, and then testing in turn with numerical simulation of the full non-linear
system. This enables understanding of the basic range of dynamics of the model. We then set up the formalism for analysing fragmentation and numerically solve to
show stochastic synchronicity. Using the three cluster {\it ansatz} we show how this behaviour arises from a zero mode.
The paper concludes with a discussion and outline of future work. Appendices provide more detail
on definitions and lengthy derivations.

\section{The deterministic frustrated two-network model: 'Blue vs Red'}\label{sec:det-BvsR}
\subsection{Model set-up and two cluster ansatz}
\label{lin-technique}
We briefly summarise here the deterministic Blue vs Red model from \cite{KallZup2015} but relegate further details relevant to this paper to Appendix A.
Consider $N$ Blue agents in a network given by adjacency matrix ${\cal B}_{ij},\ ( i,j \in \{1,\dots,N\}= {\cal B} )$, and $M$ Red agents 
in a network given by adjacency matrix ${\cal R}_{ij},\ (i,j \in \{1,\dots, M \} = {\cal R})$. Undirected networks are only considered here.
Each Blue agent $i \in {\cal B}$, has an associated phase $\beta_i$ giving the position in a limit cycle; similarly $\rho_i$ is the position in the limit cycle of a Red agent
$i\in{\cal R}$.
The `Blue vs Red model' is given by the system of equations:
\begin{equation} \label{BR-eq}
\begin{split}
\dot{\beta}_i = \omega_i - \sigma_B \sum_{j\in {\cal B}}{\cal B}_{ij} \sin(\beta_i-\beta_j) 
- \zeta_{BR} \sum_{j \in {\cal R}} {\cal A}^{(BR)}_{ij} \sin(\beta_i-\rho_j-\phi) ,\;\;  i \in {\cal B} \\
\dot{\rho}_i = \nu_i - \sigma_R \sum_{j \in {\cal R}} {\cal R}_{ij} \sin(\rho_i-\rho_j) 
- \zeta_{RB} \sum_{j \in {\cal B}} {\cal A}^{(RB)}_{ij} \sin(\rho_i-\beta_j-\psi), \;\;  i\in {\cal R}. 
\end{split}
\end{equation}
The matrices ${\cal A}^{(BR)}$ and ${\cal A}^{(RB)}$ represent the external links of Blue to Red agents (size $N\times M$)
and Red to Blue agents (size $M\times N$) respectively. For the remainder of this work we assume that they are the transpose of each other, 
so that the network between Blue and Red is symmetric.  
The $(N+M) \times (N+M)$ adjacency matrix for the corresponding external Blue-Red connections, labeled ${\cal M}$ has the following block off-diagonal form,
\begin{eqnarray*}
{\cal M} = \left( \begin{array}{cc}
0 & {\cal A}^{(BR)}\\
{\cal A}^{(RB)}& 0
\end{array} \right).
\end{eqnarray*}
The quantities $\omega_i, \nu_i$ 
give the natural frequencies of the associated Blue and Red agents respectively, and are typically drawn from some probability distribution.
Finally, $\sigma_B, \sigma_R, \zeta_{BR}, \zeta_{RB}$ are coupling constants, respectively, for intra-Blue, intra-Red, Blue to Red and Red to Blue. 
Asymmetry between Blue and Red potentially exists in the coupling constants and frustrations: $\zeta_{BR}$ need not equal $\zeta_{RB}$ and $\phi$ need not equal $\psi$. 

In general, the system in Eq.(\ref{BR-eq}) can only be solved numerically. To gain analytic insight in a region of greatest relevance - 
given that Blue and Red may deem internal phase synchronisation
ideal - we explore a fixed point given by the {\it ansatz}:
\begin{eqnarray}
\beta_i  =  B + b_i, \;\; \rho_j  =  P + p_j, \;\; \forall \; \{i,j\} \in \{ {\cal B},{\cal R}\}. \label{fixpoint1}
\end{eqnarray}
The variables $B, b_i, P, p_j$ are all time dependent, but $b_i,p_j$ are small fluctuations $b_i^2 \approx 0$,  $p^2_j \approx 0$. 
We remark that through Eq.(\ref{fixpoint1}) we now have a system of $N+M$ defining equations, but $N+M+2$ variables, however
$B$ and $P$ representing the centroids of the Blue and Red phases respectively are defined by 
\begin{eqnarray}
B= {1\over{N}} \sum_{i\in {\cal B}} \beta_i , \;\; P= {1\over{M}} \sum_{i \in {\cal R}} \rho_i \label{BPdef}
\end{eqnarray}
which completes the specification of the system. Consequently $\sum_{i\in{\cal B}} b_i = \sum_{i\in{\cal R}} p_i = 0$.
One of our main quantities of interest is the difference between each network's global phases, given by:
\begin{eqnarray}
\alpha \equiv B-P  \label{fixpoint-alph}
\end{eqnarray}
We refer to the phase locking within Blue, $\beta_i \approx \beta_j\; \forall\; i,j \in {\cal B}$, 
and Red $\rho_i \approx \rho_j\; \forall \; i,j \in {\cal R}$, as {\it internal} or {\it local locking}, and the
phase locking of Blue externally with to Red, $\beta_i \approx \rho_j \; \forall \;\{i,j\} \in \{{\cal B},{\cal R}\}$, as {\it external} or {\it global phase locking}.
However, if in this state $\alpha\neq 0$ then we may speak of external {\it frequency} locking. In this case, the distinguishability of the Blue and Red agents means we have two clusters visualising all oscillators as moving points on the unit circle, hence our name for the {\it ansatz}.

We now linearise the system using Eqs.(\ref{fixpoint1},\ref{fixpoint-alph}), keeping only terms first order in a Taylor expansion in
$b_i$ and $p_j$ giving
\begin{equation}\label{linsys}
\begin{split}
\dot{b}_i+\dot{B} \approx  \Omega_i - \sigma_B \sum_{j \in {\cal B}}L^{(B)}_{ij}b_j  - \zeta_{BR} \cos(\alpha-\phi) \sum_{j \in {\cal B} \cup {\cal R}} L^{(BR)}_{ij} v_j, \;\; i \in {\cal B}\\
\dot{p}_i+\dot{P} \approx \Omega_i - \sigma_R \sum_{j \in {\cal R}}L^{(R)}_{ij}p_j   - \zeta_{RB} \cos(\alpha+\psi) \sum_{j \in {\cal B} \cup {\cal R}} L^{(RB)}_{ij} v_j, \;\; i \in {\cal R}.
\end{split}
\end{equation}
where
\begin{eqnarray*}
v_i = \left\{ \begin{array}{cc} b_i & i \in {\cal B} \\ p_i & i \in {\cal R} \end{array} \right. , \;\; \Omega_i = \left\{ \begin{array}{cc}  \omega_i -  \zeta_{BR} d^{(BR)}_i \sin(\alpha - \phi) & i \in {\cal B}\\ \nu_i +  \zeta_{RB} d^{(RB)}_i \sin(\alpha + \psi)  & i \in {\cal R} \end{array}
\right. 
\end{eqnarray*}
The quantities $d$ and $L$, labelled with superscripts $B, \ R, \ BR$ and $RB$
represent respectively the {\it degree} and corresponding {\it Laplacian} matrices for the
Blue, Red, Blue-Red and Red-Blue networks using the matrix ${\cal M}$. These have been defined explicitly in \cite{KallZup2015} and in Appendix A; for brevity we do not repeat them here apart from pointing out that the Laplacian $L=D-A$, the difference of the degree and adjacency matrices \cite{Boll98}, whose eigenvalues
are important in the stability properties of many coupled dynamical systems \cite{Pec-Car98}. 

\subsection{Decoupling in the Blue and Red Laplacian eigen-basis}
\label{sec:decouple}
For analytical purposes we assume that 
\begin{eqnarray}
\sum_{j \in {\cal B} \cup {\cal R}} L^{(BR)}_{ij} v_j \approx \sum_{j \in {\cal B} \cup {\cal R}} L^{(RB)}_{ij} v_j \approx 0,
\label{interactionsaregone}
\end{eqnarray}
so that the equations for the fluctuations $b_i$ and $p_j$ in Eq.(\ref{linsys}) may be decoupled. This approximation does not completely hold even in regimes where linearisation might otherwise hold but we shall identify these points and where
behaviours deviate from expectations based on this.

We are thus able to employ the properties of graph Laplacians to decouple the resulting system of defining equations.
Specifically the Laplacians, $L^{(B)}$ and $L^{(R)}$ both contain a complete spanning set of orthonormal eigenvectors for the $N$ and $M$
dimensional subspaces, labeled by $e^{(B,r)}$, ($r=0,1,\dots, N-1 \in {\cal B}_E$) and $e^{(R,r)}$, ($r=0,1,\dots, M-1 \in {\cal R}_E$), respectively. 
Eigenvalues are denoted  $\lambda^{(B)}$ and  $\lambda^{(R)}$. We distinguish between indices in node space $\{ {\cal B}, {\cal R} \}$, 
and those in eigen-mode space $\{ {\cal B}^E, {\cal R}^E \}$ (which has the same dimensionality)
by reserving labels $\{i,j\}$ for expressions involving graph nodes, and labels $\{r,s\}$ for expressions involving Laplacian eigen-modes.

Importantly, the spectrum of eigenvalues is bounded by an eigenvalue zero that has degeneracy according to the number of components of the
graph \cite{Boll98}. We assume here that Blue and Red networks each consist of one component. Up to normalisation, the corresponding zero eigenvectors ${\vec e}^{(B,0)}$
and ${\vec e}^{(R,0)}$ consists of all unit valued entries.
Thus, $B$ and $P$ as defined in Eq.(\ref{BPdef}) are the zero-mode projections of the phases $\beta_i$ and $\rho_j$. 
Analogously, we denote by $x_r$ and $y_s$ the projections of $b_i$ and $p_i$ on the Blue and Red non-zero eigenvectors. We give explicit expressions in Appendix A.
The eigenvectors corresponding to low-lying Laplacian eigenvalues are known generally to expose underlying structures in a network \cite{Fied73,Ding-et-al01}.

Applying the approximation Eq.(\ref{interactionsaregone}) and the eigenvector projections in Eq.(\ref{linsys}), and taking advantage of the orthonormality of the eigenvectors, yields the system,
\begin{equation}\label{integrablesystem}
\begin{split}
\dot{x}_r = q^{(B)}_r(x_r,\alpha), \;\; r \in {\cal B}^E / \{0\}, \;\; \dot{y}_s = q^{(R)}_s(y_s,\alpha), \;\; s \in {\cal R}^E / \{0\}\\
\dot{B} = \bar{\omega} -\frac{\zeta_{BR}d^{(BR)}_T}{N}\sin(\alpha-\phi), \;\; \dot{P} = \bar{\nu} +\frac{\zeta_{RB}d^{(BR)}_T}{M}\sin(\alpha+\psi),
\end{split}
\end{equation}
where,
\begin{eqnarray}
q_r^{(B)}(x_r,\alpha) &\equiv& \omega^{(r)}-\sigma_B \lambda^{(B)}_r x_r - \zeta_{BR}d^{(BR)}_r\sin(\alpha-\phi), \nonumber \\
q_s^{(R)}(y_s,\alpha) &\equiv& \nu^{(s)}-\sigma_R \lambda^{(R)}_s y_s + \zeta_{RB}d^{(RB)}_s \sin(\alpha+\psi), 
\label{drifts}
\end{eqnarray}
and $\omega^{(r)}$ and $d^{(BR)}_r$ the projections onto the $r-$th eigenvector, and $\bar{\omega}$ the average over ${\cal B}$ 
(with $d^{(RB)}_s$ the projections onto the $s-$th eigenvector for Red).

The advantage of the structure of the system in Eq.(\ref{integrablesystem}) is that it allows now $\alpha$ to be solved first, and then
used in the forcing terms of the normal mode equations which are otherwise linear and solvable in their own right.
Taking the difference of $\dot{B}$ and $\dot{P}$ gives the
dynamics of $\alpha$,
\begin{eqnarray}
\dot{\alpha} = - V'(\alpha), \;\; \textrm{where}\;\; V(\alpha) = -\mu \alpha -\sqrt{S^2+C^2}\cos\left( \alpha -  \varrho \right), \ \mu =(\bar{\omega}-\bar{\nu})
\label{alphaeq}
\end{eqnarray}
for $\varrho = \tan^{-1} \left( S/C\right)$ and,
\begin{eqnarray*}
C  \equiv  d^{(BR)}_T \left(\frac{ \zeta_{BR} \cos\phi}{N} +\frac{ \zeta_{RB} \cos\psi}{M}\right) ,\;\; S \equiv d^{(BR)}_T \left(  \frac{ \zeta_{BR}\sin\phi}{N}  -  \frac{\zeta_{RB} \sin\psi   }{M}\right).
\end{eqnarray*}
The substitution, $\alpha = \varrho + 2 \tan^{-1} \left( \vartheta + \frac{\sqrt{S^2+C^2}}{\mu} \right)$, leads to the solution
\begin{eqnarray} 
\alpha(t) = \varrho+ 2 \tan^{-1} \left\{ \sqrt{\frac{S^2+C^2}{\mu^2}} + \sqrt{\frac{ {\cal K}}{\mu^2} } \tanh \left( \frac{\sqrt{{\cal K}}}{2} (\textrm{const} - t)  \right) \right\}
\label{alphasol}
\end{eqnarray}
where $\textrm{const} = \frac{2}{\sqrt{{\cal K}}} \textrm{tanh}^{-1}\left\{ \frac{\mu}{\sqrt{{\cal K}}}\vartheta_0 \right\}$, encodes the initial conditions for $\alpha$, and ${\cal K}$ is given by
\begin{eqnarray}\label{eq:special-K}
{\cal K} = C^2 + S^2 -\mu^2. 
\end{eqnarray}
The solutions to the normal modes are given in Appendix A. Ultimately their dynamics depends on the behaviour of $\alpha(t)$ which in turn is governed by
the potential $V(\alpha)$ in Eq.(\ref{alphaeq}). This is commonly referred to as a \textit{tilted periodic} \cite{Lin-Kos-Sch2001} or \textit{tilted Smoluchowski-Feynman ratchet} \cite{Reimann2002} potential. Importantly, we have periodicity, $V'(\alpha) = V'(\alpha+ 2\pi)$, and the tilt refers to the constant forcing term in $V'(\alpha)$ given by 
the difference of frequency averages $\mu$. The sign of ${\cal K}$ is critical. If ${\cal K} > 0$ the solution asymptotes to the value $\varrho+ \sin^{-1} \left( \frac{\mu}{\sqrt{S^2 + C^2}} \right)$ mod $2 \pi$. If ${\cal K} < 0$ then the $\tanh$ in Eq.(\ref{alphasol}) becomes a 
tangent, and the solution is oscillatory with period $\frac{2 \pi}{\sqrt{ |{\cal K}| }}$. 
Correspondingly, for ${\cal K} > 0$, $V(\alpha)$ is a series of local maxima (unstable fixed points at $\alpha = \pi + \varrho - \sin^{-1} \left( \frac{\mu}{\sqrt{S^2 + C^2}} \right)+2 \pi n, \; n \in \mathbb{Z} $) and local minima (stable fixed points at $\alpha = \varrho+ \sin^{-1} \left( \frac{\mu}{\sqrt{S^2 + C^2}} \right)+2 \pi n, \; n \in \mathbb{Z}$), on a landscape which has an overall slope according to the sign of $\mu$.
For ${\cal K} = 0$ the hills and valleys of the potential become points of inflection, and hence unstable fixed points. For ${\cal K} <0$, the potential loses all of its fixed points, even the unstable ones.

\section{The frustrated two-network model with noise}
For each graph node of the Blue and Red networks, we apply additive noise $\Lambda^{(B)}_i(t), \Lambda^{(R)}_i(t)$ (effectively a time-dependence for the frequencies)
to the full Blue-Red system, Eqs.(\ref{BR-eq}). 
The noise may be decomposed in terms of eigenvectors of the Blue/Red Laplacians
\begin{equation}\label{eq:noise-terms-sums}
\begin{split}
\Lambda_i^{(B)}(t)=\sum_{r\in{\cal B}_E }{\gamma_r^{(B)}e_i^{(B,r)}\eta_r^{(B)}(t)},\;\;i\in\mathcal{B}, \\
\Lambda_i^{(R)}(t)=\sum_{r\in{\cal R}_E }{\gamma_r^{(R)}e_i^{(R,r)}\eta_r^{(R)}(t)},\;\;i\in\mathcal{R}
\end{split}
\end{equation}
where $\eta_r^{(B)}$ and $\eta_r^{(R)}$ are uncorrelated Gaussian White Noise (GWN) terms with $\gamma_r^{(B)},\gamma_r^{(R)}\in \{0,1\}$
and variance $\Omega > 0$, namely $\langle\eta_r\rangle = 0$ and
$\langle \eta_r(t)\eta_{r'}(t')\rangle = \Omega \delta_{rr'} \delta(t-t')$. We also refer to $\Omega$ as the noise strength.
Then analogously to Sec.\ref{sec:det-BvsR}, we obtain the following Langevin system
\begin{equation}\label{eqn:integrablesystem-w-noise}
\begin{split}
\dot{x}_s &= q_s^{(B)}(x_s,\alpha)+\gamma_s^{(B)}\eta_s^{(B)}, \;\; s \in {\cal B}^E / \{0\}\\
\dot{y}_s &= q_s^{(R)}(y_s,\alpha)+\gamma_s^{(R)}\eta_s^{(R)}, \;\; s \in {\cal R}^E / \{0\}\\
\dot{\alpha} &= -V'(\alpha)+\gamma_0^{(B)}\eta_0^{(B)} -\gamma_0^{(R)}\eta_0^{(R)},
\end{split}
\end{equation}
where the variables and parameters are as used in Eq.\eqref{integrablesystem}. 

As advertised, we seek to test the consequences of noise applied separately, and in combination, to zero and normal modes.
This is achieved by selecting either zero or one for the $\gamma$ parameters. We consider three cases:
\begin{itemize}
\item{$\gamma_0^{(B)}=\gamma_0^{(R)}=0$, $\gamma_r^{(B)}=\gamma_s^{(R)}=1$, $\{r,s \} \in \{ {\cal B}^E / \{0\}, {\cal R}^E / \{0\}  \}$ has noise applied to normal modes;} 
\item{$\gamma_0^{(B)}=\gamma_0^{(R)}=1$, $\gamma_r^{(B)}=\gamma_s^{(R)}=0$, $\{r,s \} \in \{ {\cal B}^E / \{0\}, {\cal R}^E / \{0\}  \}$ has noise applied to zero modes; and}
\item{$\gamma_r^{(B)}=\gamma_s^{(R)}=1$, $\{r,s \} \in \{ {\cal B}^E , {\cal R}^E \}$ has all noise applied to all modes}.
\end{itemize}

\subsection{Fokker-Planck equations}
We denote by $\calP$ the probability that a random variable $X$ lies in the range $x\leq X\leq x+\mathrm{d}x$ in the stochastic system of Eq.\eqref{eqn:integrablesystem-w-noise}.
The initial state $(\vec{X},\vec{Y},A)=(\vec{x}',\vec{y}',\alpha')$ at time $t=0$
is
$$\calP(\vec{x},\vec{y},\alpha,0)=\delta(\vec{x}-\vec{x}')\delta(\vec{y}-\vec{y}')\delta(\alpha-\alpha')$$
where $\vec{x}$ and $\vec{y}$ represent the vectors, in the Laplacian basis, of components $x_r$ and $y_r$, and $\alpha$ represents the difference between the zero modes. 
From the Langevin equations Eq.(\ref{eqn:integrablesystem-w-noise}) using the It\^{o} interpretation one can immediately construct the Fokker-Planck equation for the 
evolution of the \textit{joint} probability density \cite{Risken,Schuss} $\calP (\vec{x},\vec{y},\alpha,t)$. Nevertheless, apart from being quite unwieldy, the joint distribution is not helpful for our requirements. 
Instead, when we come to consider probability density functions, we shall rely on the \textit{marginal} and \textit{conditional} densities of the zero, and normal modes respectively (see Chap. 1 of \cite{Kloeden} and Chap. 2 of \cite{Risken}). To enable this we rely on the fact that although $\alpha$ appears as a common term in the normal mode Langevin equations, the zero mode Langevin equation for $\alpha$ is independent of the modes $x_r$ and $y_s$. 

Intuitive as the Fokker-Planck equation is, it is difficult to determine the stability of the non-linear dynamical system \cite{Zup2013}. Because of the linearisation for Eq.(\ref{eqn:integrablesystem-w-noise}), we may not trust behaviour for arbitrarily large $x_r, y_r$ and $\alpha$
and we lack clear bounds for the basin of attraction for the phase synchronised fixed point. Thus solutions ${\cal P}$ exhibiting significant tails
in these regimes imply instability only insofar as there is a non-negligible probability that the non-linear terms will be non-vanishing. 
For the deterministic Kuramoto model \cite{Zup2013}
this requires that $\frac{\omega^{(r)}}{\sigma\lambda_r}$ must be small to be close to the linearised regime; for the stochastic Kuramoto system
with additive Gaussian noise the requirement is for small variance $\frac{\Omega}{2\sigma\lambda_r}$ of the Fokker-Planck density.  

\subsection{Example networks, frequencies and couplings}
To illustrate our otherwise general solutions and compare to numerical
simulation we consider an example of Blue agents forming a hierarchy and Red agents on a random network.
As in \cite{KallZup2015}, for ${\cal B}$ we consider a tree graph, namely a complete 4-ary tree, thus setting $N=21$.
We use a random Erd\"os-R\'enyi network also of $N=21$ with link of probability $0.4$
for ${\cal R}$.
The two networks, and further details of their interconnection, are shown in Appendix C in Fig.\ref{fig:BvsR-networks}.

The specific frequencies of each agent were drawn from a uniform distribution between
zero and one, $\omega_i,\nu_j\in [0,1]$ and are also plotted in Appendix C. In the examples 
used here, the average frequencies turn out to be $\bar{\omega}=0.503, \bar{\nu}=0.551$, 
giving a slightly negative $\mu=-0.048$. Thus if 
cross-couplings were set to zero the Red population would lap Blue over time. This is reflected in a negative slope for the tilted ratchet potential $V(\alpha)$.

Finally, as in \cite{KallZup2015} we choose the couplings
\begin{eqnarray}
\sigma_B=8, \sigma_R=0.5, \zeta_{BR}=\zeta_{RB} = 0.4. \label{coupls}
\end{eqnarray}
For the networks in Fig.\ref{fig:BvsR-networks}, these will give high internal phase synchronisation - the measure for which is given momentarily - but
allow for some changes in dynamics as the frustrations $\phi$ and $\psi$ are varied. In \cite{KallZup2015}, we also set $\psi=0$ but varied $\phi$.
For the deterministic system, this means a change in the sign of ${\cal K}$ from positive to negative at a point that, to four significant figures, is $\phi=0.9498\pi$. Thus in the vicinity
of this point, in the absence of
noise, for $\phi$ less than this value the system stabilises to Blue a fixed angle ahead of Red, and for $\phi\geq 0.95\pi$ Blue and Red remain internally
phase synchronised but lapping each other with a period that decreases as $\phi$ gets larger. Eventually the periodicity with respect
to $\phi$ and $\psi$ implicit in Eq.(\ref{eq:special-K}) will manifest itself.
When we explore below numerical examples on either side of ${\cal K}=0$, at typical points $\phi=0.5\pi$ and $\phi=0.95\pi$, we ask: 
how does noise change the deterministic behaviours?

\subsection{Measures of synchronisation}
To measure the degree of synchronisation within a given population we use
local forms of Kuramoto's order parameter \cite{Kura1984}
\begin{eqnarray*}
O_B = \frac{1}{N}\left|  \sum_{j\in {\cal B}} e^{i\beta_j} \right|, \;\; O_R = \frac{1}{M} \left| \sum_{j\in {\cal R}} e^{i\rho_j} \right|, O_{R_k} = \frac{1}{M_k} \left| \sum_{j\in {\cal R}_k} e^{i\rho_j} \right|,\;\; k \in \{1,2\}.
\end{eqnarray*}
The third, $O_{R_k}$, examines the degree of synchronisation within the Red population for two sub-clusters, where ${\cal R}_1$ 
has Red agents interacting with a Blue agent.
We emphasise that in many of our examples in the following the total system order parameter 
$1/(N+M) | \sum_{i\in{\cal B}} e^{i\beta_i} +\sum_{j\in{\cal R}} e^{i\rho_j}|$
will be far from the value one.

\section{Noisy normal modes}
\subsection{Analytical considerations}
We consider first the case $\gamma_0^B=\gamma_0^R=0$. 
Consequently, $\alpha$ becomes deterministic with solution Eq.\eqref{alphasol}, tending to a fixed value as $t\to\infty$ or periodic, 
according to the sign of ${\cal K}$ in Eq.\eqref{eq:special-K}. As $\alpha$ is no longer a random variable all the normal mode Langevin equations in Eq.(\ref{eqn:integrablesystem-w-noise}) are independent. Thus the corresponding Fokker-Planck equation can be decomposed into a product of $N+M-2$ densities,
$$\calP(\vec{x},\vec{y},t)=\prod_{r\in{\cal B}^E / \{0\}}{\calP^{(B,r)}(x_r,t)}\prod_{s\in{\cal R}^E / \{0\}}{\calP^{(R,s)}(y_s,t)}$$
leading to a decoupling into separate Fokker-Planck equations for each mode
\begin{equation}\label{eq:FP-norm-mode}
\begin{split}
&\frac{\partial \calP^{(B,r)}}{\partial t}=\frac{\Omega}{2}\frac{\partial^2}{\partial x_r^2}\calP^{(B,r)}-\frac{\partial}{\partial x_r}\{q_r^{(B)}(x_r,\alpha(t))\calP^{(B,r)}\},\;r\in{\cal B}^E / \{0\}\\
&\calP^{(B,r)}(x_r,0)=\delta(x_r-x_r').
\end{split}
\end{equation}
We obtain a similar set of Fokker-Planck equations for the non-zero modes $y_s$. These equations are the direct analogue of 
the system solved in section 3.2.1 of \cite{Zup2013} for the pure Kuramoto model.

Using a simple variable transformation as given in Sec.(1.8.3.6) of \cite{Polyanin2002} we obtain the following time dependent solution for Eq.\eqref{eq:FP-norm-mode}.
\begin{equation}
\label{eq:FP-norm-mode-sol}
\calP^{(B,r)}(x_r,t)=\sqrt{\frac{\sigma_B\lambda_r^{(B)}}{\Omega\pi (1-\re^{-2\sigma_B\lambda_r^{(B)}t})}}\exp \left(-\frac{\sigma_B\lambda_r^{(B)}}{\Omega }\frac{z(t)^2}{1-\re^{-2\sigma_B\lambda_r^{(B)}t}}\right),
\end{equation}
where
$$z(t)=x_r-x_r'-\int_0^t{\mathrm{d}\tau\,q_r^{(B)}(x_r',\alpha(\tau))\re^{\sigma_B\lambda_r^{(B)}(\tau-t)}}.$$
The statistical properties of mean, mode and variance are straightforwardly extracted from this result, where the mean
is
%
%
equivalent to the
solution to the deterministic equation (given in Appendix A). We note from Sec.\ref{sec:det-BvsR} that if $\mathcal{K}<0$ then this solution decays to a time periodic solution and furthermore we can see that the variance approaches $\Omega/(2\sigma_B\lambda_r^{(B)})$ for large time. If however we have that $\mathcal{K}>0$, then as noted in Sec.\ref{sec:det-BvsR}, $\alpha$ tends to a stable fixed point and the solution approaches a steady-state form:
%
\begin{equation*}
\calP_{\mathrm{st}}^{(B,r)}(x_r)=\lim_{t\to\infty}\calP^{(B,r)}(x_r,t)=\sqrt{\frac{\sigma_B\lambda_r^{(B)}}{\Omega\pi}}\exp \left(-\frac{\sigma_B\lambda_r^{(B)}}{\Omega }[x_r-\lim_{t\to\infty}\langle x_r \rangle]^2\right).
\end{equation*}

Basically, we have a Gaussian distribution whose mean is contingent on the deterministic behaviour of $\alpha$: decaying to a constant over time for ${\cal K}>0$, 
or decaying to an oscillation for ${\cal K}<0$. We may anticipate several regimes of behaviour here. For large couplings $\sigma_B$ and/or $\sigma_R$, or 
large Laplacian modes, where `large' means
on the scale of the noise variance $\Omega$, we obtain densities close to
the deterministic solution with small variance - and thus a narrow Gaussian smearing around the behaviour either at fixed $\alpha$ for ${\cal K}>0$, or
periodic $\alpha$ for ${\cal K}<0$. However, once the couplings become too small or alternatively we examine low-lying Laplacian modes while ${\cal K}>0$, the variance becomes large consistent with a long tail in the density. This is one mechanism by which modes may lie outside the basin of attraction with non-zero probability. The other
mechanism is simply that the centre of the distribution lies away from the origin so that the tails again fall outside the basin. The consequences of this are not described in the
approximation: non-linearities may switch on and the system may appear to fragment - with clusters jumping around the unit circle in relation to the main system of frequency locked Blue and Red agents.  However, these same non-linearities that led to the initial appearance of synchronisation recur and the system may resynchronise. The solution itself does not describe this process because
of its limitation to the linearised regime. Note that such jumps would appear in $\alpha$ but would not have any particular periodicity if ${\cal K}>0$ because the 
escape from the basin of attraction is purely probabilistic.
Contrastingly, in the regime of ${\cal K}<0$ only periodic jumps in $\alpha$ would be expected. However, here because of the oscillation in the mean of
the density in Eq.(\ref{eq:FP-norm-mode-sol}) there may be slight deviations as, correspondingly, the tail of the distribution reaches larger values of $x_r$ or $y_r$.
Thus non-periodic fragmentations may appear in individual Laplacian modes on top of the basic oscillation with respect to $\alpha$.

\subsection{Numerical simulations}
We now solve the full system Eq.(\ref{BR-eq}) with noise added and $\gamma_0^B=\gamma_0^R=0$ and $\gamma_r^B=\gamma_s^R=1$, for all $r\in\mathcal{B}^E/\{0\}$ and $s\in\mathcal{R}^E/\{0\}$. 
Unless otherwise stated we set $\Omega=1$ but comment on behaviours at lower values.
Numerical solution is performed in Mathematica implementing the noise Eq.\eqref{eq:noise-terms-sums} using the ``ItoProcess'' and ``RandomFunction'' capabilities, taking 50 paths of the simulation, step sizes of 0.025 over the time interval 0 to 1000. These simulations were run for varying values of the frustration parameter $\phi$ and the variance $\Omega$ while keeping initial conditions fixed. The results were then aggregated by taking the mean, median and upper and lower quartiles of $O_{B,R}$ and $\alpha$. The first five paths of each simulation were plotted against the relevant deterministic solutions, which were produced via use of the ``NDSolve'' function using the ``Stiffness Switching'' method in Mathematica.

\begin{figure}
\centering
\subfloat[][Order Parameters for Red and Blue]{\includegraphics[width=.9\linewidth]{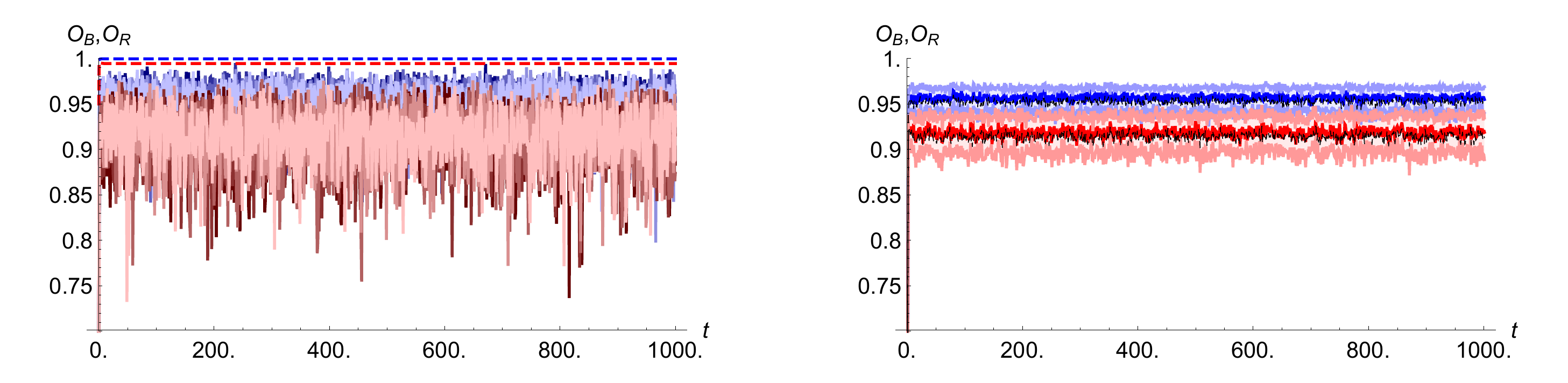}
\label{subfig:Normal-Order12}}

\subfloat[][Deterministic and Langevin simulations of $\alpha$]{\includegraphics[width=.9\linewidth]{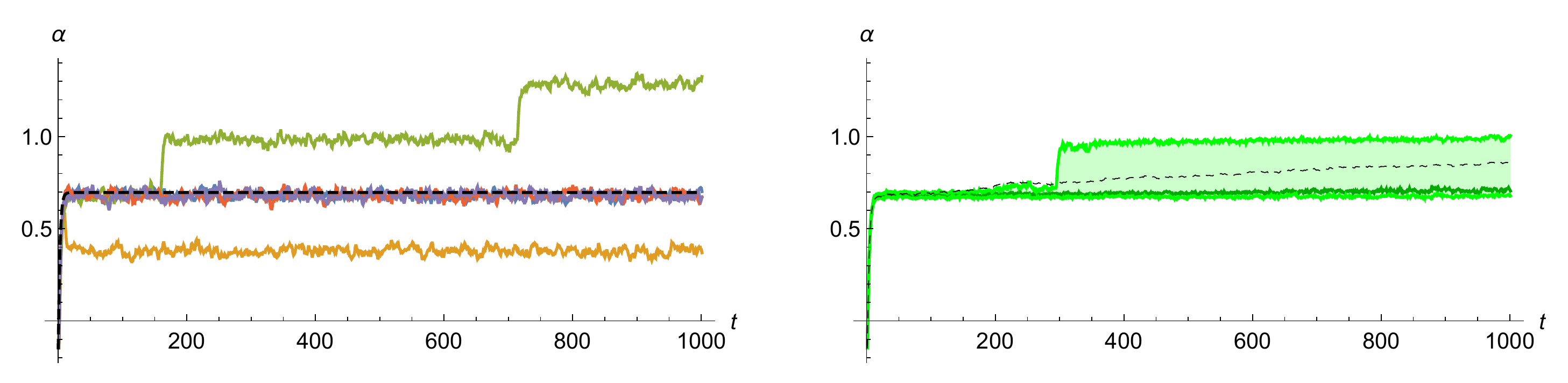}
\label{subfig:Normal-Alpha12}}

\subfloat[][Simulations of the $r=1,10,20$ modes]{\includegraphics[width=.9\linewidth]{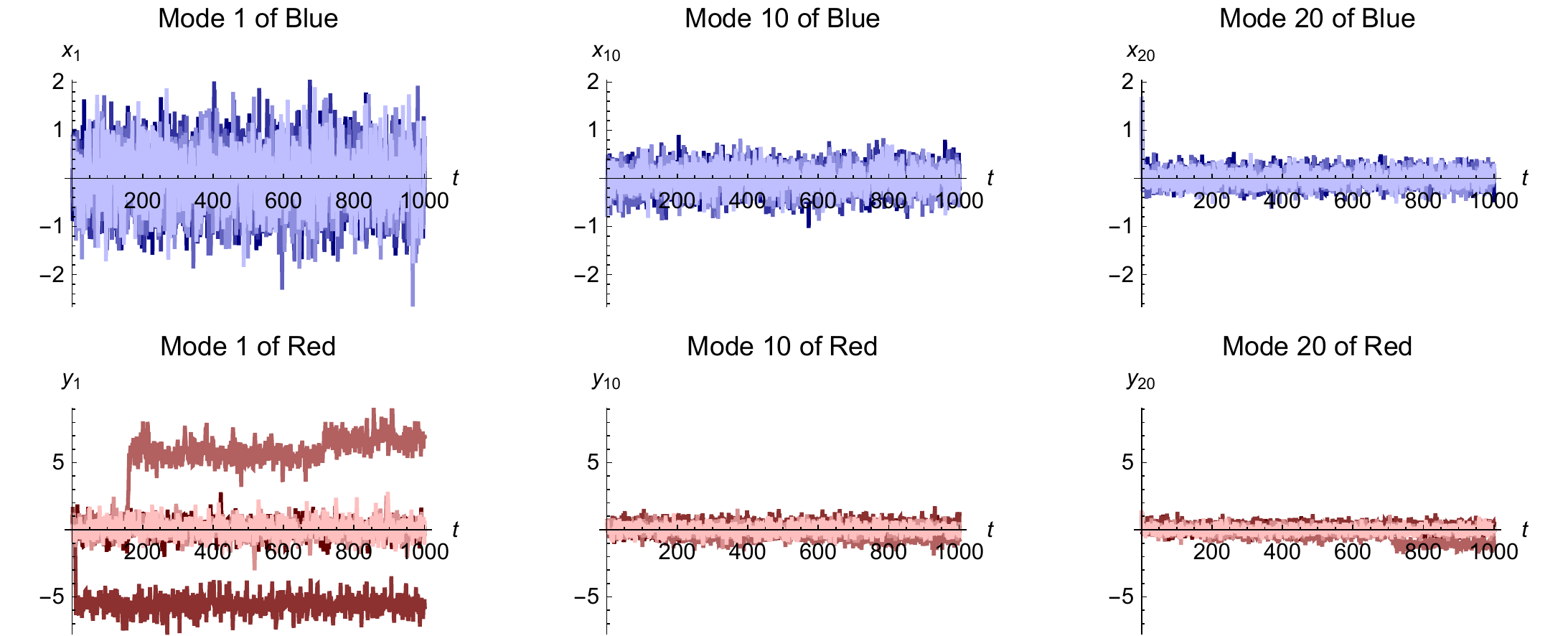}
\label{subfig:Normal-Modes12}}

\caption{Simulations of  Eq.\eqref{BR-eq} with and without noise for the Blue tree network
and Red random network, with white noise with $\sqrt{\Omega}=1$ on the normal modes $x_r$ and $y_r$ and parameters $\sigma_B=8,\,\sigma_R=0.5,\,\zeta_{BR}=\zeta_{RB}=0.4,\,\phi=0.5\pi,\,\psi=0$: Fig.\protect\subref{subfig:Normal-Order12} (Left) displays the deterministic local synchronisation order parameter $O_B$ (dashed blue) and $O_R$ (dashed red) and five paths of the Langevin local synchronisation order parameter $O_B$ and $O_R$, respectively, (Right) displays the median (blue/red), upper and lower quartile (light blue/red) and mean (dashed black) of 50 simulation of  $O_B$ and $O_R$, respectively. Fig.\protect\subref{subfig:Normal-Alpha12} (Left) displays deterministic $\alpha$ (Eq.\eqref{fixpoint-alph}) and 5 paths of the Langevin simulations for $\alpha$, (Right) displays the median (green), upper and lower quartile (light green) and mean (dashed black) of 50 simulation of $\alpha$. Fig.\protect\subref{subfig:Normal-Modes12} contains plots of the $r=1,10,20$ modes for 5 paths of the Langevin simulations. The top row contains the Blue modes and the bottom row contains the Red modes.}
\label{fig:Normal-Frac05-StdDev-1}
\end{figure}

In Fig.\ref{fig:Normal-Frac05-StdDev-1}, we  show the behaviour as a function of time of $O_{B,R}$, $\alpha$ and individual modes
$r=1, 10$ and $20$ going up the Laplacian spectrum, with $\phi=0.5\pi$, namely ${\cal K}>0$. The order parameters $O_{B,R}$ show reasonably high internal synchronisation, though $O_R$ shows stronger fluctuations than $O_B$. The plot for $\alpha$ shows constant behaviour, but with discrete jumps. However, a plot of the median value of $\alpha$  (middle, right plot) shows no indication of
periodicity. The behaviour of $y_1(t)$ explains the picture: some instances shows jumps, consistent with this lowest mode leaving the basin of attraction but then
returning. The Blue network mode, $x_1(t)$, shows no such behaviour - and higher modes correspondingly are absent of such behaviour.
All this is consistent with Eq.(\ref{eq:FP-norm-mode-sol}) in the regime with ${\cal K}>0$. 
If the noise strength $\Omega$ were reduced the jumps in $\alpha$ and the red Laplacian modes would disappear and the fluctuations in $O_R$ would reduce; for brevity we omit such plots here.

The fact that the Red modes show fragmentation
at stronger noise is noteworthy. For these cases, the variables controlling the variance of the Gaussian in Eq.(\ref{eq:FP-norm-mode-sol})
is the product of the internal coupling and Laplacian eigenvalue. For the lowest mode $r=1$, numerically
these turn out to be nearly the same, $\sigma_B \lambda_r^{(B)}=1.36, \sigma_R \lambda_r^{(R)}=1.36$
(by deliberate choice of $\sigma_B$, to balance the smaller lowest eigenvalues
for the tree network compared to the random - see Fig.\ref{spectra}). 
The significant difference between the Blue and Red densities here lies in the numerical values of the quantities $d^{(BR)}_r$ and
$d^{(RB)}_r$ appearing in $z(t)$ in  Eq.(\ref{eq:FP-norm-mode-sol}) and in turn in the drift terms $q^{(B)}_r$ and $q^{(R)}_r$ in Eqs.(\ref{drifts}).
These quantities will recur throughout our analysis. For the $r=1$ modes these are of order of magnitude $O(10^{-15})$ for Blue and
$O(10^{-1})$ for Red. The impact of these values here is in locating the centre of the densities: for Blue close to the origin, for Red
further away. Thus, though both modes' densities exhibit similar diffusivity, that for Red is further from the origin so that the tail of the 
density may fall outside the basin of attraction - triggering the onset of nonlinearities and fragmentation as described.

\begin{figure}
\centering
\subfloat[][Order Parameters for Red and Blue]{\includegraphics[width=.9\linewidth]{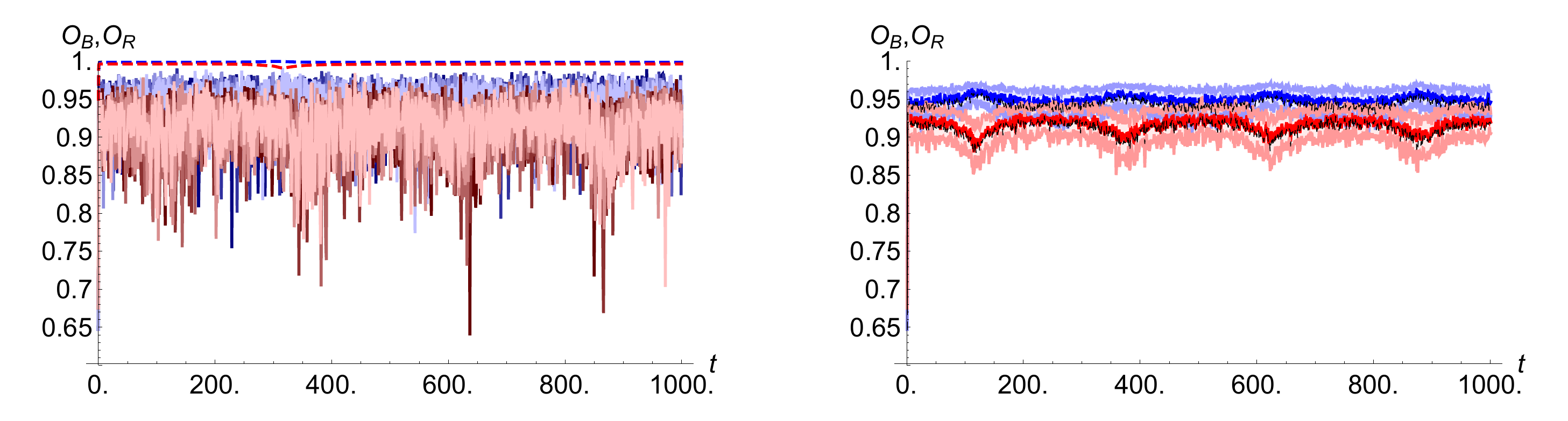}
\label{subfig:Normal-Order14}}

\subfloat[][Deterministic and Langevin simulations of $\alpha$]{\includegraphics[width=.9\linewidth]{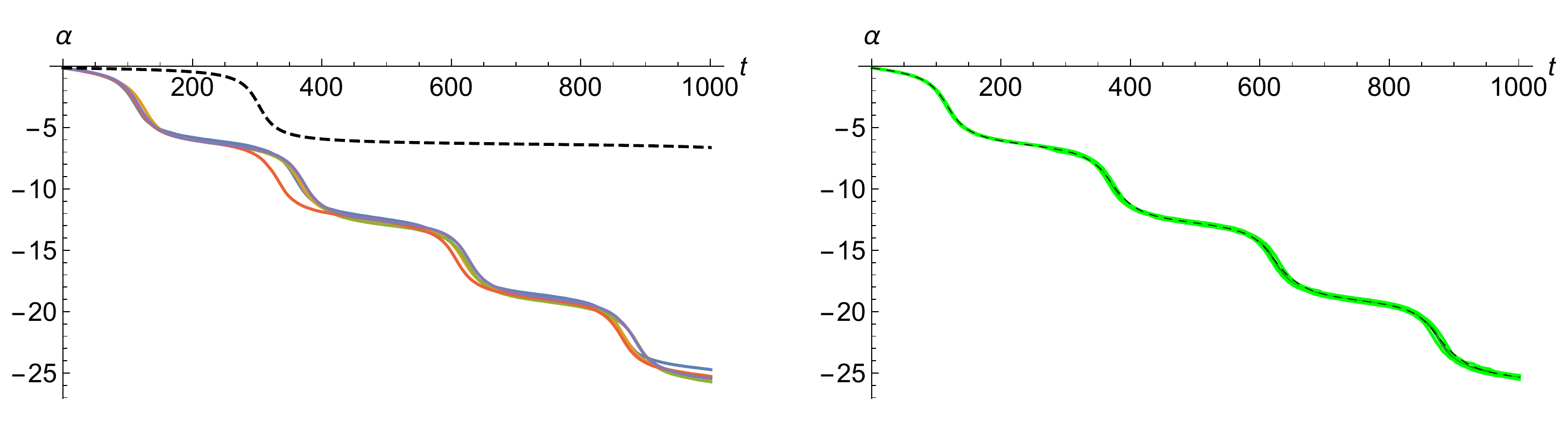}
\label{subfig:Normal-Alpha14}}

\subfloat[][Simulations of the $r=1,10,20$ modes]{\includegraphics[width=.9\linewidth]{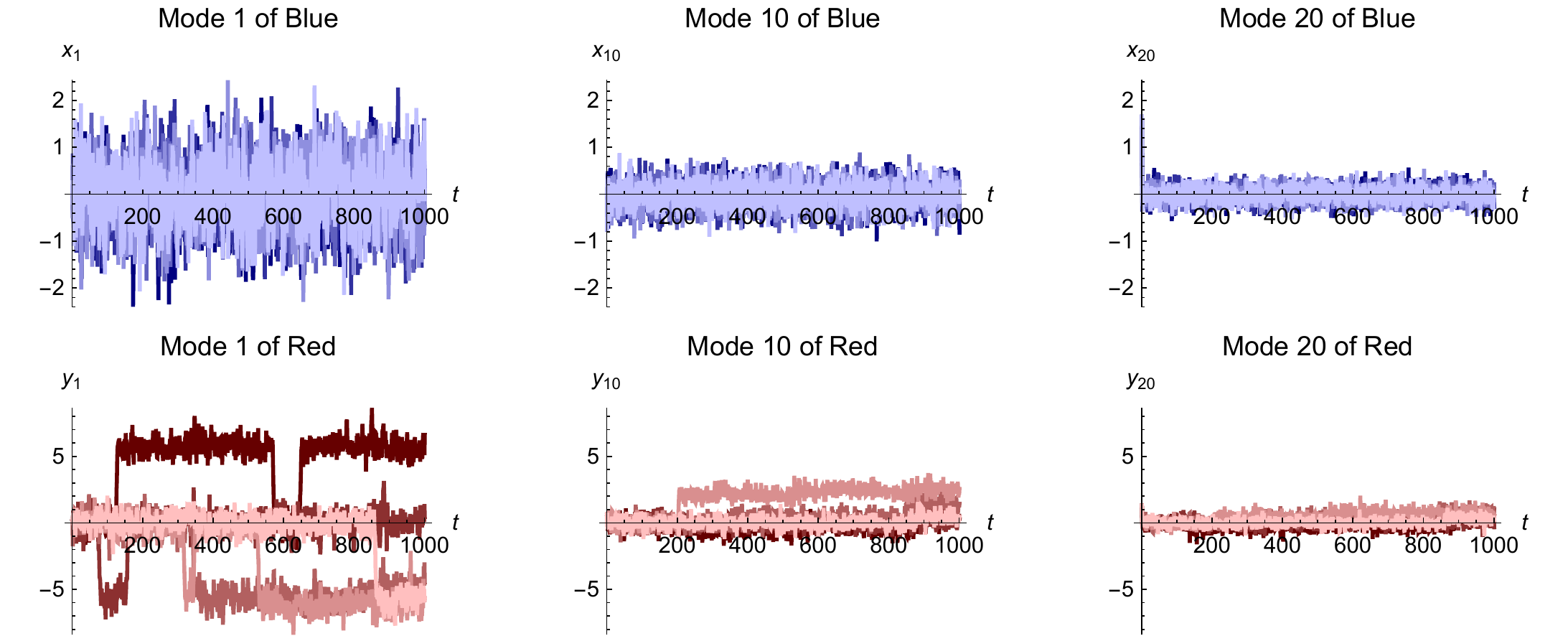}
\label{subfig:Normal-Modes14}}

\caption{Simulations of Eq.\eqref{BR-eq} with and without noise for the Blue tree network
and Red random network, with white noise with $\sqrt{\Omega}=1$ on the normal modes $x_r$ and $y_r$ and parameters $\sigma_B=8,\,\sigma_R=0.5,\,\zeta_{BR}=\zeta_{RB}=0.4,\,\phi=0.95\pi,\,\psi=0$: Fig.\protect\subref{subfig:Normal-Order14} (Left) displays the deterministic local synchronisation order parameter $O_B$ (dashed blue) and $O_R$ (dashed red) and five paths of the Langevin local synchronisation order parameter $O_B$ and $O_R$, respectively, (Right) displays the median (blue/red), upper and lower quartile (light blue/red) and mean (dashed black) of 50 simulation of  $O_B$ and $O_R$, respectively. Fig.\protect\subref{subfig:Normal-Alpha14} (Left) displays deterministic $\alpha$ (Eq.\eqref{fixpoint-alph}) and 5 paths of the Langevin simulations for $\alpha$, (Right) displays the median (green), upper and lower quartile (light green) and mean (dashed black) of 50 simulation of $\alpha$. Fig.\protect\subref{subfig:Normal-Modes14} contains plots of the $r=1,10,20$ modes for 5 paths of the Langevin simulations. The top row contains the Blue modes and the bottom row contains the Red modes.}
\label{fig:Normal-Frac095-StdDev-1}
\end{figure}

Now we set $\phi=0.95\pi$ where, deterministically, ${\cal K}=0^-$ by which we mean marginally below the value required
for a steady state solution. In Fig.\ref{fig:Normal-Frac095-StdDev-1} we show the analogue of Fig.\ref{fig:Normal-Frac05-StdDev-1}.
We observe a distinct periodicity for both order parameters $O_{B,R}$ and $\alpha$, though
the periodicity in the order parameters is best detected in the mean and median. 
At weaker noise values the fluctuations in the order parameters is naturally reduced (again, not shown for brevity).
On the other hand, the behaviour of $\alpha$ is rather smooth and consistent even for individual paths. 
The period of the oscillation is shorter than that for the deterministic case - shown as a dashed line in  Fig.\ref{fig:Normal-Frac095-StdDev-1} (b) -
though for weak noise the periodicity is consistent with that of the deterministic case.
Nonetheless, from order parameters and $\alpha$ we may conclude that periodically
the Red cluster undergoes a rapid rotation with respect to Blue. This leads to subtle effects in the order parameters: Blue appears to improve in synchronisation, $\langle O_B \rangle \rightarrow 1$, as a consequence of the red agents interacting with them pass by leading to a `bunching up' of phases. Subsequently the red population rotates away so that those Red agents interacting with Blue momentarily `splays , caught between their partners and Blue, hence the slight drop in Red's degree of synchronisation in the plot of $\langle O_R\rangle $ (top, left). 
The individual Laplacian modes, particularly $y_1(t)$ show that some jumping - fragmentation - is occuring, but also the fluctuations are understandably stronger. Indeed, decreasing $\Omega$ removes the fragmentation effects allowing the cycling of 
Red to be visible in $y_1(t)$.

This behaviour is consistent with what we might expect from our analytical considerations: the underlying deterministic periodicity, overlaid now with occasional fragmentation because the tail of the associated Fokker-Planck density cyclically visits regions of large values for $y_s(t)$ thus allowing for 
{\it periodic} fragmentation of the system.
The differences between individual paths for $\alpha$ confirm a degree of stochasticity so that deterministic dynamics based on the linearisation here cannot wholly explain the behaviour. The most glaring surprise is the faster period of the system but nevertheless, for some stochasticity, remarkably robust.
Because the order parameters reach relatively low values, for example $O_R\approx 0.65$, the system regularly reaches regions where a two-cluster (Blue and Red) linearisation no longer holds. Therefore we should not look  in Eq.(\ref{eq:FP-norm-mode-sol}) for an account of the dynamics after the point of fragmentation because
our approximation $\sum_{j \in {\cal B} \cup {\cal R}} L^{(BR)}_{ij} v_j\approx 0$ breaks down. This leads to additional interactions in the equation
for $\alpha$ that modify its {\it deterministic} behaviour - triggering an extra cycle compared to the linearised prediction. However, these same
nonlinearities lead the system to resynchronise so that the approximation takes hold again, leading to another deterministic cycle in $\alpha$ according to the present two-cluster
simplification. Thus we may say that the linearisation breaks down half the time. 

\section{Noisy zero modes}
\label{sec:5}
In this case the noise is projected solely on the zero modes, thus
$\gamma_0^B=\gamma_0^R=1$ and $\gamma_r^B=\gamma_s^R=0$, for all $r\in\mathcal{B}^E/\{0\}$ and $s\in\mathcal{R}^E/\{0\}$.
Thus the first two normal mode Langevin equations in Eq.\eqref{eqn:integrablesystem-w-noise} become deterministic and $x_s$ and $y_s$ only experience stochastic effects through the nonlinear Langevin equation for $\alpha$. In this case we first construct the marginal density of $\alpha$, and use this to give the corresponding marginal densities of the non-zero normal modes.

\subsection{Analytical considerations - stochastic tilted ratchets and $\alpha$}\label{sec:5.2}
The marginal density for $\alpha$, $\calP (\alpha,t)$, is the solution to the following Fokker-Planck equation,
\begin{eqnarray}
\frac{\partial \calP}{\partial t}= \left\{ \Omega \frac{\partial^2}{\partial \alpha^2} +{\frac{\partial}{\partial\alpha}V'(\alpha)}  \right\}\calP, 
\label{ratcheteq}
\end{eqnarray} 
where $V(\alpha)$ is given in Eq.(\ref{alphaeq}). It is also advantageous to consider the \textit{probability current} ${\cal J} (\alpha,t)$, which can be obtained by re-expressing the Fokker-Planck equation in the form of a probability conservation/continuity equation,
\begin{eqnarray}
\frac{\partial \calP}{\partial t}+ \frac{\partial {\cal J}}{\partial \alpha}=0 \;\; \Rightarrow {\cal J} =  - \left\{ \Omega \frac{\partial}{\partial \alpha} +V'(\alpha)  \right\}\calP.
\label{continEQ}
\end{eqnarray}
As previously stated, the Langevin equation with such $V(\alpha)$ is referred to as a stochastic periodic tilted ratchet \cite{Wellens, Khangjune, Challis}. Unlike most diffusive processes which have zero average velocity for the steady-state, tilted ratchet processes are overwhelmingly 
influenced by the tilt
\cite{Risken}.

Traditionally \cite{Reimann2002}, one solves for the steady-state density $\calP_{st}$ 
from a Pearson equation by applying a vanishing boundary condition $\calP_{st}(\alpha) \rightarrow 0$ as $\alpha \rightarrow \pm \infty$. However in this case, due to the periodicity, one finds that the resulting density is \textit{non-normalisable}. 
Alternatively, we may construct the so-called \textit{reduced marginal density} and \textit{reduced probability current}, 
$\hat{\calP}(\alpha,t)$ and $\hat{ {\cal J}}(\alpha,t)$, given explicitly by
\begin{eqnarray}\label{eq:reduced-probabilities}
\hat{\calP}(\alpha,t) \equiv \sum^{\infty}_{n = -\infty}\calP (\alpha+2 \pi n ,t), \;\; \hat{\cal J}(\alpha,t) = \sum^{\infty}_{n = -\infty}{\cal J} (\alpha+2 \pi n ,t)
\end{eqnarray}
where $\alpha \in (-\pi, \pi]$. Due to the linearity of the Fokker-Planck equation, the reduced marginal density also obeys Eq.(\ref{ratcheteq}), but with the following boundary and normalisation conditions
\begin{eqnarray}
\hat{\calP}(-\pi,t)=\hat{\calP}(\pi,t), \;\; \int^{\pi}_{-\pi} d\alpha \hat{\calP}(\alpha,t) = 1.
\label{ratchetBC}
\end{eqnarray}
From Chap.9 of \cite{Strat67}, the steady-state  density satisfying Eq.(\ref{ratchetBC}) is
\begin{eqnarray}
\hat{\calP}_{st}(\alpha) = \frac{\kappa_1}{e^{-\frac{2 \pi \mu}{\Omega}}-1} e^{- \frac{V(\alpha)}{\Omega} } \int^{\alpha+2 \pi}_{\alpha}d \varphi  e^{\frac{V(\varphi)}{\Omega}}, 
\label{integralmess}
\end{eqnarray}
where the normalisation constant is \cite{Strat67}
\begin{eqnarray}
\kappa_1 = - \frac{\sinh \left( \frac{\pi \mu}{\Omega} \right)}{2 \pi^2 \left| I_{i \mu}\left( \frac{\sqrt{S^2+C^2}}{\Omega} \right) \right|^2}.
\label{constantkappa}
\end{eqnarray} 
and $I_{ i \mu}$ is the modified Bessel function of \textit{imaginary order}. 

Additionally, the linearity of the continuity expression in Eq.(\ref{continEQ}) means that the reduced probability current is given similarly by
\begin{eqnarray*}
\hat{{\cal J}}(\alpha,t) =  - \left\{ \Omega \frac{\partial}{\partial \alpha} +V'(\alpha)  \right\}\hat{\calP}(\alpha,t).
\end{eqnarray*}
Moreover, the linearity of the integration operation means that the average velocity, $\langle \dot{\alpha} \rangle$, may be 
computed through the reduced probability current:
\begin{eqnarray}
\langle \dot{\alpha} \rangle \equiv \int^{\pi}_{-\pi}d \alpha \hat{{\cal J}}(\alpha,t) = - \int^{\pi}_{-\pi}d \alpha \Omega \kappa_1 = \frac{ \Omega \sinh \left( \frac{\pi \mu}{\Omega} \right)}{ \pi \left| I_{i \mu}\left( \frac{\sqrt{S^2+C^2}}{\Omega} \right) \right|^2}
\label{avdotalpha}
\end{eqnarray}
Through Eq.(\ref{avdotalpha}) we see that $\langle \dot{\alpha} \rangle$ is an odd function of $\mu $
highlighting the role of the tilt in tilted ratchet potentials.

\begin{figure}
\centering
\includegraphics[width=0.9\linewidth]{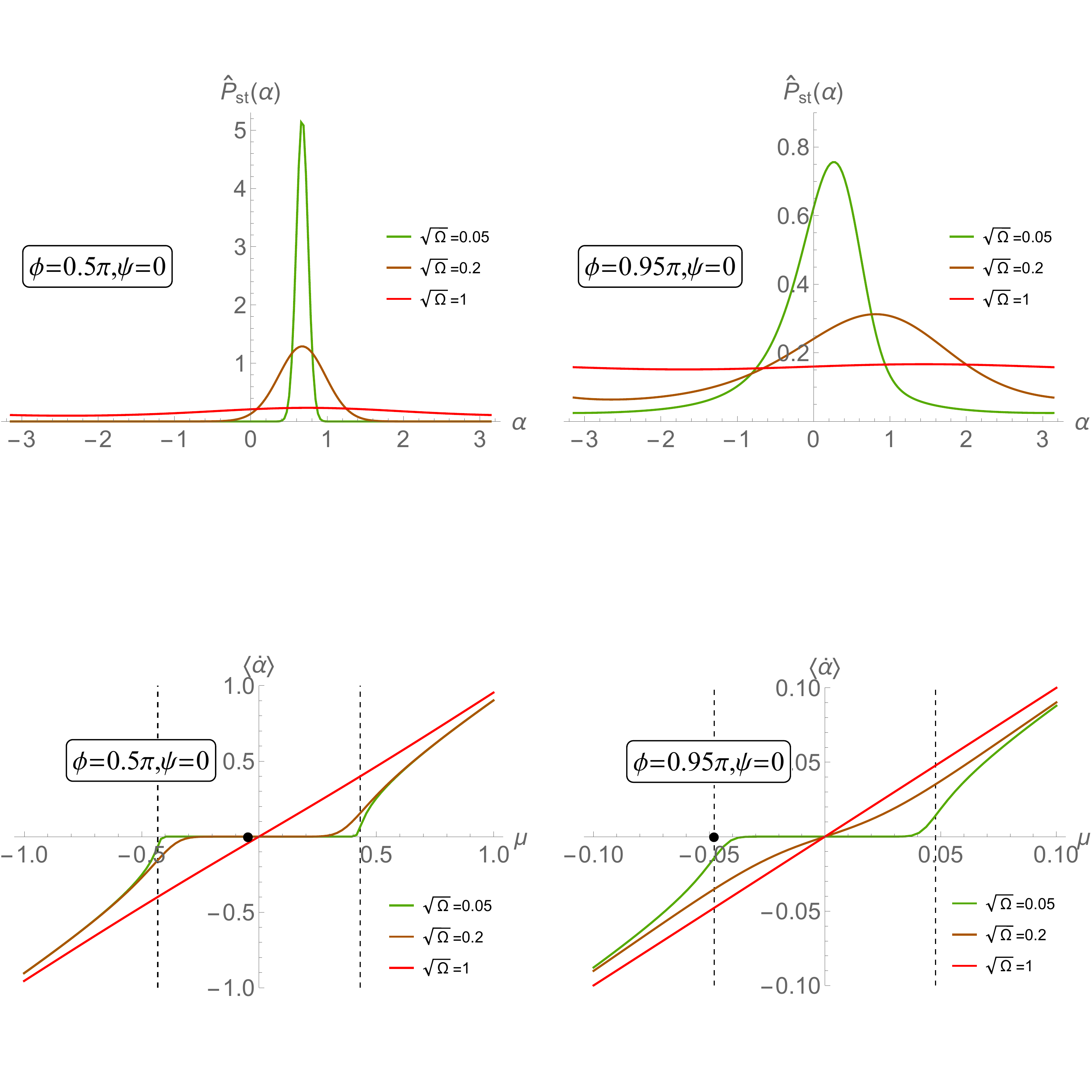}

\caption{Plots of $\hat{\calP}_{st}(\alpha)$, from Eq.(\ref{integralmess}) (top row), and $\langle \dot{\alpha} \rangle$, from Eq.(\ref{avdotalpha}) (bottom row) for $\phi=0.5\pi, \psi=0$ (left column) and
$\phi=0.95\pi, \psi=0$ (right column). Colours from green to red correspond to increasing noise strength $\sqrt{\Omega}$ values of $0.05$, $0.2$ and $1$. In the bottom row, the $\mu\equiv \bar{\omega}-\bar{\nu}$ values leading to ${\cal K} =0$ are indicated by the vertical dashed lines so that inside the lines ${\cal K}>0$.
The dot indicates the actual value of $\mu$ for the example system in the paper.}
\label{fig:Ratchet}
\end{figure}

In upper panels of Fig.\ref{fig:Ratchet} we plot the density Eq.(\ref{integralmess}) for various parameter values drawing upon the tree-vs-random network example. 
For $\phi=0.5\pi$ curves are indistinguishable from standard Gaussian noise, with the broadening as $\Omega$ increases. For $\phi=0.95\pi$ things are different. For weak noise $\sqrt{\Omega}=0.05$ the density is noticeably non-zero with negligible gradient at $\alpha=\pm\pi$, a consequence of the periodic boundary conditions. Note the changed vertical scale: the density at its
peak is still low showing weak localisation. At strong noise, $\sqrt{\Omega}=1$, localisation is destroyed. Underlying these results is the shape of the potential $V(\alpha)$: for $\phi=0.5\pi$ the potential, though sloped
slightly negatively for the particular value of $\mu$ in the numerical example, has significant wells so that localisation is possible; only at stronger noise levels (the green
curve in the upper left hand plot) is there a significant probability that the system falls outside the well. Contrastingly, for $\phi=0.95\pi$ there is no convex well so that even
low noise levels allow for the system to run out though at these values
of $\Omega$ there is enough `stalling' around the fixed point that the system spends finite time in the vicinity. 

These behaviours are reflected now in $\langle \dot{\alpha} \rangle $ in the lower plots of
Fig.\ref{fig:Ratchet}. Here we relax the value of $\mu=\bar\omega-\bar\nu$; the actual value we use
is a tiny negative value, $\mu=-0.048$, indicated by the dot in Fig.\ref{fig:Ratchet}. We see that in the left hand lower plot, with $\phi=0.5\pi$, large average frequency difference $\mu$ (thus large tilt) or large 
noise $\Omega$ are required for $\langle \dot{\alpha} \rangle \neq 0$, so that $\alpha$ is stochastically time-varying because of the system running down the ratchet
potential. Since ${\cal K}>0$ we are dealing with a stable fixed point and yet the system spends a significant amount of time outside the basin of attraction. The sign of the slope of the tilt clearly determines the direction of the running of $\alpha$. 
Contrastingly, at $\phi=0.95\pi$ the range of values of $\mu$ for which $\langle \dot{\alpha} \rangle = 0$ is much narrower; note the change of horizontal scale in this plot, one
tenth of the size of the range compared to the lower value of $\phi$. The actual value of $\mu=-0.048$ in the numerical example lies just outside the vertical dashed lines which indicates the
region where ${\cal K}$ is positive.
Therefore even at weak noise with $\sqrt{\Omega}=0.05$ we have $\langle \dot{\alpha} \rangle \neq 0$ so that the system runs down the tilted potential.

The behaviour in the bottom row plots of Fig.\ref{fig:Ratchet} when ${\cal K}>0$ but the average velocity $\langle \dot{\alpha} \rangle\neq 0$ is known as \textit{metastability} and is addressed using Freidlin-Wentzell (FW) theory \cite{Deville2012, Freidlin, Berglund}. Qualitatively FW theory states that a dynamical system perturbed by small amounts of Gaussian noise spends the majority of its time in the immediate vicinity of a particular  deterministic well of attraction. In rare, exponentially long instances, the system then makes the transition to another deterministic well of attraction. In Appendix B we show how this may be quantified further.


\subsection{Steady state conditional probabilities for normal modes}
As stated, the normal modes $x_r$ and $y_s$ are not subject to explicit noise but develop stochastic behaviour through dependence on $\alpha$ in the drift terms, $q^{(B)}_r(x_r,\alpha)$ and $q^{(R)}_s (y_s,\alpha)$. In this case (see Chap.2 of \cite{Risken} for example), the conditional probability for $x_r$ given some value of $\alpha$, labelled $P(x_r | \alpha)$, is given simply as 
\begin{eqnarray*}
\calP(x_r | \alpha) = \delta \left(x_r - \frac{\omega^{(r)} -\zeta_{BR} d^{(BR)}_r \sin(\alpha - \phi) }{\sigma_B \lambda^{(B)}_r} \right)\\
= \frac{\delta\left( \alpha- \phi-\sin^{-1}\left(\frac{\omega^{(r)}-\sigma_B\lambda_r^{(B)}x_r}{\zeta_{BR} d^{(BR)}_r} \right) \right) +\delta\left( \alpha- \phi -\pi+\sin^{-1}\left(\frac{\omega^{(r)}-\sigma_B\lambda_r^{(B)}x_r}{\zeta_{BR} d^{(BR)}_r} \right) \right) }{\frac{\zeta_{BR}d^{(BR)}_r}{\sigma_B \lambda^{(B)}_r}\left| \cos \left[ \sin^{-1} \left( \frac{\omega^{(r)}-\sigma_B\lambda_r^{(B)}x_r}{\zeta_{BR} d^{(BR)}_r} \right) \right] \right|}.
\end{eqnarray*}
We notice that the conditional probability is already $2 \pi$-periodic in $\alpha$, thus we do not need to perform the reduction-operation of Eq.(\ref{eq:reduced-probabilities}) for the $\alpha$ argument. Therefore, applying the marginal probability for $\alpha$, we obtain the marginal probability for $x_r$ by way of
\begin{eqnarray}\label{marginalnormal}
\begin{split}
\calP(x_r) =& \int_{-\pi}^{\pi}\mathrm{d}\alpha\,\calP(x_r|\alpha)\hat{\calP}_{st}(\alpha)\\
=&\,
\frac{\hat{\calP}_{st}\left(\phi+\sin^{-1}\left(\frac{\omega^{(r)}-\sigma_B\lambda_r^{(B)}x_r}{\zeta_{BR} d^{(BR)}_r}\right)\right)
+\hat{\calP}_{st}\left(\phi+\pi-\sin^{-1}\left(\frac{\omega^{(r)}-\sigma_B\lambda_r^{(B)}x_r}{\zeta_{BR} d^{(BR)}_r}\right)\right)}{\dfrac{\sqrt{[\zeta_{BR} d^{(BR)}_r]^2-[\omega^{(r)}-\sigma_B\lambda_r^{(B)}x_r]^2}}{\sigma_B \lambda^{(B)}_r}}.
\end{split}
\end{eqnarray}
We straightforwardly see from Eq.(\ref{marginalnormal}) that to obtain a real-valued density, $x_r$ must satisfy the bound
\begin{equation}\label{eq:modes-bounds}
\left|x_r-\frac{\omega^{(r)}}{\sigma_B\lambda_r^{(B)}}\right|\le\left|\frac{\zeta_{BR}d_r^{(BR)}}{\sigma_B\lambda_r^{(B)}}\right|.
\end{equation}
Recall that when stability is satisfied, deterministically $x_r$ will reach the value $\{ \omega^{(r)} -\zeta_{BR} d^{(BR)}_r \sin(\alpha^*-\phi) \}/(\sigma_B\lambda_r^{(B)})$, for steady state value of $\alpha$ given in Eq.(\ref{alphasol}). Given that the sine function is bounded by $\pm 1$, the maximum amount $x_r$ may \textit{stochastically} vary from this value is given by the above bound which is
now sensitive to the Blue-Red coupling strength and network degree. 
We emphasise that this bounding arises due to there being no direct noise applied to the normal mode equations - the noise enters explicitly through $\alpha$. In the case where noise is applied to \textit{both} the normal and zero mode equations, in Appendix D, no such bounding occurs.

\begin{figure}
\centering
\includegraphics[width=.9\linewidth]{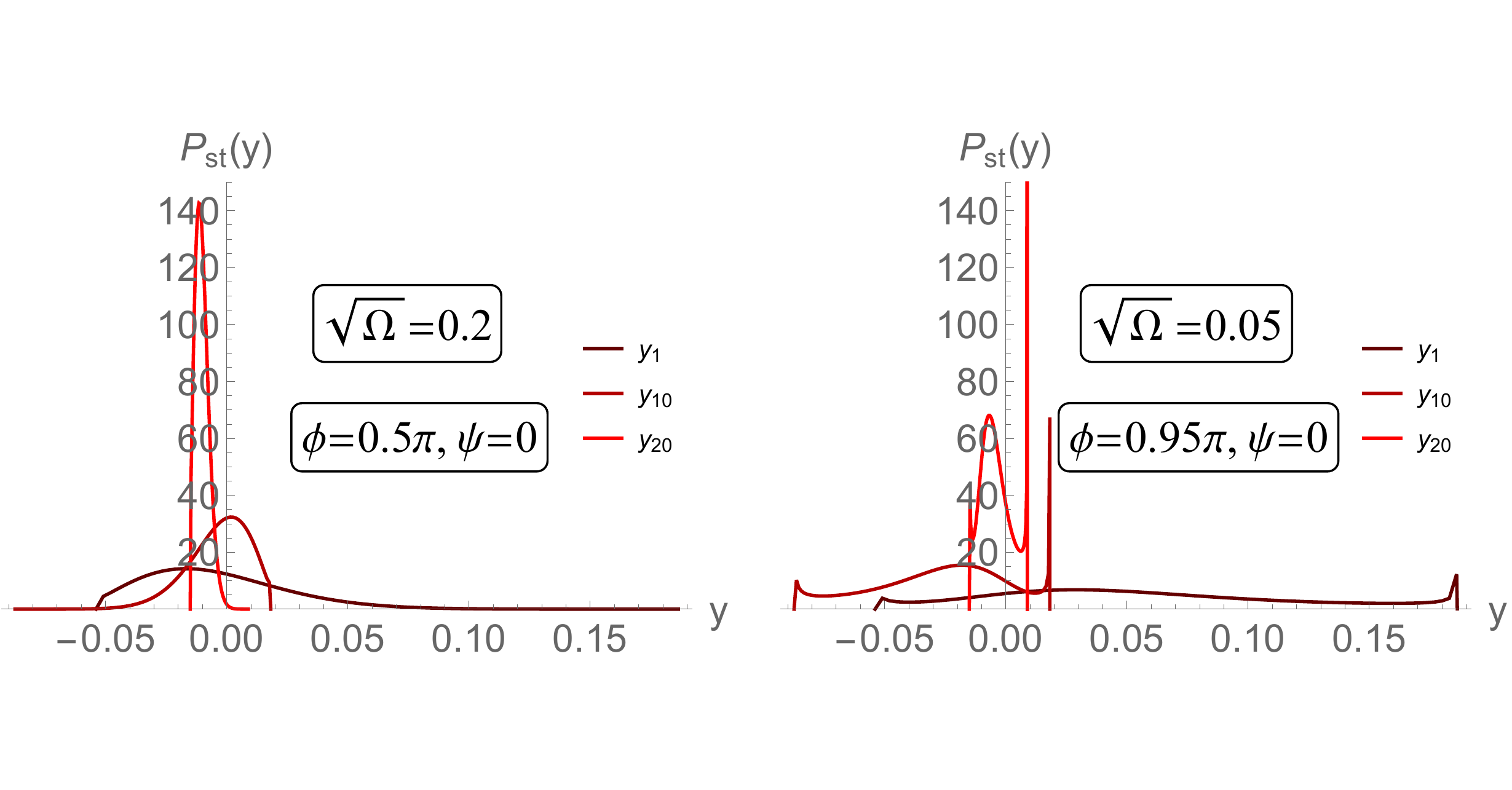}
\includegraphics[width=.9\linewidth]{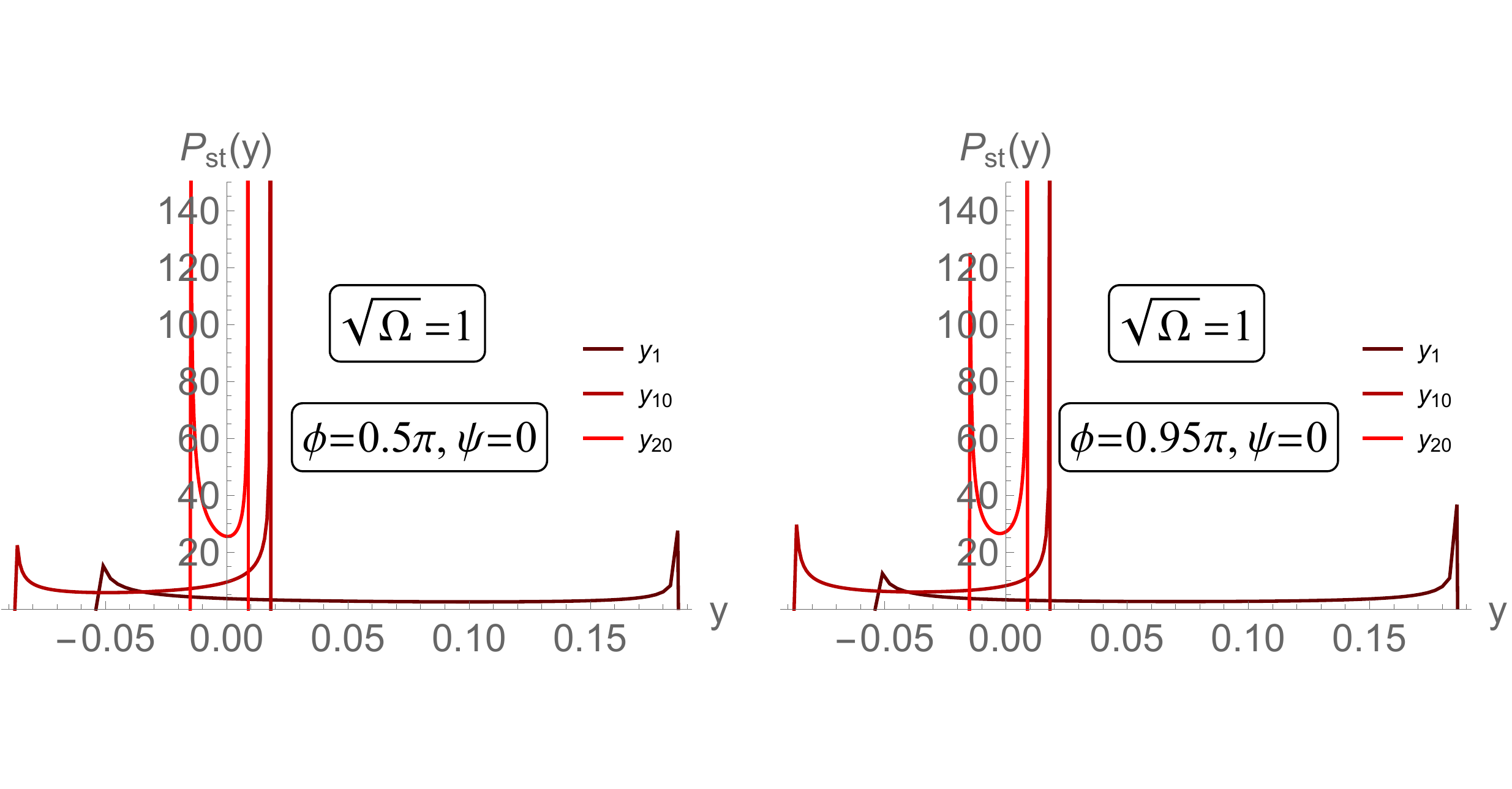}
\caption{
Plots of ${\cal P}_{st}(y_r)$ from Eq.(\ref{marginalnormal}) for the red network for parameters $\phi=0.5 \pi$, $\psi=0$ (left) 
and $\phi=0.95 \pi$, $\psi=0$ (right),
and $\sqrt{\Omega}=0.05$ and $1$ for the left and right plots respectively.
}
\label{fig:xsx3}
\end{figure}

Similar expressions for $y_s$ follow, with labels $B$ and $R$ swapped. We plot examples for this case, the densities for $y_s$ in  
Fig.\ref{fig:xsx3} for the same modes
that we explored in the numerical study for noisy zero modes. Because of the small values of low-lying eigenvalues $\lambda_r^{(B)}$
the densities for Blue turn out to be compressed into a tiny domain which are not visually useful; we return to this aspect in the numerical results.

The left column of Fig.\ref{fig:xsx3} represents behaviours of ${\cal P}(y_r)$ for ${\cal K}>0$ but with noise on $\alpha$. The right column
plots are for ${\cal K}<0$. The overwhelming feature of these densities is the sharp cusp at certain values of $y$; beyond these points the imaginary
part is non-zero and the real part vanishes. We may call these `dead zones' in that there appears to be zero probability of the system existing in these states.
In fact, the densities diverge delta-function-like to infinity at the edges here, a consequence of the first step in their
derivation from $\calP(y_r | \alpha) $ or, later, in the singular denominator in Eq.(\ref{marginalnormal}). As a consequence, the densities imply
quite strong `localisation' of the normal modes but with quite different characteristics across the range of $\Omega$ and $\phi$.
This localisation diminishes with increasing $\Omega$ but increases with higher Laplacian eigenvalue. In some cases, the density is highly asymmetric with a sharper cusp on one side than the other. But this localisation may subtly change
according to the mode and the strength of noise: observe that for $\sqrt{\Omega}=0.2$ and $\phi=0.5\pi$ (top left) the densities
are peaked in a typical fashion around the values $y\approx 0$. These peaks approximately correspond to their deterministic fixed point $\{\nu^{(s)} +\zeta_{RB} d^{(RB)}_s \sin(\alpha^*+\psi)  \}/\sigma_R\lambda_r^{(R)}$. For $\phi=0.95\pi$ this peaking remains but the cusps become stronger. Deterministically the system has no fixed point (see Fig.\ref{fig:Normal-Frac095-StdDev-1}), but $\alpha$ does spend a long time approximately near what is \textit{approximately} an unstable fixed point. Such weak localisation is a consequence of the noise kicking the solution off the ratchet potential's approximate unstable fixed point intermittently. Contrastingly,
for $\sqrt\Omega=1$ the densities are flat between the cusps - suggestive of delocalisation. The dead zones imply that the system cannot exist with any probability for values of $y$ beyond the cusp - a form of localisation in its own right.

Underlying all this is the stochastic behaviour of $\alpha$: only for the top left case of Fig.\ref{fig:xsx3}, as we see in the corresponding plot of Fig.\ref{fig:Ratchet}, 
are the densities for $\alpha$ localised around a fixed value and $\langle \dot{\alpha} \rangle=0$ so that the normal modes in turn localise
but away from the deterministic fixed point.
For the bottom left case of Fig.\ref{fig:xsx3}, stronger noise allows the $\alpha$ state to stochastically escape the wells in the ratchet potential, 
so that $\langle \dot{\alpha} \rangle\neq 0$ and, correspondingly normal modes, especially the lowest ones (with small Laplacian eigenvalue), do not localise
to a point but to a region. In the right column plots of Fig.\ref{fig:xsx3} the tilted ratchet potential has no wells, $\alpha$ runs down the tilt, therefore none of the normal modes may localise to the fixed point. For weak noise and a high Laplacian mode - the $y_{20}$ line in the upper right hand plot of Fig.\ref{fig:xsx3}
- there is some peaking in the density because, from the corresponding point on Fig.\ref{fig:Ratchet}, $\langle \dot{\alpha} \rangle$ is only slightly non-zero
so that $\alpha$ `catches' temporarily.

Overall then we expect behaviour such as stochasticity in $\alpha$, order parameters and the period itself for ${\cal K}<0$,
and indication of dead zones in the normal modes.

\subsection{Numerical simulations}
\begin{figure}
\centering

{\includegraphics[width=.9\linewidth]{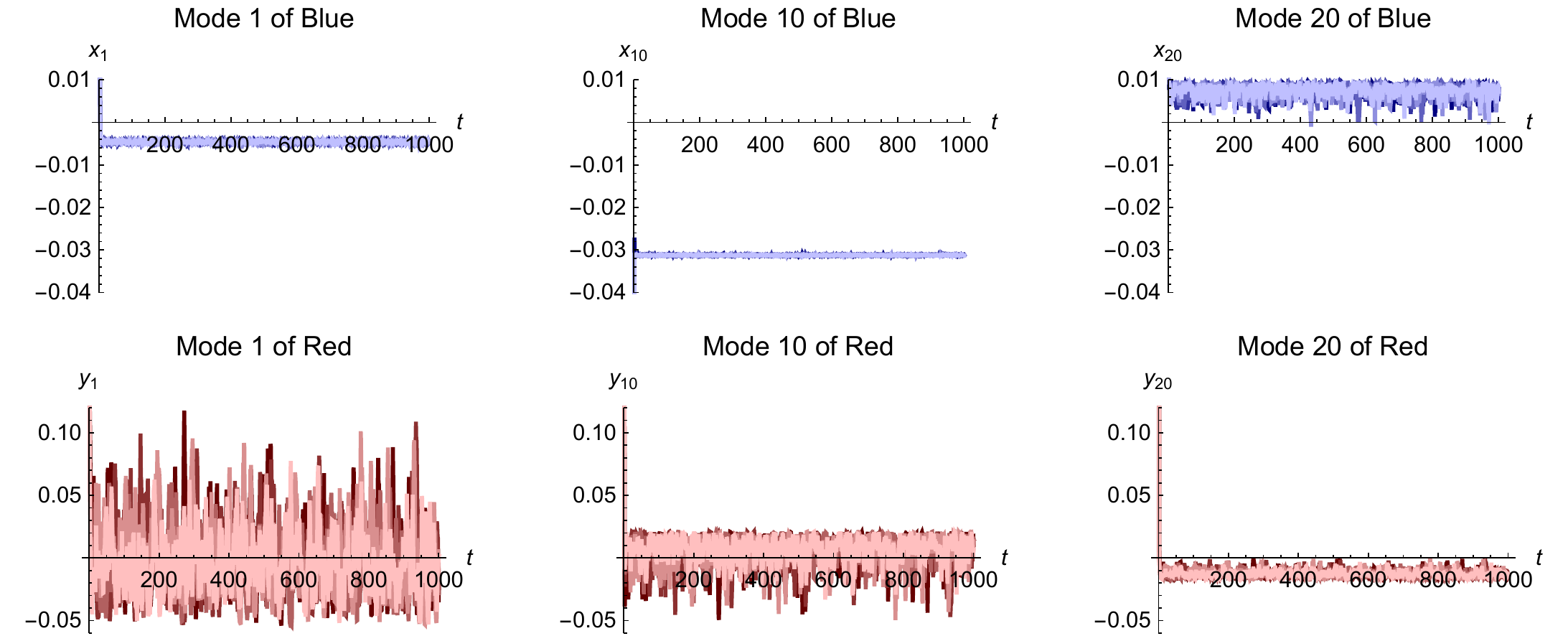}
\label{subfig:Zero-Modes12}}

\caption{Simulations of 5 paths of the $r=1,10,20$ modes of Eq.\eqref{BR-eq} with white noise of $\sqrt{\Omega}=1$ on the zero modes $B$ and $P$ and parameters $\sigma_B=8,\,\sigma_R=0.5,\,\zeta_{BR}=\zeta_{RB}=0.4$ and $\phi=0.5\pi,\,\psi=0$.}
\label{fig:Zero-Frac05-StdDev-1}
\end{figure}

In Fig.\ref{fig:Zero-Frac05-StdDev-1} we forego the order parameters and $\alpha$ where results are much as n previous cases,
but focus on the Laplacian modes for $r=1,10,20$ for $\phi=0.5\pi$.

We note that the variance in fluctuations for the Blue mode $x_{20}$ is {\it higher} than for lower modes, while
for Red modes there is the usual pattern of suppression going up the spectrum. 
Additionally, we see a clear asymmetry in the fluctuations, with a `flatter' profile on the positive side
for $x_{20}$ compared to the negative side. This is reversed and amplified for $y_1$. 
This is consistent with the shape of the densities seen in Fig.\ref{fig:xsx3}: the upper right hand plot shows distinct bias
of $y_1$ on the positive side with a longer tail in that direction. This matches the larger but less frequent fluctuations in the positive direction of $y_1$ in
the lower left hand plot of Fig.\ref{fig:Zero-Frac05-StdDev-1}. We can be more quantitative than just inspection of the plotted densities.
Recall the numerical values of the projections $d^{(BR)}_r$ and $d^{(RB)}_r$ for the specific case here, where the former are significantly smaller
than the latter. Mode-by-mode, the predicted bounds in Eq.\eqref{eq:modes-bounds} for the 1st, 10th and 20th modes of Blue are $O(10^{-15})$, $O(10^{-17})$ and $O(10^{-3})$, respectively. Similarly for Red the predicted bounds on the 1st, 10th and 20th modes are $O(10^{-1})$, $O(10^{-2})$ and $O(10^{-2})$, respectively.
This pattern is consistent with the analytic results showing a sharp vanishing of the probability density
outside specific ranges of $x_r$ and $y_r$, the stronger suppression for Blue compared to Red and the relative diffuseness
of the $r=20$ mode of Blue compared to those of lower $r$.

\begin{figure}
\centering
\subfloat[][Order Parameters for Red and Blue]{\includegraphics[width=.9\linewidth]{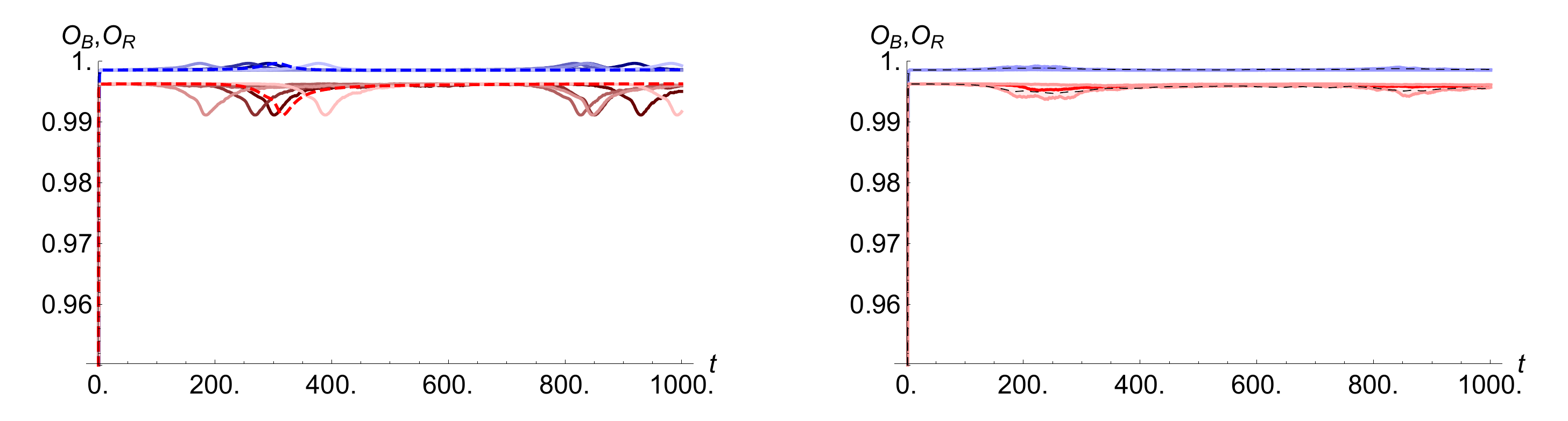}
\label{subfig:Zero-Order13}}

\subfloat[][Deterministic and Langevin simulations of $\alpha$]{\includegraphics[width=.9\linewidth]{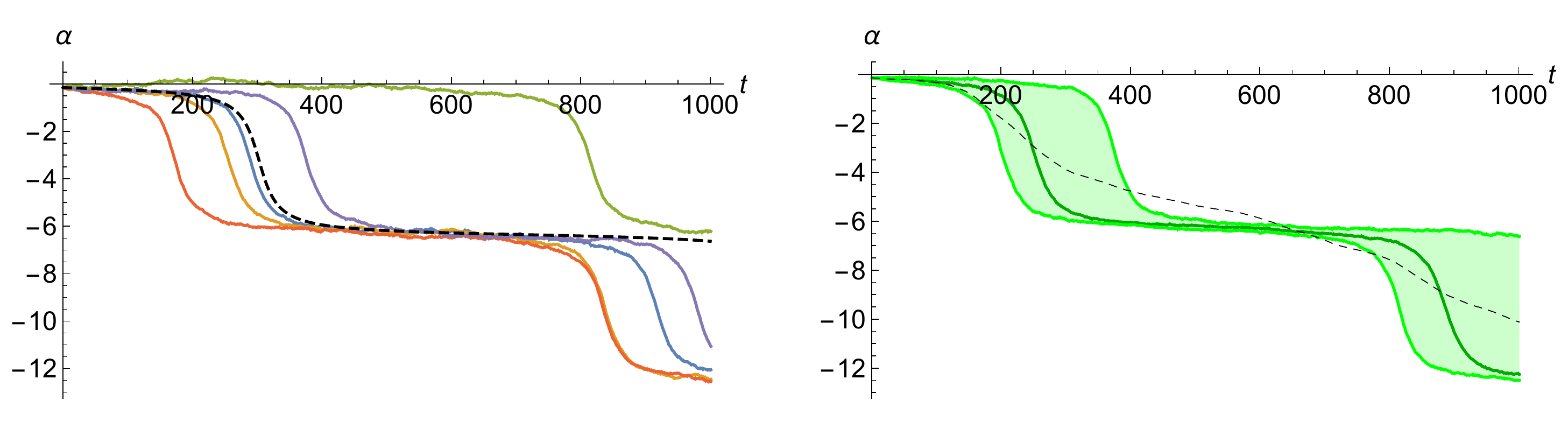}
\label{subfig:Zero-Alpha13}}

\subfloat[][Simulations of the $r=1,10,20$ modes]{\includegraphics[width=.9\linewidth]{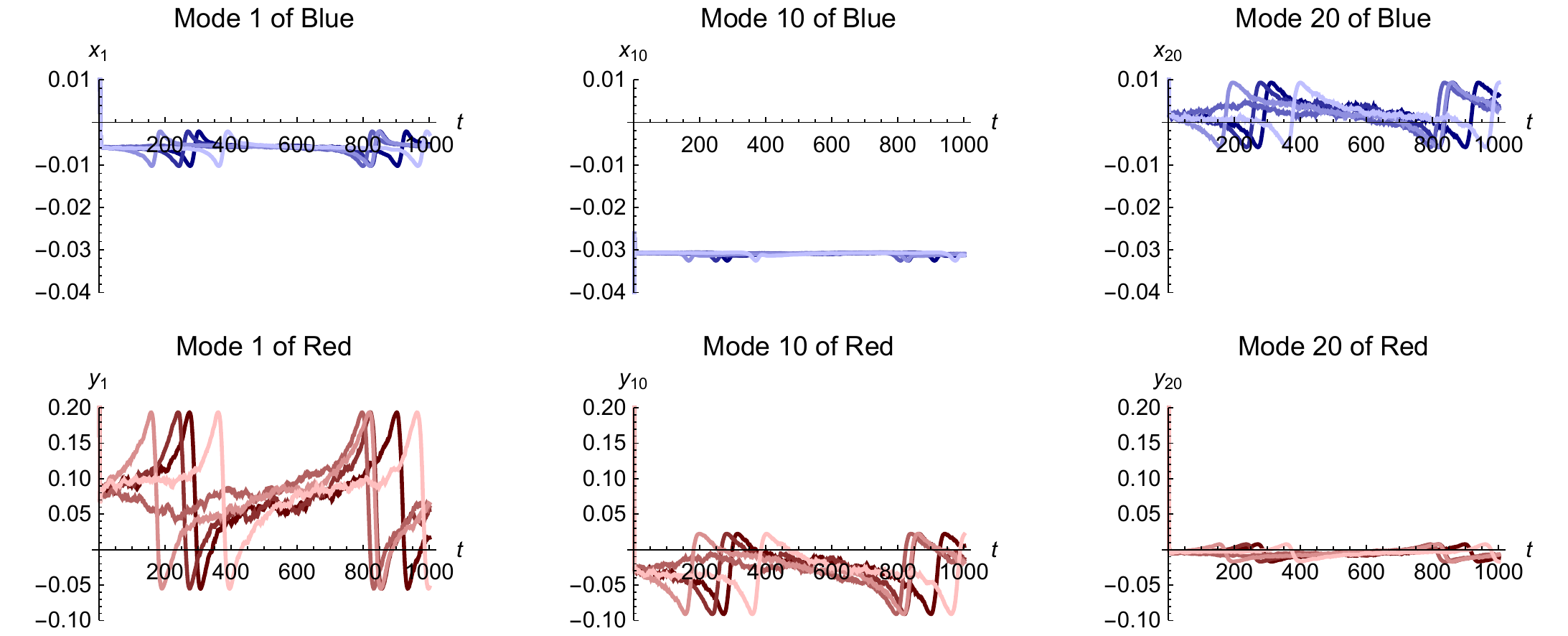}
\label{subfig:Zero-Modes13}}

\caption{Simulations of  Eq.\eqref{BR-eq} with and without noise for the Blue tree network
and Red random network, with white noise with $\sqrt{\Omega}=0.05$ on the zero modes $B$ and $P$ and parameters $\sigma_B=8,\,\sigma_R=0.5,\,\zeta_{BR}=\zeta_{RB}=0.4,\,\phi=0.95\pi,\,\psi=0$: Fig.\protect\subref{subfig:Zero-Order13} (Left) displays the deterministic local synchronisation order parameter $O_B$ (dashed blue) and $O_R$ (dashed red) and five paths of the Langevin local synchronisation order parameter $O_B$ and $O_R$, respectively, (Right) displays the median (blue/red), upper and lower quartile (light blue/red) and mean (dashed black) of 50 simulation of  $O_B$ and $O_R$, respectively. Fig.\protect\subref{subfig:Zero-Alpha13} (Left) displays deterministic $\alpha$ (Eq.\eqref{fixpoint-alph}) and 5 paths of the Langevin simulations for $\alpha$, (Right) displays the median (green), upper and lower quartile (light green) and mean (dashed black) of 50 simulation of $\alpha$. Fig.\protect\subref{subfig:Zero-Modes13} contains plots of the $r=1,10,20$ modes for 5 paths of the Langevin simulations. The top row contains the Blue modes and the bottom row contains the Red modes.}
\label{fig:Zero-Frac095-StdDev-005}
\end{figure}

Now we turn to $\phi=0.95\pi$ but for {\it weak} noise, $\sqrt{\Omega}=0.05$ in Fig.\ref{fig:Zero-Frac095-StdDev-005}.
The periodicity is consistent with ${\cal K}<0$, however its value varies. This matches the expectation from Fig.\ref{fig:Ratchet} where noise over the ratchet potential with no stable wells 
smears the period of the cycle for $\alpha$, as discussed. The effect is also visible in the order parameters and the modes $x_r$ and $y_r$.
We comment, but do not provide plots, that for $\Omega=1$ there is no periodicity remaining in the order parameters, where $O_R$ is fluctuating significantly more than $O_B$.  Individual paths for $\alpha$ are stochastically time-varying
with strong negative drift with only little hint of (rapid)
periodicity in the median over 50 paths. 
The Laplacian modes again show fluctuations $x_{20}$ is larger than the lower modes, consistent with estimates from
the cutoff in the density of Eq.\eqref{eq:modes-bounds}.

\begin{figure}
\centering
{\includegraphics[width=.9\linewidth]{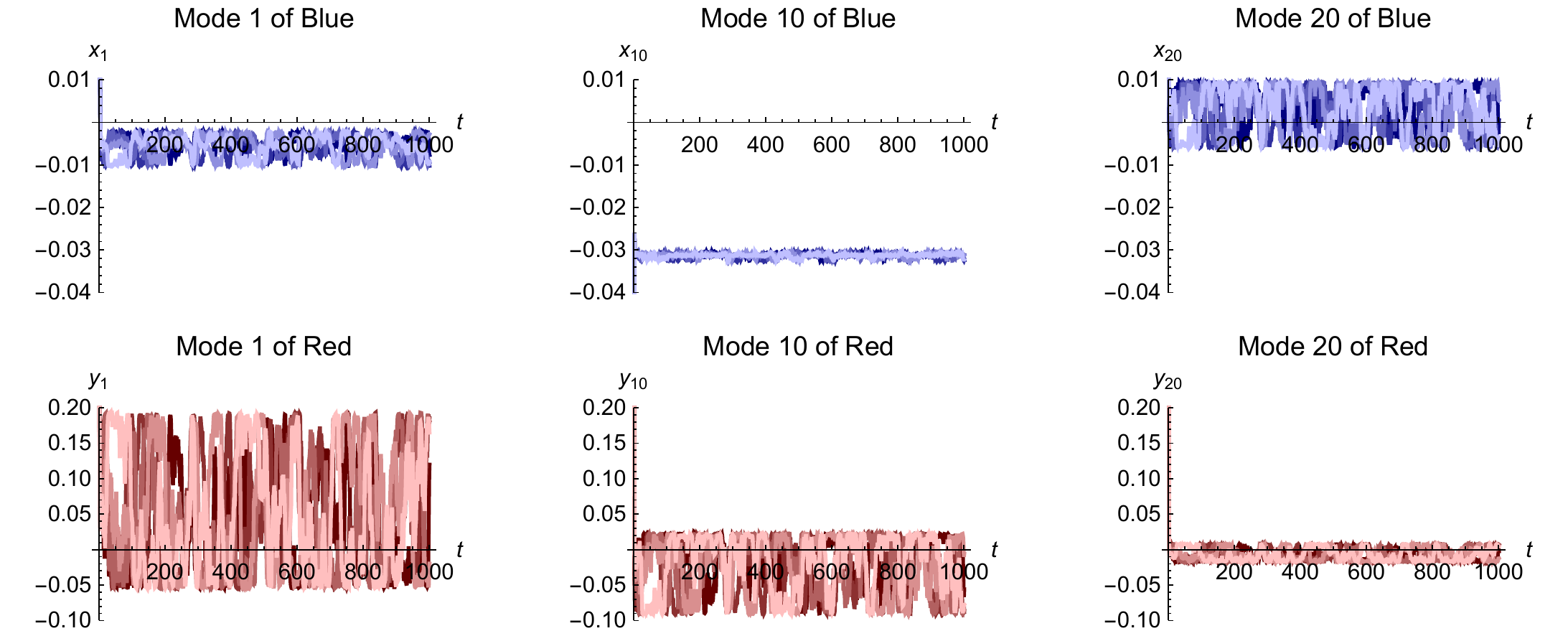}
\label{subfig:Zero-Modes14}}
\caption{
Simulations of 5 paths of the $r=1,10,20$ modes of Eq.\eqref{BR-eq} with white noise of $\sqrt{\Omega}=1$ on the zero modes $B$ and $P$ and parameters $\sigma_B=8,\,\sigma_R=0.5,\,\zeta_{BR}=\zeta_{RB}=0.4$ and $\phi=0.95\pi,\,\psi=0$.
}
\label{fig:Zero-Frac095-StdDev-1}
\end{figure}

Finally, for $\sqrt\Omega=1$ in Fig.\ref{fig:Zero-Frac095-StdDev-1} we forego the order parameters and $\alpha$
which show no periodicity and strong fluctuations, with negative drift for $\alpha$.
We show rather the Laplacian modes which again show fluctuations $x_{20}$ is larger than the lower modes, consistent with estimates from
the cutoff in the density of Eq.\eqref{eq:modes-bounds}. Moreover the fluctuations in Red modes show strong boundedness on
both sides, consistent with our analytical results in Fig.\ref{fig:xsx3}.

Appendix D shows the result of noise applied to all modes revealing a composite of all the behaviours seen thus far, 
in particular the regimes distinguished by the sign of ${\cal K}$.

\section{Stochastic synchronisation in Red fragmented system}
\subsection{The case of three clusters: fragmentation of Red}
We now consider the Red population fragmented into two sub-populations, on sub-graphs ${\cal R}_1$ and ${\cal R}_2$ each with $M_1$ and $M_2$ nodes respectively, with $M_1+M_2=M$. The deterministic version of this case was explained in detail in \cite{KallZup2015} so the
following will be concise. Details are in Appendix E.

The defining equations Eqs.(\ref{BR-eq}) may be rewritten replacing the sum after $\zeta_{BR}$ with $\sum_{a=1}^2 \sum_{j\in {\cal R}_a} {\cal M}_{ij}$
and that after $\sigma_R$ by $\sum^2_{a=1} \sum_{j \in {\cal R}_a} {\cal R}_{ij}$
and interaction matrix ${\cal M}$ involving three off-diagonal blocks describing connectivity from ${\cal B}$ to ${\cal R}_1$, and ${\cal R}_1$ and ${\cal R}_2$. The full equations are given in \cite{KallZup2015}.

The analogues to Eq.(\ref{fixpoint1}) for centroid decomposition are
\begin{eqnarray}
\beta_i = B + b_i,\; i \in {\cal B},\;\; \rho_{j_1} = P_1 + p^{(1)}_{j_1}, \;  j_1 \in {\cal R}_1,\;\;  \rho_{j_2} = P_2 + p^{(2)}_{j_2}, \;  j_2 \in {\cal R}_2 \label{fixpoint2} 
\end{eqnarray}
where the centroids $B, P_1$ and $P_2$  of each of the three populations are defined by the average of $\beta_i$ and the two groups of $\rho_i$ in ${\cal B}, {\cal R}_1$
and ${\cal R}_2$.
Correspondingly, the central quantities of interest in this regime are the difference between each network's centroid
\begin{eqnarray}
 B-P_1\equiv \alpha_{B R_1},\;\; P_1 - P_2 \equiv \alpha_{R_1 R_2} \label{3-cent-def}
\end{eqnarray}
and the quantity $\alpha_{B R_2}$ is given as the following linear sum,
\begin{eqnarray*}
 B-P_2\equiv \alpha_{B R_2} = \alpha_{B R_1} + \alpha_{R_1 R_2}.
\end{eqnarray*}

The linearised three cluster system is adapted from Eq.(\ref{linsys}), with the sum after $\sigma_R$ 
replaced with $\sum_{j \in {\cal R}_a} L^{(R_a)}_{ij} p_j^a$, $a=1,2$ and
where the $\Omega$ vector contains three distinct parts:
\begin{eqnarray*}
\Omega_i &=& \left\{  \begin{array}{ll}
\omega_i -\zeta_{BR} \sin(\alpha_{B R_1}-\phi) d^{(B R_1)}_i  & i \in {\cal B}\\
\nu^{(1)}_i + \zeta_{RB} \sin(\alpha_{B R_1} + \psi) d^{(R_1 B)}_i - \sigma_R \sin(\alpha_{R_1 R_2}) d^{(R_1 R_2)}_i & i \in {\cal R}_1 \\
\nu^{(2)}_i + \sigma_R \sin(\alpha_{R_1 R_2}) d^{(R_2 R_1)}_i & i \in {\cal R}_2
\end{array} \right.
\end{eqnarray*}
The fluctuations $b_i,p_i^{(a)}$  now form the vector $v$. There are additional terms which need to
be set to zero; these are given in Appendix E.
A complete spanning set of orthonormal eigenvectors can be constructed. Those for ${\cal B}^E$ remain untouched, while the eigen-space ${\cal R}^E$ is partitioned into ${\cal R}^E_1$ and ${\cal R}^E_2$. The corresponding eigenvectors are labeled $e^{(R_a,r)}$, $(r = 0, 1, \dots, M_a-1 \in {\cal R}^E_a)$ for $a=\{1,2\}$
and eigenvalues $\lambda^{(R_a)}_r$, with that for $r=0$ vanishing \cite{Boll98}. The fluctuations are expanded in normal modes, as before,
with now the Red system represented as $y^{(a)}_r$.
This brings us to the three cluster version of the zero-mode equations
\begin{eqnarray}\label{sortofintegrablesystem}
\dot{\alpha}_{B R_1} &=& - \frac{\partial}{\partial \alpha_{B R_1}} \tilde{V}(\alpha_{B R_1}, \alpha_{R_1 R_2}) \nonumber \\
\dot{\alpha}_{R_1 R_2} &=& - \frac{(M_1+M_2)\sigma_R d^{(R_1 R_2)}_T}{M_1 M_2}\left\{ \sin(\alpha_{R_1 R_2}) - \tilde{F}(\alpha_{B R_1})  \right\}
\end{eqnarray}
where,
\begin{eqnarray*}
\tilde{V}(\alpha_{B R_1}, \alpha_{R_1 R_2}) &\equiv& -\tilde{\mu}(\alpha_{R_1 R_2})\alpha_{B R_1} -\sqrt{\tilde{S}^2 +\tilde{C}^2}\cos( \alpha_{B R_1} - \tilde{\varrho}) ,\\
\tilde{\mu}(\alpha_{R_1 R_2}) &\equiv& \bar{\omega} - \bar{\nu}^{(1)} + \frac{\sigma_R d^{(R_1 R_2)}_T}{M_1} \sin(\alpha_{R_1 R_2})\\
\tilde{C} & \equiv&  d^{(BR_1)}_T \left(\frac{ \zeta_{BR} \cos\phi}{N} +\frac{ \zeta_{RB} \cos\psi}{M_1}\right) \\ 
\tilde{S} &\equiv& d^{(BR_1)}_T \left(  \frac{ \zeta_{BR}\sin\phi}{N}  -  \frac{\zeta_{RB} \sin\psi   }{M_1}\right) \\
\tilde{F}(\alpha_{B R_1}) &\equiv&  \frac{ M_2 \left\{ M_1  (\bar{\nu}^{(1)}-\bar{\nu}^{(2)}) +  \zeta_{RB}d^{(BR_1)}_T  \sin(\alpha_{B R_1} + \psi) \right\}}{ (M_1+M_2)\sigma_R d^{(R_1 R_2)}_T} ,
\end{eqnarray*}
and $\tilde{\varrho} =  \tan^{-1} ( \tilde{S}/\tilde{C})$. We see that the equation for $\dot{\alpha}_{B R_1}$ in Eq.(\ref{sortofintegrablesystem}) is a generalisation of $\dot{\alpha}$ in Eq.(\ref{alphaeq}), where $\tilde{V}$ is again a tilted ratchet and $\tilde{F}$ is also the derivative with respect to $\alpha_{R_1 R_2}$
of a tilted ratchet. However, the tilt $\tilde{\mu}(\alpha_{R_1 R_2})$ in Eq.(\ref{sortofintegrablesystem}) depends on the sine of $\alpha_{R_1 R_2}$. 
Fixed points of the deterministic system Eq.(\ref{sortofintegrablesystem}) may be found but a time-dependent solution equivalent to Eq.(\ref{alphasol}) is, to the best of our knowledge, out of reach. We give in \cite{KallZup2015} the equations for $x_r$, $y^{(1)}_{s_1}$ and $y^{(2)}_{s_2}$.

\subsection{Applying noise to Blue}
In \cite{KallZup2015} we explored a regime where, as a consequence of a tight internal coupling of the Blue population, an
increase of cross-couplings triggered fragmentation of Red. Essentially here the competitive interaction between the populations overwhelms the
capacity for one of them to achieve internal coherence. We now explore here how noise {\it specifically on one population}, in this case
the Blue population, may change the dynamic for the other.
We therefore add
\begin{eqnarray*}
\Lambda_i^{(B)}=\sum_{r\in{\cal B}_E }{e_i^{(B,r)}\eta_r^{(B)}},\;\;i\in\mathcal{B}
\end{eqnarray*}
to the equation for the $\beta_i$. This projects a single uncorrelated GWN term to eigenmodes of the Blue network.

In illustrating behaviours here we continue to work with the Blue hierarchy versus the Red random network, 
with internal couplings as before $\sigma_B=8, \sigma_R=0.5$. This time, to enhance the competitive effect and test its limits against internal synchronisation, we set $\phi=\psi=\frac\pi{4}, \zeta_{BR}=\zeta_{RB}$,
and vary $\zeta_{BR}$. We simulate up to time $t=400$ increasing $\zeta_{BR}$ up to 4.5 within the range where, deterministically
we know \cite{KallZup2015} that dynamics kick in for $\alpha_{R_1 R_2}$ (namely there is no solution to the steady-state equation). In other words, 
we test how noise
triggers departure from deterministic equilibrium.

\begin{figure}[h]
\centering
\subfloat[][Blue network order parameters]{\includegraphics[width=.48\linewidth]{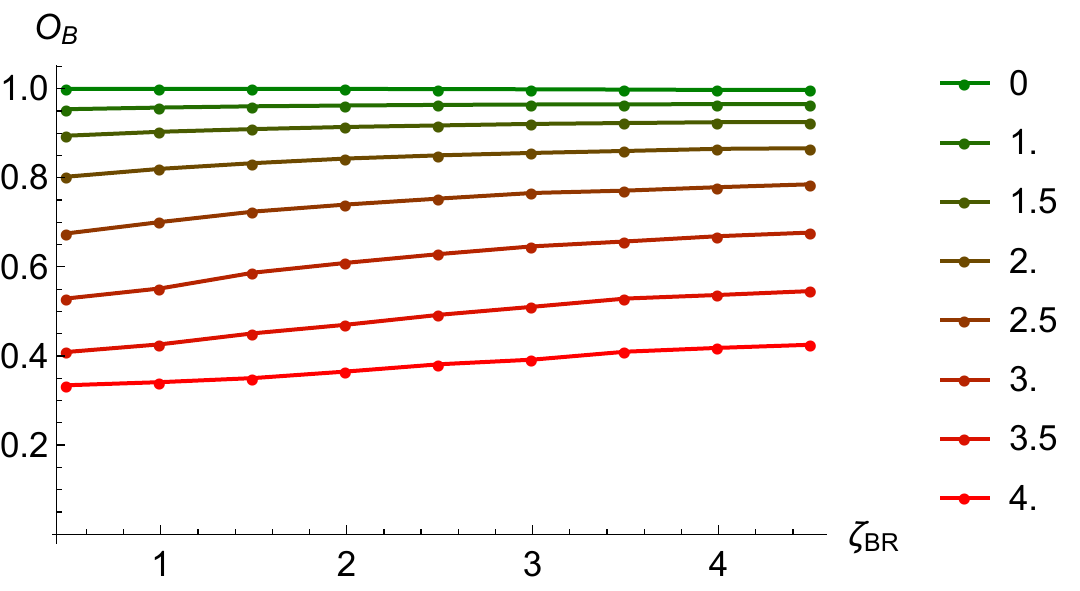}
\label{subfig:3cluster-Blue-order}}
~\subfloat[][Red-1 network order parameters]{\includegraphics[width=.48\linewidth]{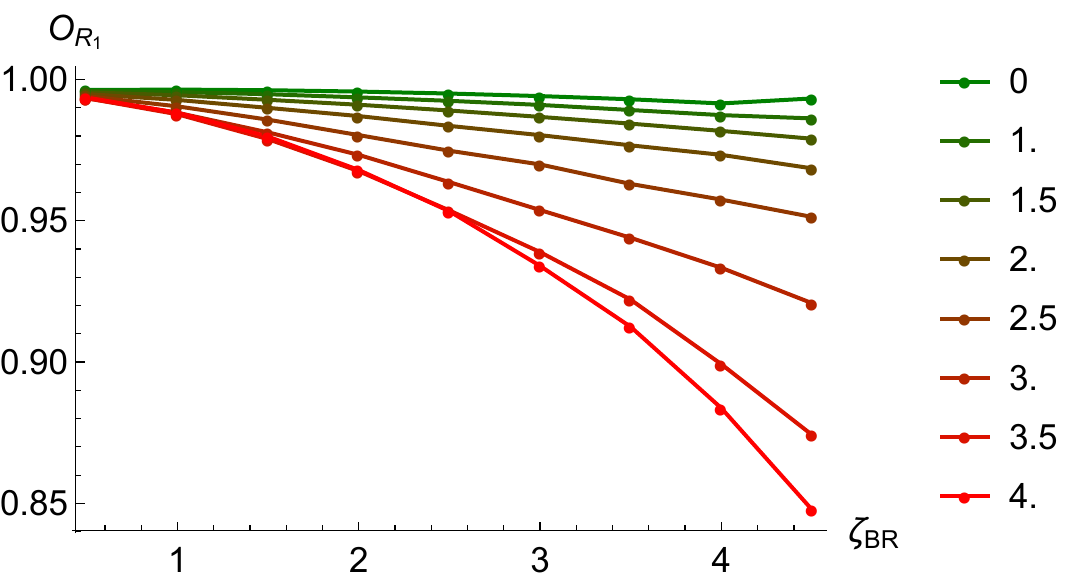}
\label{subfig:3cluster-Red1-order}}

\subfloat[][Red-2 network order parameters]{\includegraphics[width=.48\linewidth]{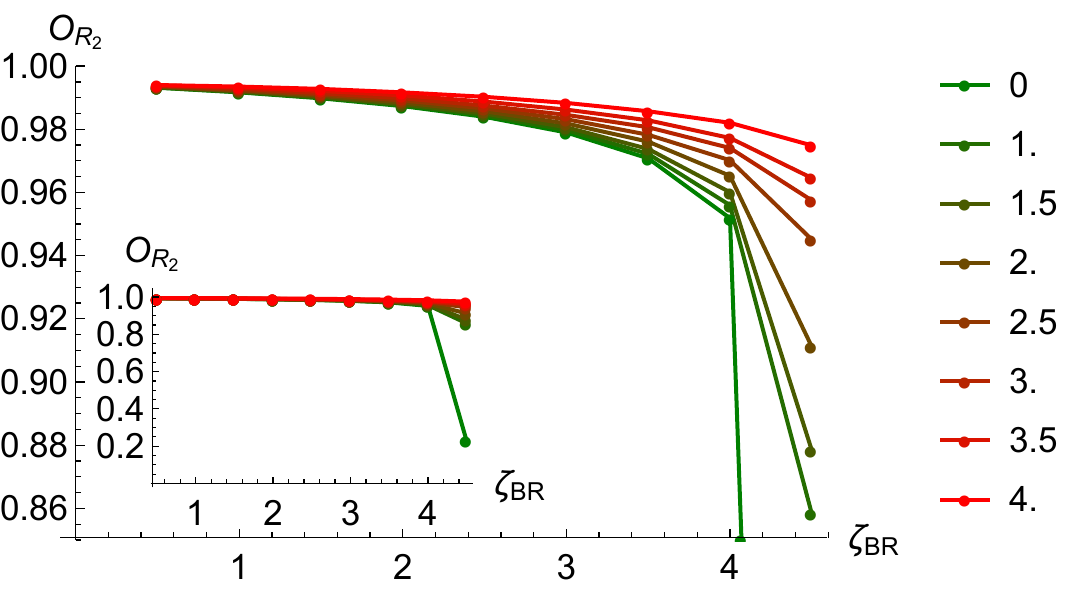}
\label{subfig:3cluster-Red2-order}}

\caption{Order parameters at large time obtained from simulations of the full stochastic system Eq.\eqref{BR-eq} for the Blue tree network
and Red random network, with varying Noise and inter-network coupling. The inter-network coupling varies from $\zeta_{BR}=0.5$ to $4.5$ in steps of $0.5$ and the standard deviation varies from $\sqrt{\Omega}=1$ to $4$ in steps of $0.5$. The orders for the deterministic system (i.e. $\sqrt{\Omega}=0$) are also given. Each point is given by the average of 125 simulations of the stochastic system for each $\zeta_{BR}$ and $\sqrt{\Omega}$ value.}
\label{fig:3-cluster-orders}
\end{figure}

In Fig.\ref{fig:3-cluster-orders} we plot the order parameters $O_B$, $O_{R_1}$ and $O_{R_2}$ averaged over the last 50 time steps and over 125 paths, as a function of increasing $\zeta_{BR}$ for different noise strengths $\Omega$.
These should be compared to plots for the deterministic case in Fig.7 of \cite{KallZup2015}, reproduced in Fig.\ref{fig:3-cluster-orders}
in the green coloured curves. We observe, unsurprisingly, in the two top plots that, firstly as cross-coupling increases there is little impact on the order 
within ${\cal B}$ and ${\cal R}_1$, while, secondly, increasing noise on Blue decreases the overall degree of synchronisation of both Blue (most dramatically) and
${\cal R}_1$ (less severely). Clearly, the more strongly coupled Red agents are to Blue, the more the stochastic behaviour of the latter impacts on the former, destroying synchronisation. 

The novel behaviour is seen in the lower plot of  Fig.\ref{fig:3-cluster-orders} showing the order parameter for ${\cal R}_2$. We emphasise the colour code, with
green low noise and red higher noise. The curves show a pattern the reverse of the top right plot, with noise {\it enhancing} the degree of synchronisation for
${\cal R}_2$. This is the stochastic synchronisation we anticipated. 

Physically, the noise effectively loosens the Blue population, working against its tight internal coupling. This loosening of Blue 
has a second order effect on Red, so that the previously incompatible competition and internal Red goals become more compatible with increasing noise on Blue.

\begin{figure}[h]
\centering
\subfloat[][Stable $\alpha_{BR_1}$ values]{\includegraphics[width=.48\linewidth]{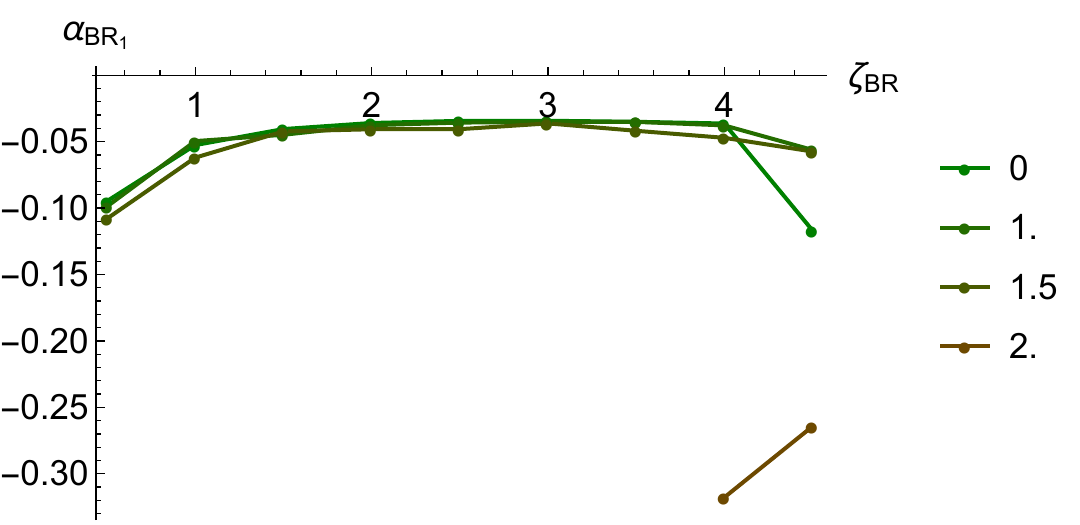}
\label{subfig:3cluster-aBR1}}
~\subfloat[][Stable $\alpha_{R_1R_2}$ values]{\includegraphics[width=.48\linewidth]{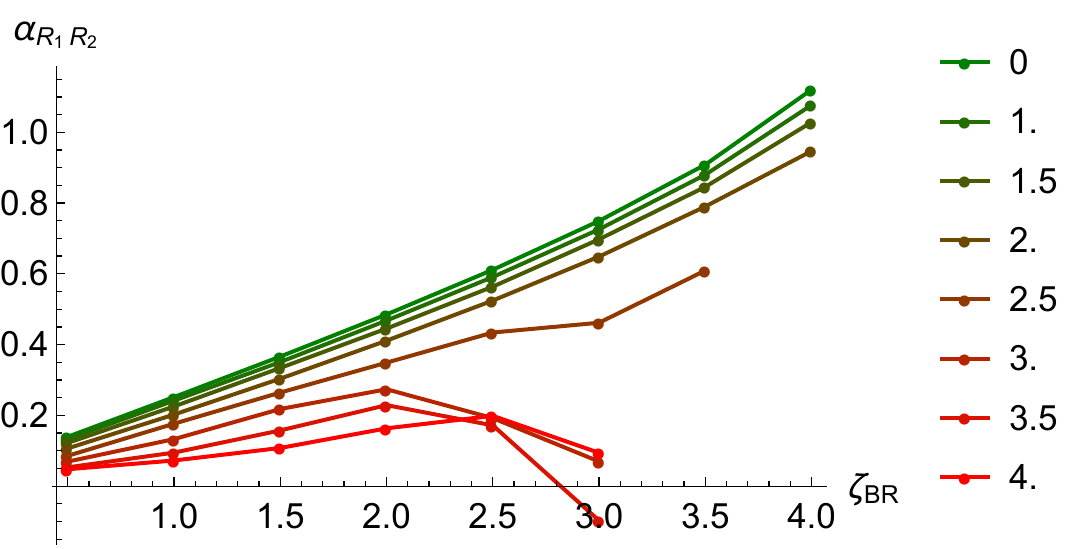}
\label{subfig:3cluster-aBR}}

\caption{Stable $\alpha_{BR_1}$ and $\alpha_{R_1R_2}$ values obtained from simulations of the full stochastic system Eq.\eqref{BR-eq} for the Blue tree network
and Red random network, with varying Noise and inter-network coupling. The inter-network coupling varies from $\zeta_{BR}=0.5$ to $4.5$ in steps of $0.5$ and the standard deviation varies from $\sqrt{\Omega}=1$ to $4$ in steps of $0.5$. Each point is given by the average of 125 simulations of the stochastic system for each $\zeta_{BR}$ and $\sqrt{\Omega}$ value. Points are not present if the diagram if for the simulation has been deemed to provide a ``dynamic'' solution. }
\label{fig:3-cluster-alphas}
\end{figure}

We may explore the impact of this now in the angles between the centroids. Note that we must discern cases where the angles may be
fixed in time against situations where `slippage' occurs and ${\cal R}_2$ rotates through the circle in relation to the potentially
locked ${\cal B}$ and ${\cal R}_2$. In Fig.\ref{fig:3-cluster-alphas} we show the average values of the two angles between centroids but only 
plot a point where we deem that the simulation shows a dynamical value (based on measuring the slope of the trajectory as a function of time in the simulation).

We can see in the left hand plot of Fig.\ref{fig:3-cluster-alphas} for $\alpha_{BR_1}$ that for $\sqrt{\Omega}\ge 2.5$ all solutions have been deemed dynamic;
only low $\Omega$ lead to a fixed negative value for $\alpha_{BR_1}$ - so that Red is marginally leading - that gently rises and then drops off. Note that
the $\Omega=0$ deterministic behaviour is correctly accounted for by the linearisation, as shown in \cite{KallZup2015}.
Contrastingly, there is a growing angle between ${\cal R}_1$ and ${\cal R}_2$ with increasing $\zeta_{BR}$. With increasing noise
this angle decreases confirming the improvement of synchronisation within the Red population: up to a certain $\zeta_{BR}$ the angle is
fixed in time, so there is frequency synchronisation within Red (as well as higher synchronisation within ${\cal R}_2$ while ${\cal R}_1$ is more splayed,
as seen in Fig.\ref{fig:3-cluster-orders}). As noise increases this angle diminishes so that Red approaches phase synchronisation. 
At $\sqrt{\Omega}=2.5$ we see beyond $\zeta_{BR}=3.5$ that $\alpha_{R_1 R_2}$ is no longer constant. For larger $\Omega$, there is a decrease in the threshold value of $\zeta_{BR}$ at which frequency synchronisation within Red breaks down. 

Combining then the views from Fig.\ref{fig:3-cluster-orders} and Fig.\ref{fig:3-cluster-alphas}, the stochastic synchronisation evident in the order
parameter $O_{R_2}$ comes at a price, where frequency synchronisation within the Red population eventually breaks down so that $\alpha_{R_1 R_2}$ is no longer
constant in time. Curves such as these thus provide a means of determing the balance between the conflicting requirements of internal synchronisation,
synchronisation of the competitor population and fulfillment of the strategy of being collectively ahead or behind.

\subsection{Reduced density functions: zero modes}
Though the physical intuition behind the stochastic synchronisation is clear, can the analytical formalism based on linearisation account quantitatively
for this behaviour?
For the linearised system in the three cluster {\it ansatz} the Langevin equations are
\begin{eqnarray}\label{3zeroLang}
\dot{\alpha}_{B R_1} &=& - \frac{\partial}{\partial \alpha_{B R_1}} \tilde{V}(\alpha_{B R_1}, \alpha_{R_1 R_2}) + \eta^{(B)}_0 \\
\dot{\alpha}_{R_1 R_2} &=&- \frac{(M_1+M_2)\sigma_R d^{(R_1 R_2)}_T}{M_1 M_2}\left\{ \sin(\alpha_{R_1 R_2}) - \tilde{F}(\alpha_{B R_1})  \right\}.
\end{eqnarray}
Here only the Blue eigenmodes are explicitly influenced by GWN terms. We explore how far the analysis of the zero mode
may account for the stochastic synchronisation behaviour.

Solving for the reduced marginal densities of the zero modes is not as straightforward as in Sec.\ref{sec:5} as now we have a system of Langevin equations to consider given by Eq.(\ref{3zeroLang}).  We begin by constructing the reduced conditional densities associated with the zero modes.
For the reduced density associated with $\alpha_{B R_1}$, we notice that for fixed $\alpha_{R_1 R_2}$, the Langevin equation for $\dot{\alpha}_{B R_1}$ in Eq.(\ref{3zeroLang}) is of the same form as the corresponding expression for $\dot{\alpha}$ in Eq.(\ref{eqn:integrablesystem-w-noise}). Taking advantage of this we have,
\begin{eqnarray}\label{condalphaBR1}
\begin{split}
\hat{{\cal P}}^{(t)}_{st}(\alpha_{B R_1}| \alpha_{R_1 R_2}) &=&
\frac{ \tilde{\kappa}_1(\alpha_{R_1 R_2}) e^{- \frac{2 \tilde{V}(\alpha_{B R_1},\alpha_{R_1 R_2})}{\Omega} }}{e^{-\frac{4 \pi \tilde{\mu}(\alpha_{R_1 R_2})}{\Omega}}-1} 
\int^{\alpha_{BR_1}+2\pi}_{\alpha_{BR_1}} d \varphi e^{ \frac{2 \tilde{V}(\varphi,\alpha_{R_1 R_2})}{\Omega} },
\end{split}
\end{eqnarray}
where the superscript $(t)$ stands for \textit{tilted-ratchet}, and $\tilde{\kappa}_1(\alpha_{R_1 R_2})$ is the normalisation constant. 
The integrals here may be computed in terms of Bessel functions but this form is more
computationally efficient.

Now $\dot{\alpha}_{R_1 R_2}$ in Eq.(\ref{3zeroLang}) has noise enter entirely through $\alpha_{B R_1}$. 
As before, the reduced conditional density in this case is simply given by the dirac $\delta$-function expression,
\begin{eqnarray}\label{condalphaR1R2}
\hat{{\cal P}}^{(\delta)}_{st}(\alpha_{R_1 R_2}| \alpha_{B R_1}) = \frac{ \left|\cos \left[ \sin^{-1}  \tilde{F}(\alpha_{B R_1})  \right]\right|}{2} \delta \left( \sin( \alpha_{R_1 R_2}) - \tilde{F}(\alpha_{B R_1}) \right) 
\end{eqnarray}
where the superscript $(\delta)$ signifies the delta function nature of this conditional density. In order to obtain the reduced marginal densities, we consider the following reduced joint density identity,
\begin{eqnarray}
\hat{{\cal P}}_{st}(\alpha_{B R_1}, \alpha_{R_1 R_2}) &=& \hat{{\cal P}}^{(t)}_{st}(\alpha_{B R_1}| \alpha_{R_1 R_2})\hat{{\cal P}}^{(t)}_{st}(\alpha_{R_1 R_2}) \nonumber \\
&=& \hat{{\cal P}}^{(\delta)}_{st}(\alpha_{R_1 R_2}| \alpha_{B R_1})\hat{{\cal P}}^{(\delta)}_{st}(\alpha_{B R_1})\nonumber \\
\Rightarrow \frac{\hat{{\cal P}}^{(t)}_{st}(\alpha_{ R_1R_2})}{\hat{{\cal P}}^{(\delta)}_{st}(\alpha_{BR_1 })} &=& \frac{\hat{{\cal P}}^{(\delta)}_{st}(\alpha_{R_1 R_2}| \alpha_{B R_1})}{ \hat{{\cal P}}^{(t)}_{st}(\alpha_{B R_1}| \alpha_{R_1 R_2})} \label{jointdensities}
\end{eqnarray}
where we have similarly placed $(t)$ and $(\delta)$ superscripts on the marginal densities to distinguish them. Taking the integral 
of Eq.(\ref{jointdensities}) over $\alpha_{ R_1 R_2}$ on the interval $(-\pi, \pi )$, and remembering that $\int^{\pi}_{-\pi}d \alpha_{ R_1 R_2} \hat{{\cal P}}^{(t)}_{st}(\alpha_{ R_1R_2}) =1$, we obtain the first marginal density associated with $\alpha_{B R_1}$,
\begin{eqnarray}\label{marginalBR1}
\begin{split}
\hat{{\cal P}}^{(\delta)}_{st}(\alpha_{BR_1}) = \frac{1}{\int^{\pi}_{-\pi} d\alpha_{R_1R_2} \frac{\hat{{\cal P}}^{(\delta)}_{st}(\alpha_{R_1 R_2}| \alpha_{B R_1})}{ \hat{{\cal P}}^{(t)}_{st}(\alpha_{B R_1}| \alpha_{R_1 R_2})}}\\
=  \frac{2 }{ \frac{1}{\hat{{\cal P}}^{(t)}_{st}\left(\alpha_{B R_1}\left|\sin^{-1}  \tilde{F}(\alpha_{B R_1})   \right) \right.} +\frac{1}{\hat{{\cal P}}^{(t)}_{st} \left(\alpha_{B R_1}\left|\pi-\sin^{-1}  \tilde{F}(\alpha_{B R_1}) \right. \right)} }\\
= \hat{{\cal P}}^{(t)}_{st}\left(\alpha_{B R_1}\left|\sin^{-1}  \tilde{F}(\alpha_{B R_1})   \right) \right.
\end{split}
\end{eqnarray}
where we have employed the properties of the delta function in Eq.(\ref{condalphaR1R2}) to complete the integral and the denominators
of the two terms in the intermediate step can be shown to be equal.

Eq.(\ref{marginalBR1}) leads to densities (not plotted here) that peak around the fixed point, are Gaussian-like, flatten for large noise
and sharpen for increasing $\zeta_{BR}$. This aligns with the results in the left hand plot of Fig.\ref{fig:3-cluster-alphas}, essentially smearing
the deterministic value of $\alpha_{BR_1}$ up to where noise destroys localisation of the system, 
so the state leaves the basin of attraction and no constant angle is
permitted.

More interesting is the remaining reduced marginal density for $\alpha_{R_1 R_2}$, given explicitly by
\begin{eqnarray}\label{marginalR1R2}
\begin{split}
\hat{{\cal P}}^{(t)}_{st}(\alpha_{ R_1R_2}) = \int^{\pi}_{-\pi} d \alpha_{ B R_1} \hat{{\cal P}}^{(\delta)}_{st}(\alpha_{R_1 R_2}| \alpha_{B R_1})\hat{{\cal P}}^{(\delta)}_{st}(\alpha_{ BR_1})\\
= \frac{(M_1 + M_2) \sigma_R d^{(R_1 R_2)}_T |\cos (\alpha_{R_1 R_2})|}{2 M_2 \zeta_{RB} d^{(BR_1)}_T  \sqrt{1-\tilde{Y}^2(\alpha_{R_1 R_2})}} \\
\times \left\{ \hat{{\cal P}}^{(\delta)}_{st}(-\psi+ \sin^{-1} \tilde{Y}(\alpha_{ R_1 R_2}) )+ \hat{{\cal P}}^{(\delta)}_{st}(\pi -\psi- \sin^{-1} \tilde{Y}(\alpha_{ R_1 R_2}) )  \right\}
\end{split}
\end{eqnarray}
for,
\begin{eqnarray*}
\tilde{Y}(\alpha_{R_1 R_2}) = \frac{\frac{(M_1+M_2)}{M_2} \sigma_R d^{(R_1 R_2)}_T \sin(\alpha_{R_1 R_2}) - M_1 (\bar{\nu}^{(1)}-\bar{\nu}^{(2)})}{\zeta_{RB} d^{(BR_1)}_T}.
\end{eqnarray*}

\begin{figure}
\centering
\includegraphics[width=1\linewidth]{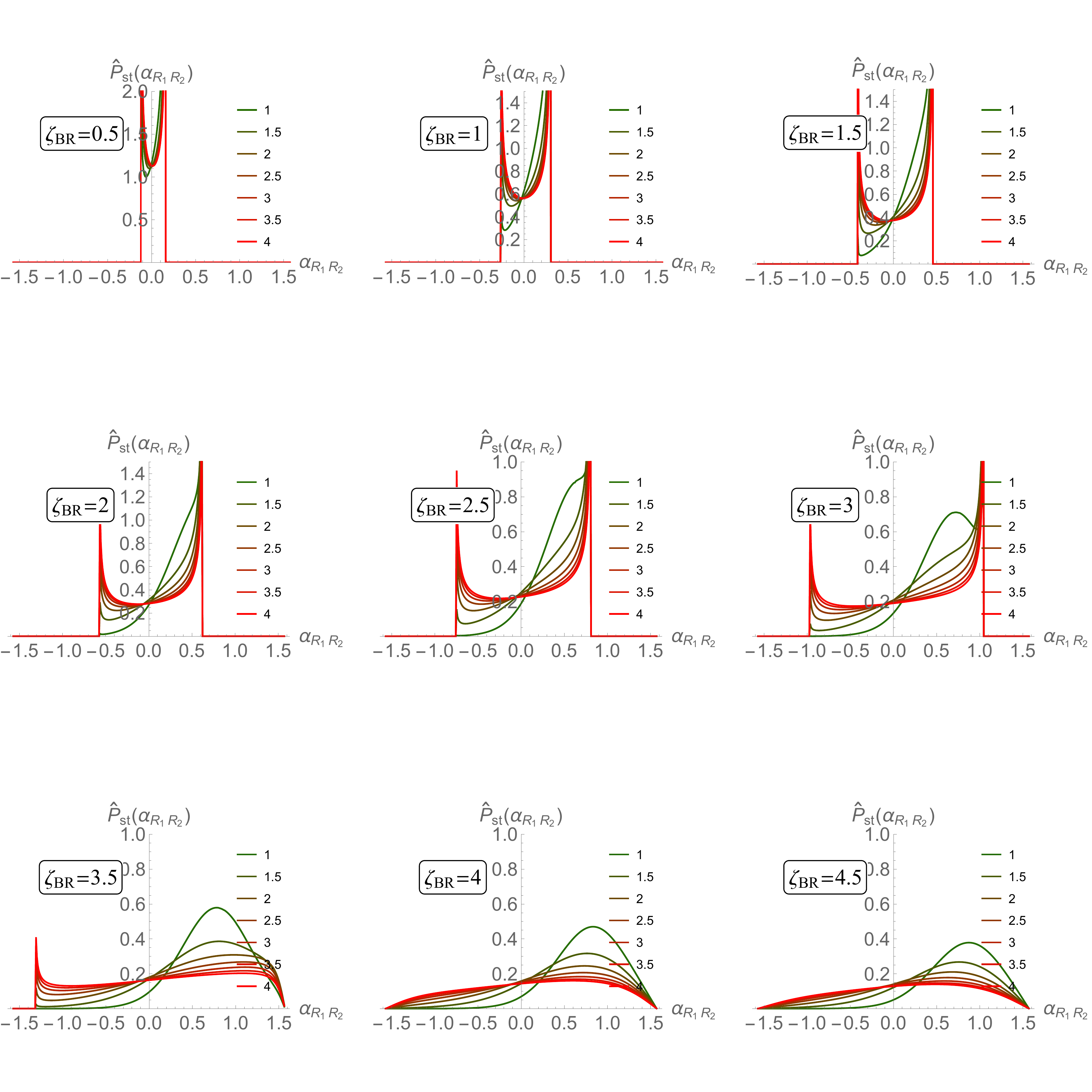}

\caption{Plots of Eq.(\ref{marginalR1R2}) for the Blue tree network
and Red random network, with varying Noise and inter-network coupling. As with Fig.\ref{fig:3-cluster-alphas}, the  inter-network coupling varies from $\zeta_{BR}=0.5$ to $4.5$ in steps of $0.5$ and the standard deviation varies from $\sqrt{\Omega}=1$ to $4$ in steps of $0.5$.}
\label{fig:AlphaR1R2densities}
\end{figure}

We plot Eq.(\ref{marginalR1R2}) in Fig.\ref{fig:AlphaR1R2densities} and see stark bounding of the densities.
They are not Gaussian. For $\zeta_{BR}$ values of 0.5 to 3, they are bound from both sides from the term $\sqrt{1-\tilde{Y}^2(\alpha_{ R_1 R_2})}$ in the denominator of Eq.(\ref{marginalR1R2}). At $\zeta_{BR} = 3.5$, the density is bound on the left but is equal to zero on the right at $\pi/2$ due to the 
$|\cos(\alpha_{R_1 R_2})|$ term in the numerator of Eq.(\ref{marginalR1R2}). 
Finally at $\zeta_{BR} = \{4,4.5\}$ both families of densities are equal to zero at $-\pi/2$ and $\pi/2$ due to the $|\cos(\alpha_{R_1 R_2})|$ term. It must be noted that all of the above densities are reflected at $\alpha_{R_1 R_2} = \pi/2$. We do not show this range for the densities, at least for steady state considerations, as Fig.\ref{subfig:3cluster-aBR} shows that it is unphysical.

The plots with $\zeta_{BR}$ varying from 0.5 to 3 do very well in explaining Fig.\ref{subfig:3cluster-aBR} in that all the densities are bound within 
a narrow basin with exactly zero probability for escape, consistent with a steady state solution. Also, densities for low noise are sharply peaked around the fixed point at positive $\alpha_{R_1 R_2}$; as noise increases for fixed $\zeta_{BR}$ the densities flatten around the origin, but do not lose their boundedness (most clearly seen for $\zeta_{BR}=1.5$). 
This flattening means that the expected value of $\alpha_{R_1 R_2}$ shifts from the fixed point to the origin,
exactly what we see in the right hand plot of Fig.\ref{subfig:3cluster-aBR}. 
Significantly, the densities for large $\zeta_{BR}$ and $\Omega$ in Fig.\ref{fig:AlphaR1R2densities} show a flattening and loss of
double-boundedness from $\zeta_{BR} > 3$.
Thus, as noise strengthens there is an increase in likelihood of $\alpha_{R_1R_2}$ taking values outside the basin of attraction triggering dynamic behaviour. This is consistent with Fig.\ref{subfig:3cluster-aBR}. However, the absence of sharp change in behaviour here, suggests a
 limitation of the linearisation.

\begin{figure}
\centering
\includegraphics[width=0.5\linewidth]{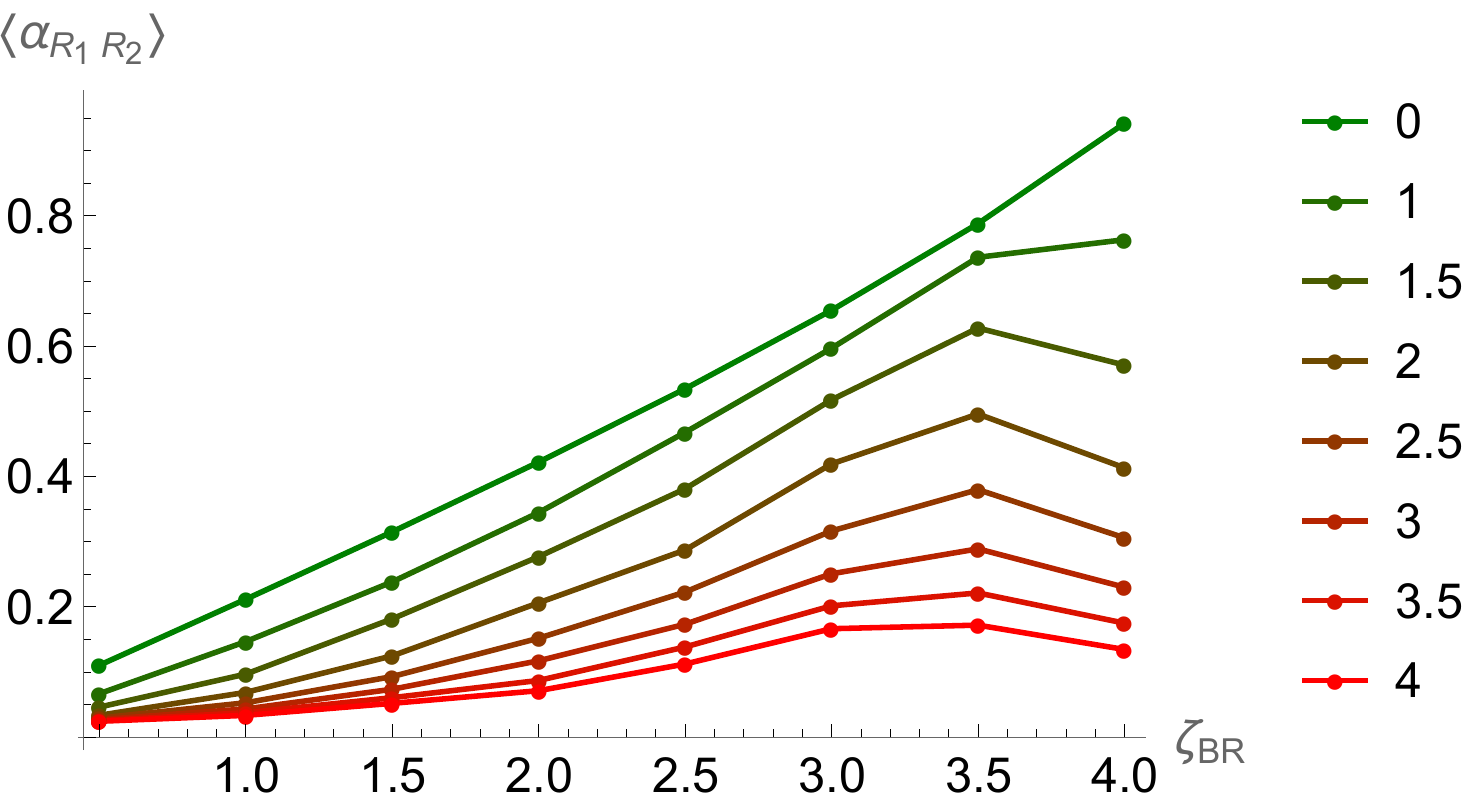}

\caption{Plots of Eq.(\ref{expectedR1R2}) for the Blue tree network
and Red random network, with varying Noise and inter-network coupling. As usual, the  inter-network coupling varies from $\zeta_{BR}=0.5$ to $4$ in steps of $0.5$ and the standard deviation varies from $\sqrt{\Omega}=1$ to $4$ in steps of $0.5$.}
\label{fig:AlphaBR1expected}
\end{figure}

Finally, in Fig.\ref{fig:AlphaBR1expected} we plot
\begin{eqnarray}
\langle \alpha_{R_1R_2} \rangle = \frac{\int^{\varphi^+}_{\varphi_-}d\alpha_{R_1R_2} \hat{{\cal P}}^{(t)}_{st}(\alpha_{ R_1R_2}) \alpha_{ R_1R_2}}{\int^{\varphi^+}_{\varphi_-}d\alpha_{R_1R_2} \hat{{\cal P}}^{(t)}_{st}(\alpha_{ R_1R_2})}
\label{expectedR1R2}
\end{eqnarray}
where $\varphi^+$ and $\varphi^-$ are given by the allowable regions of $\sqrt{1-\tilde{Y}^2(\alpha_{ R_1 R_2})}$ in $( - \pi/2,\pi/2 )$ and $\hat{{\cal P}}^{(t)}_{st}(\alpha_{ R_1R_2})$ is given by Eq.(\ref{marginalR1R2}).

We compare the linearisation based Fig.\ref{fig:AlphaBR1expected} with the numerical simulation in Fig.\ref{subfig:3cluster-aBR}. We see that 
Fig.\ref{fig:AlphaBR1expected} emulates Fig.\ref{subfig:3cluster-aBR} specifically in the decrease of the expected value of the angle with increasing noise values (red lines are below green lines). The results also correctly reproduce the slight increase in
$\alpha_{R_1 R_2}$ with $\zeta_{BR}\approx 3.5$ followed by decrease for higher $\Omega$. 
However, the analytical approach does not detect a sharp threshold for the onset of dynamics, so that curves in Fig.\ref{fig:AlphaBR1expected} are given
for the entire range of $\zeta_{BR}$ in contradistinction to Fig.\ref{subfig:3cluster-aBR}. 
Additionally, Fig.\ref{fig:AlphaBR1expected} does not pick up the interesting crossing of the large noise ($\sqrt{\Omega}=\{3,3.5,4\}$) lines for the $\zeta_{BR}$ values from 2 to 3. 

\subsection{Back to the order parameter}
$O_{R_2}$ is straightforwardly computed in the linearised
approximation after observing that its magnitude may be written as a sum over cosine of phase differences
\begin{eqnarray*}
|O_{R_2}| = \frac{1}{M_2^2} \sum_{j,j'\in{\cal R}_2} \cos^2(p_j^{(2)}-p_{j'}^{(2)}).
\end{eqnarray*}
With linearisation, this may be expanded in ${\cal R}_2$ Laplacian modes, so that using completeness and orthonormality (analogous to a computation in \cite{Kall2014}
for the ordinary Kuramoto model), we obtain the representation
\begin{eqnarray}
|O_{R_2}| = 1 - \frac{1}{M_2} \sum_{r\in{\cal R}_2^E/\{0\}} (y_r^{(2)}(t))^2.
\end{eqnarray}
As noise does not explicity influence the Red eigenmodes, to leading order $O_{R_2}$ only depends on the angle $\alpha_{R_1 R_2}$, which is stochastic. Thus, at steady-state
\begin{eqnarray}
|O_{R_2}| = 1 -  \frac{1}{M_2} \sum_{r\in{\cal R}_2^E/\{0\}} \left( \frac{\nu_r^{(2)}}{\sigma_R \lambda_r^{(R_2)}} + \frac{d_r^{(R_1 R_2)}}{\lambda_r^{(R_2)}}\sin\alpha_{R_1 R_2}
\right)^2.
\end{eqnarray}
We see then that, given $\langle \alpha_{R_1 R_2}\rangle \rightarrow 0^+$ monotonically as $\Omega$ increases in Fig.\ref{fig:AlphaBR1expected}, it follows that $|O_{R_2}|\rightarrow 1^-$ {\it monotonically}. Thus the manifestation of stochastic synchronisation in the order parameter for the fragmented Red subpopulation is
reproduced  in the linearised approach by the stochastic decrease in the angle between the two Red populations.


\section{Conclusions and Discussion}
We have explored the impact of Gaussian White Noise on the deterministic behaviour of a frustrated two network - or `Blue vs Red' model.
In particular observing the different ways noise may be fed to sub-structures of the network to generate quite different departures
from frequency locked synchronisation. The most remarkable of these is when noise acts on zero modes of the Laplacian - essentially the coarsest structure
of the two networks. This stochastically drives one population to lap the other in increasingly random ways. A linear approximation close to
where the populations are phase locked, or alternatively where one is at the point of fragmenting, allows analytic computations using the eigenvector decomposition of Laplacians for the sub-networks of the different populations. 

The linearisation goes quite far in explaining the numerical behaviours, particularly
where the rather beautiful theory of tilted ratchet potentials proves fundamental. Specific dependences arising from this theory lead to results that are counter-intuitive but visible in numerical simulations. These dependencies include the difference of average frequencies of the populations, which determines the direction of drift
in the ratchet potential, the internal couplings multiplied by the individual Laplacian eigenvalues for the internal networks of the two populations, which determines the diffusivity of the Fokker-Planck densities, and the projection of the inter-network degree matrix onto Laplacian eigenvectors for the populations, which figures in numerous results through the analysis.
The quantity ${\cal K}$, arising from the analysis where the populations are close to phase synchrony, unifies network, frequency and coupling information, and leaves a significant imprint on the stochastic dynamics.
As a consequence of the two network frustrated model having more structure than the Kuramoto model, linearisation goes quite far in accounting for 
thresholds for changes in behaviour with and without noise. 

The phenomenon of stochastic synchronisation is observed, whereby noise on the tightly
coupled sub-system allows another part with which it interacts to improve in synchronisation. Analytically this can be accounted for in quite some quantitative detail, though not
in quantifying the sharp threshold of dynamics. The linearisation takes us quite far, but there are evident places where missing dynamics kicks in, such as the faster period and
this transition to dynamics in the three cluster case.

Extensions of this work include the consideration of the varieties of non-Gaussian noise, and
the impact of noise on tunability of a frustrated system. 
Also, a study of mean first passage time in tilted ratchets may yield more quantitative analytical results about the onset of dynamics at critical thresholds in
the cases studied here. In view of the frustrated two-network model offering a representation of
competing teams with well-defined strategies for advantage for one over the other, there is the enticing prospect of a game theoretic treatment of this model - and its natural extension
to stochastic Game Theory.

%
%
%
\section*{Appendix A: Definitions for the linearised Blue-vs-Red model}
We define the quantities $d$ and $L$ used in the main body, drawing basic concepts from graph theory \cite{Boll98}. 
The {\it degree} of Blue agent at node $i \in {\cal B}$ is the number of links from node $i$ to other Blue agents,
\begin{eqnarray*}
d^{(B)}_i \equiv \sum_{j\in {\cal B}} {\cal B}_{ij}, \;\; i \in {\cal B}.
\end{eqnarray*}
The corresponding Blue {\it degree-matrix} ${\cal D}^{(B)}$ is a diagonal matrix with the degrees $d^{(B)}_i$ inhabiting the diagonal entries
\begin{eqnarray*}
{\cal D}^{(B)}_{ij} \equiv d^{(B)}_i \delta_{ij}, \;\;\{i,j\} \in {\cal B}. 
\end{eqnarray*}
The matrices $L$ in Eq.(\ref{linsys}) constitute a {\it graph Laplacian}, where the Laplacian for the Blue population is given by the expression
\begin{eqnarray*}
L^{(B)}_{ij}\equiv {\cal D}^{(B)}_{ij} - {\cal B}_{ij}, \;\;\{i,j\} \in {\cal B}.   
\end{eqnarray*}
Equivalent definitions for $L^{(R)}$ apply to the Red network.

Addressing the corresponding cross-network quantities, we define the degree with which a Blue agent at node $i$ connects to Red agents as,
\begin{eqnarray*}
d^{(BR)}_i \equiv \sum_{j\in {\cal B} \cup {\cal R}} {\cal M}_{ij} = \sum_{j\in {\cal R}} {\cal A}^{(BR)}_{ij},\;\; i \in {\cal B}. 
\end{eqnarray*}
We note that in this instantiation of the model ${\cal A}^{(BR)}$ is the transpose of ${\cal A}^{(RB)}$, and vice-versa (it need not be so, however), so there is symmetry between the total number of degrees between the Blue and red networks:
\begin{eqnarray}
d^{(BR)}_T = d^{(RB)}_T= \sum_{i \in {\cal B}}d^{(BR)}_i =\sum_{i \in {\cal R}}d^{(RB)}_i .
\label{totaldegree}
\end{eqnarray}
The diagonal degree matrix ${\cal D}^{(BR)}$ which encodes all the Blue to Red links is given by,
\begin{eqnarray*}
{\cal D}^{(BR)}_{ij} \equiv d^{(BR)}_i \delta_{ij},\;\; i \in {\cal B}, \;\; j \in  {\cal B} \cup {\cal R}   ,
\end{eqnarray*}
where the final $M$ diagonal entries of ${\cal D}^{(BR)}$ are zero. This finally leads to the cross-network Laplacian from Blue to Red
\begin{eqnarray*}
L^{(BR)}_{ij}\equiv {\cal D}^{(BR)}_{ij} - {\cal M }_{ij}, \;\; i \in {\cal B}, \;\; j \in  {\cal B} \cup {\cal R}  .   
\end{eqnarray*}
Similar considerations lead to an equivalent expression for the Red to Blue cross-network Laplacian $L^{(RB)}$. 

The Blue and Red network Laplacians  obey the following eigenvalue equations,
\begin{eqnarray*}
\sum_{j \in {\cal B}}L^{(B)}_{ij}e^{(B,r)}_j = \lambda^{(B)}_r e^{(B,r)}_i \;\; i \in {\cal B} , \;\; \sum_{j \in {\cal R}}L^{(R)}_{ij}e^{(R,r)}_j = \lambda^{(R)}_r e^{(R,r)}_i \;\; i \in {\cal R}.
\end{eqnarray*}
The graph eigenvalues are well-studied objects \cite{Boll98}. For instance, the \textit{zeroth}-eigenvalue is always zero valued, and the remaining eigenvalues are all real, positive, semi-definite and can be ordered as follows,
\begin{eqnarray*}
0 = \lambda^{(B)}_0  \le \lambda^{(B)}_1 \le \lambda^{(B)}_2 \le \dots \le \lambda^{(B)}_N 
\end{eqnarray*}
where we have used the Blue network as an example. The normalised zero eigenvectors are
\begin{eqnarray*}
e^{(B,0)}_i = \frac{1}{\sqrt{N}} \;\; i \in {\cal B}, \;\; e^{(R,0)}_i = \frac{1}{\sqrt{M}} \;\; i \in {\cal R}.
\end{eqnarray*}
They provide an alternate expression for the Blue and Red centroids:
\begin{eqnarray*}
B =\frac{1}{\sqrt{N}} \sum_{i \in {\cal B}}\beta_i e^{(B,0)}_i, \;\; P = \frac{1}{\sqrt{M}} \sum_{j \in {\cal R}}\rho_j e^{(R,0)}_j.
\end{eqnarray*}

Projections onto the eigenvectors are:
\begin{eqnarray*}
\omega^{(s)} = \sum_{i \in {\cal B}}\omega_i e^{(B,s)}_i, \;\; \bar{\omega} = \frac{1}{N}\sum_{i \in {\cal B}}\omega_i, \;\;d^{(BR)}_s =  \sum_{i \in {\cal B}}d^{(BR)}_i e^{(B,s)}_i,
\end{eqnarray*}
and equivalent expressions hold for the relevant Red quantities. 

Fluctuations $b_i$ and $p_j$ are expanded in the following non-zero \textit{normal-modes}:
\begin{eqnarray}
b_i = \sum_{r \in {\cal B}^E / \{0\}} x_r e^{(B,r)}_i \;\; i \in {\cal B}, \;\; p_j = \sum_{r \in {\cal R}^E / \{0\}} y_r e^{(R,r)}_j \;\; j \in {\cal R}.
\end{eqnarray}

The dynamical equations for the normal modes Eq.(\ref{integrablesystem}) may be explicitly solved to give:
\begin{equation}\label{eq:non-zero-modes}
\begin{split}
x_s = x'_s \re^{-\sigma_B \lambda^{(B)}_s t}  +\frac{ \omega^{(s)}}{\sigma_B \lambda^{(B)}_s} \left\{ 1- \re^{-\sigma_B \lambda^{(B)}_s t}  \right\} \\
- \zeta_{BR}d^{(BR)}_s \int^t_0 d \tau \re^{\sigma_B \lambda^{(B)}_s(\tau- t)} \sin(\alpha(\tau)-\phi), \;\; s \in {\cal B}_E / \{0\}, 
\end{split}
\end{equation}
where $x'_s$ is the initial condition. A similar equation applies for $y_s$. It is evident that the integral expressions in the non zero normal-mode solutions, due to the presence of two clusters, offers a slight generalisation to the equivalent expression for only one cluster (see \cite{Kall10} for an example). We note that if $\mathcal{K}<0$ then $\sin(\alpha(t)-\phi)$ is a periodic function of time with period $T:=\frac{2\pi}{\sqrt{|\mathcal{K}|}}$ and under the circumstance that the initial condition is given by
\begin{equation}
x_s'=\frac{\omega^{(s)}}{\sigma_B\lambda_s^{(B)}}-\frac{\zeta_{BR}d_s^{(BR)}}{1-\re^{-\sigma_B\lambda_s^{(B)}T}}\int^T_0 d \tau \re^{\sigma_B \lambda^{(B)}_s(\tau- T)} \sin(\alpha(\tau)-\phi),
\end{equation}
then $x_s$ is strictly time periodic with period $T$ and no exponential decay. It can then be show by standard theory \cite{Hale1969,Farkas1994} that this periodic solution is unique and globally stable. That is, for more general initial conditions $x_s'$, the solutions will decay to this time periodic solution.

\section*{Appendix B: Friedlin-Wentzel theory for the ratchet potential}
The qualitative statement from FW theory used in the main body may be quantified in a number a ways. A classic result \cite{Wellens,Hanggi} involves the integral in Eq.(\ref{avdotalpha}). Assuming that ${\cal K} >0$ and the noise is suitably small then the main contributions from $e^{ \frac{V(\alpha)}{\Omega} }$ are the local maxima of $V(\alpha)$, $\alpha^+ = \pi + \varrho - \sin^{-1}\left( \frac{\bar{\omega}-\bar{\nu}}{\sqrt{s^2+C^2}} \right)$. 
Correspondingly, the main contributions from $e^{- \frac{V(\alpha)}{\Omega} }$ are the local minima of $V(\alpha)$, $\alpha^- = \varrho + \sin^{-1}\left( \frac{\bar{\omega}-\bar{\nu}}{\sqrt{s^2+C^2}} \right)$. 
Applying the saddle-point method to these integrals (approximating the functions $\pm V(\alpha)$ as quadratic forms with maxima at $\alpha^{\pm}$) one obtains the following expression,
\begin{eqnarray}
\langle \dot{\alpha} \rangle \approx  \frac{-2 \pi \Omega e^{\frac{V(\alpha^-)-V(\alpha^+)}{\Omega} }}{\left(1+ \frac{1}{e^{-\frac{2 \pi \mu}{\Omega}-1}} \right) \int^{\infty}_{-\infty}d \vartheta e^{-\frac{V''(\alpha^-)}{2 \Omega} (\vartheta-\alpha^-) ^2} \int^{\infty}_{-\infty}d \varphi e^{\frac{V''(\alpha^+)}{2 \Omega} (\varphi-\alpha^+) ^2 }}\nonumber\\
\approx \underbrace{\sqrt{|V''(\alpha^-)V''(\alpha^+ -2\pi)|}e^{\frac{V(\alpha^-)-V(\alpha^+ -2\pi)}{\Omega}}}_{{\cal W}_1} -\underbrace{\sqrt{|V''(\alpha^-)V''(\alpha^+ )|}e^{\frac{V(\alpha^-)-V(\alpha^+ )}{\Omega}}}_{{\cal W}_2}.\nonumber\\
\label{SADDLE}
\end{eqnarray}
Here we recognise ${\cal W}_1$ as the Eyring-Kramers transition rate \cite{Kramers} from $\alpha^-$ over the left barrier $\alpha^+-2 \pi$. Similarly,  ${\cal W}_2$ is the transition rate from $\alpha^-$ over the right barrier $\alpha^+$. As explained in \cite{Wellens}, the \textit{exact} expression for $\langle \dot{\alpha} \rangle$ given by Eq.(\ref{avdotalpha}) monotonously increases as a function of noise strength $\Omega$, where $\lim_{\Omega \rightarrow \infty}\langle \dot{\alpha} \rangle = \mu$. This can be contrasted with Eq.(\ref{SADDLE}) where the saddle point approximations result in $\langle \dot{\alpha} \rangle$ possessing an optimum \textit{finite} noise strength, labeled $\Omega^*$, where $\langle \dot{\alpha} \rangle$ reaches a global maximum. However, the size of $\Omega^*$ is comparable to the height of the peaks in the potential $V(\alpha)$, which indicates a breakdown of the saddle point approximation.

\section*{Appendix C: Networks and frequencies used in numerical calculations} \label{APP_C}
\begin{figure}[h]
\centering
\subfloat[][Blue Network (Tree)]{\includegraphics[width=0.45\linewidth]{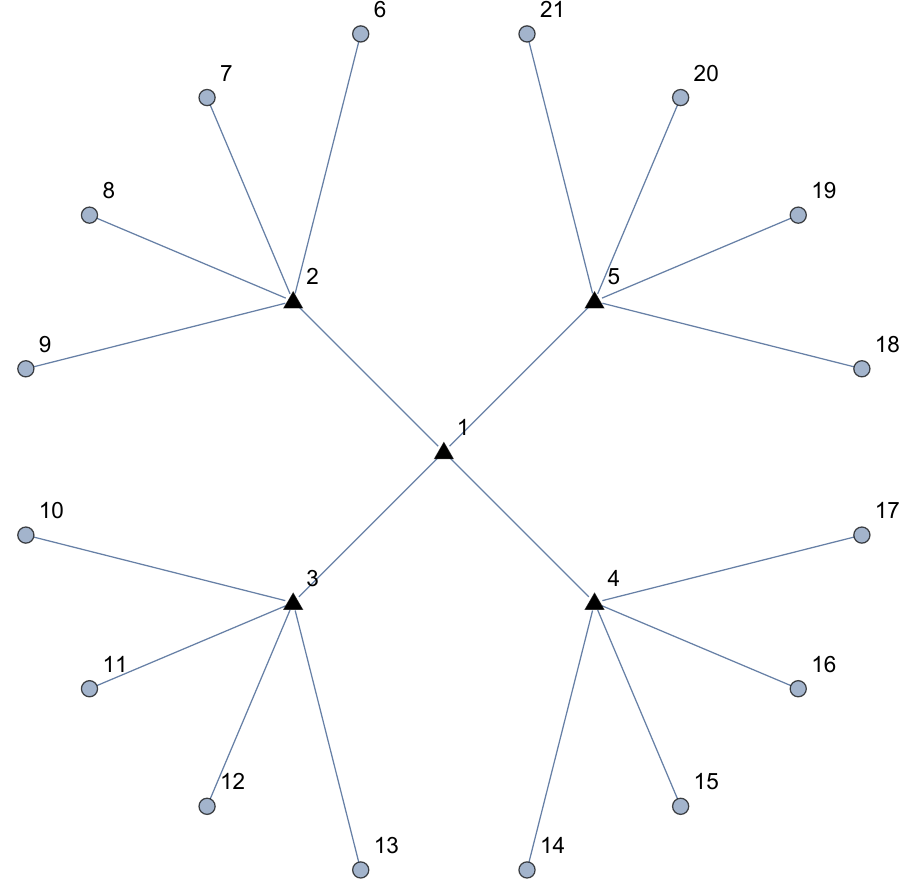}}
~
\subfloat[][Red Network (Erd\H{o}s-R\'{e}nyi)]{\includegraphics[width=0.45\linewidth]{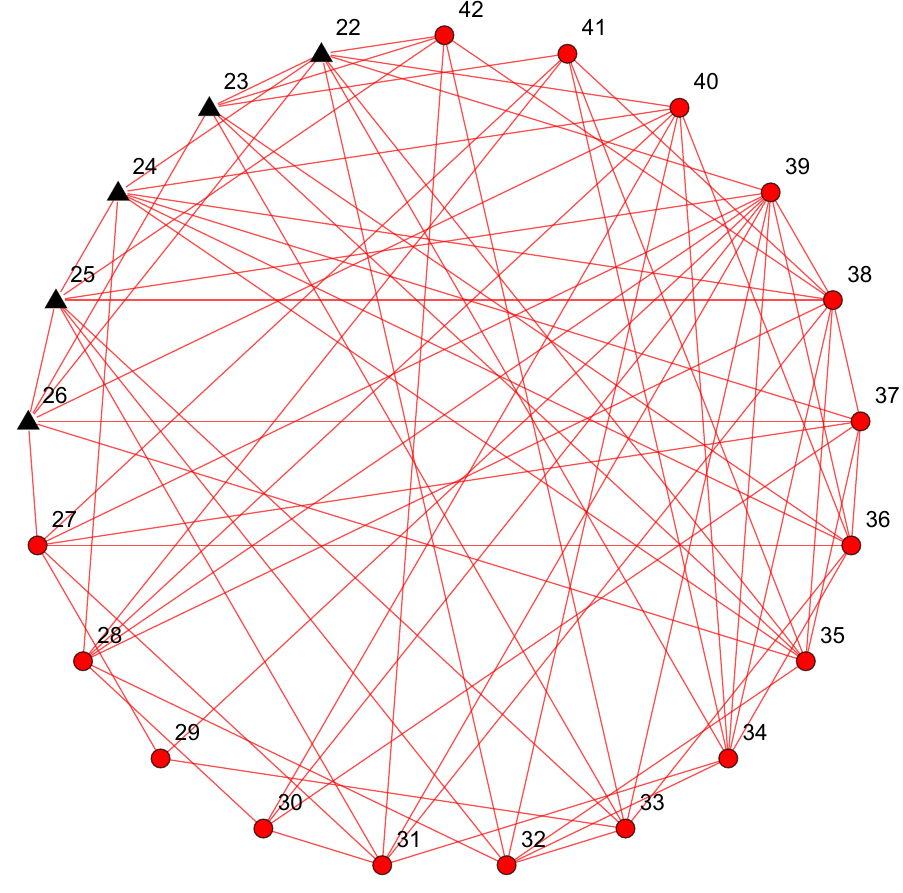}}
\caption{Blue and Red networks used in the numerical simulation of Eq.\eqref{BR-eq} with noise}
\label{fig:BvsR-networks}
\end{figure}

As mentioned in the main body, a tree network was used for Blue and a random graph for Red.
Blue-to-Red interactions are arranged such that each leaf-node of Blue ($i=6,\dots,21$) interacts with the correspondingly labelled Red node ($i=27,\dots,42$), shown as open circles in Fig.\ref{fig:BvsR-networks}. Thus
$d_i^{(BR)}=d_i^{(RB)}= 1 \ \rm{or} \ 0$, for agents engaged, respectively not engaged, with a competitor, but $d^{(BR)}_T=16$.
\begin{figure}[h]
\begin{center}
\includegraphics[height=4.5cm]{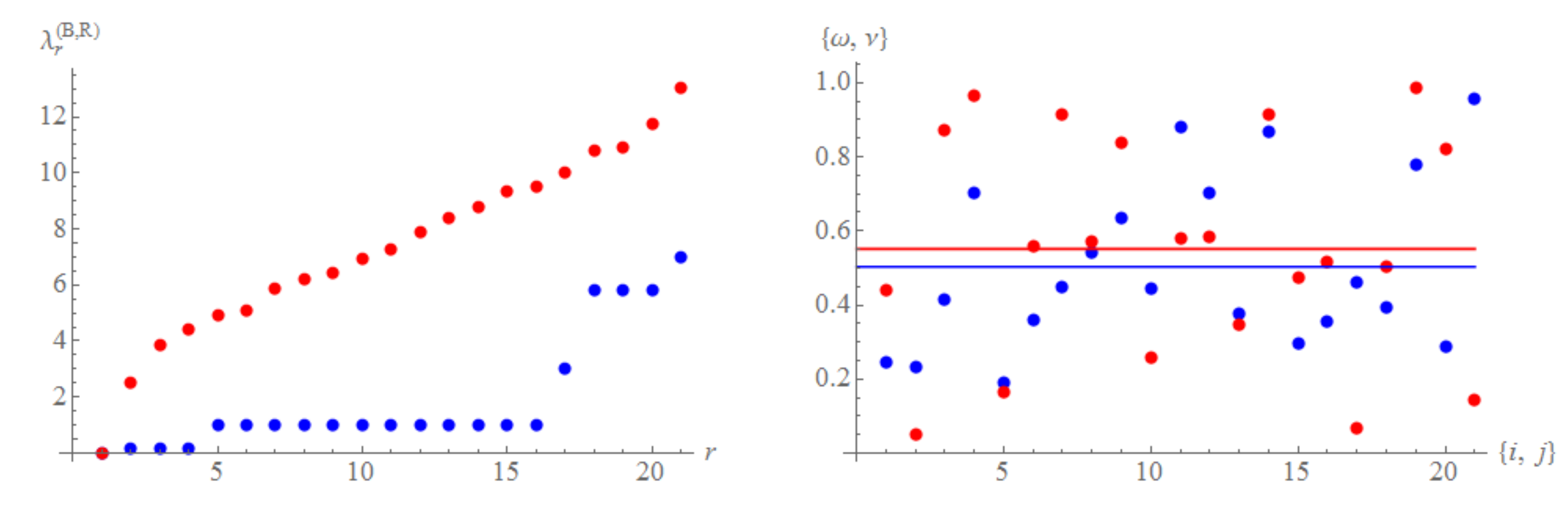}
\caption{Left: The spectrum of the graph Laplacian for Blue and Red networks coloured respectively blue and red. 
Right: The frequencies for Blue and Red agents according to the node labelling again coloured blue and red respectively, with solid lines indicating their corresponding means.} 
\label{spectra}
\end{center}
\end{figure}
Given the significance of the graph Laplacians we show their spectra and the frequencies of each agent in Fig.\ref{spectra}. Note here that there are 
many more low lying eigenvalues for $L^{(B)}$ compared to $L^{(R)}$ - a 
consequence of the comparatively poor connectivity of the tree graph
(left, Fig.\ref{spectra}).

\section*{Appendix D: Applying noise to all the modes} \label{APP_D}
Here we apply noise uniformly across all Blue and Red network Laplacian  modes, hence we consider the full system shown in Eq.\eqref{eqn:integrablesystem-w-noise}. 
We note that the results obtained in the main body for the marginal density of $\alpha$ will be instrumental in obtaining results for the probability densities of this system.

\subsection*{D.1 Analytical considerations}
Considering the steady-state regime again, it is possible to see that the marginal distribution for $\alpha$ in this case, with noise in \textrm{all} the eigenmode Langevin equations, is exactly the same as in Sec.\ref{sec:5}. Thus, due to the tilted ratchet potential, it is necessary to consider the reduced density, and we again obtain Eq.(\ref{integralmess}) for $\hat{{\cal P}}_{st}(\alpha)$. Nevertheless, the normal mode Langevin equations now possess explicit noise terms, as opposed to having stochastic behaviour introduced through the sine expressions as in Sec.\ref{sec:5}. Due to this, the conditional densities for the normal modes are now Gaussians,
\begin{eqnarray}
{\cal P}_{st}(x_r|\alpha) = \sqrt{\frac{\sigma_B \lambda^{(B)}_r}{\pi \Omega}}\exp \left\{ -\frac{\sigma_B \lambda^{(B)}_r}{ \Omega} \left[ x_r- \frac{\omega^{(r)} -\zeta_{BR} d^{(BR)}_r \sin(\alpha - \phi) }{\sigma_B \lambda^{(B)}_r} \right]^2 \right\}
\label{normalmodejoint}
\end{eqnarray}
and equivalent expressions for ${\cal P}_{st}(y_s|\alpha)$. As in Sec.\ref{sec:5}, the conditional probabilities are already $2 \pi$-periodic in $\alpha$, thus we do not need to perform the reduction-operation in Eq.(\ref{eq:reduced-probabilities}) in the $\alpha$ argument. Therefore, applying the marginal probability for $\alpha$, we can obtain the marginal probability for $x_r$ by way of
\begin{eqnarray}\label{marginalnormalall}
\calP(x_r) = \int_{-\pi}^{\pi}\mathrm{d}\alpha\,\calP(x_r|\alpha)\hat{\calP}_{st}(\alpha)
\end{eqnarray}
for which we have no explicit expression. Nevertheless, we see from this that the highly non-Gaussian properties of $\hat{\calP}_{st}(\alpha)$ will now
be modulated by the Gaussian $\calP(x_r|\alpha)$ so that we anticipate a straightforward smearing of the distinct ratchet potential behaviours for
${\cal K}>$ and $<0$.

\begin{figure}
\centering
{\includegraphics[width=.9\linewidth]{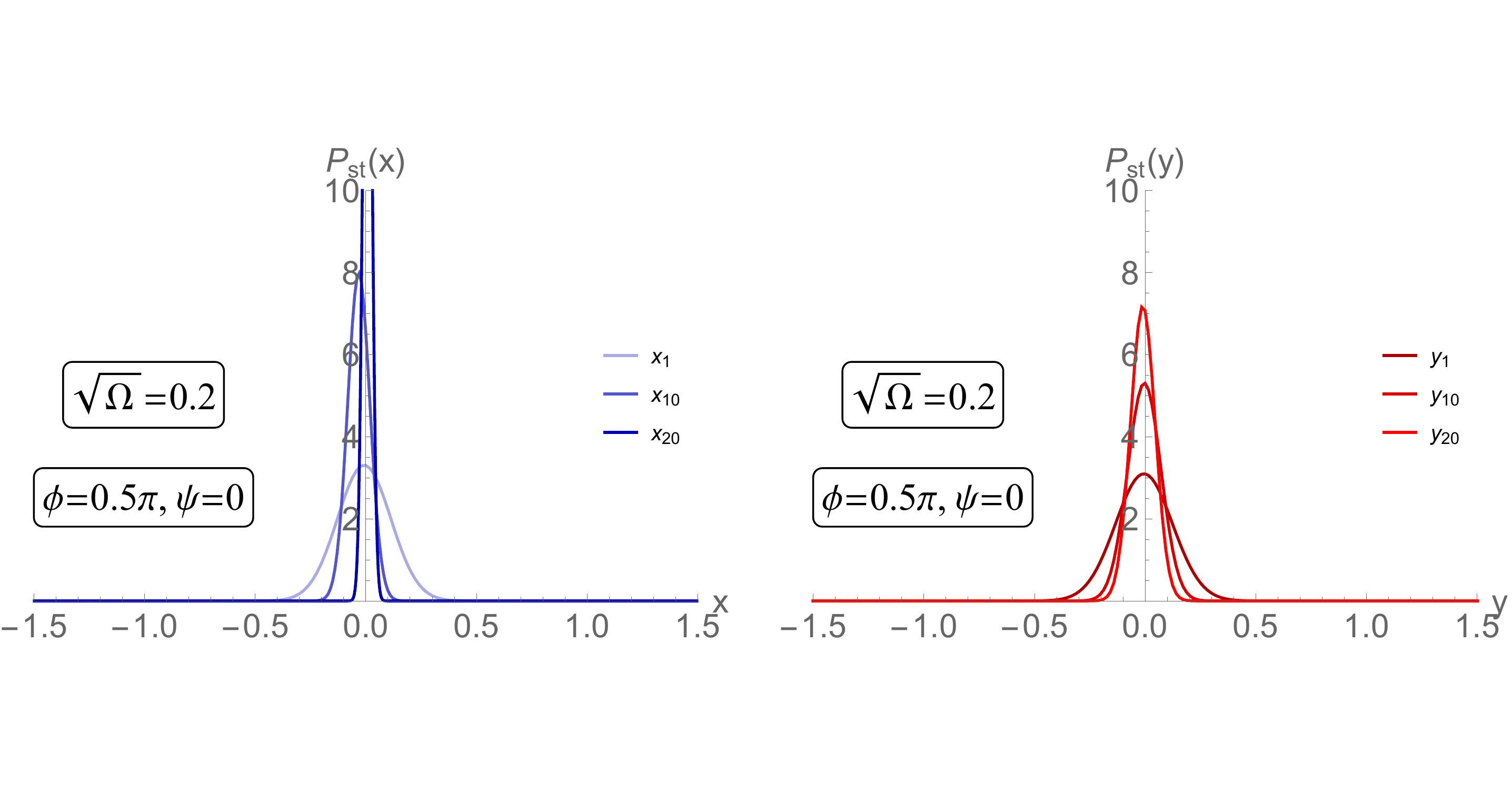}
\label{subfig:marginalnormalfigBLUE1}}
{\includegraphics[width=.9\linewidth]{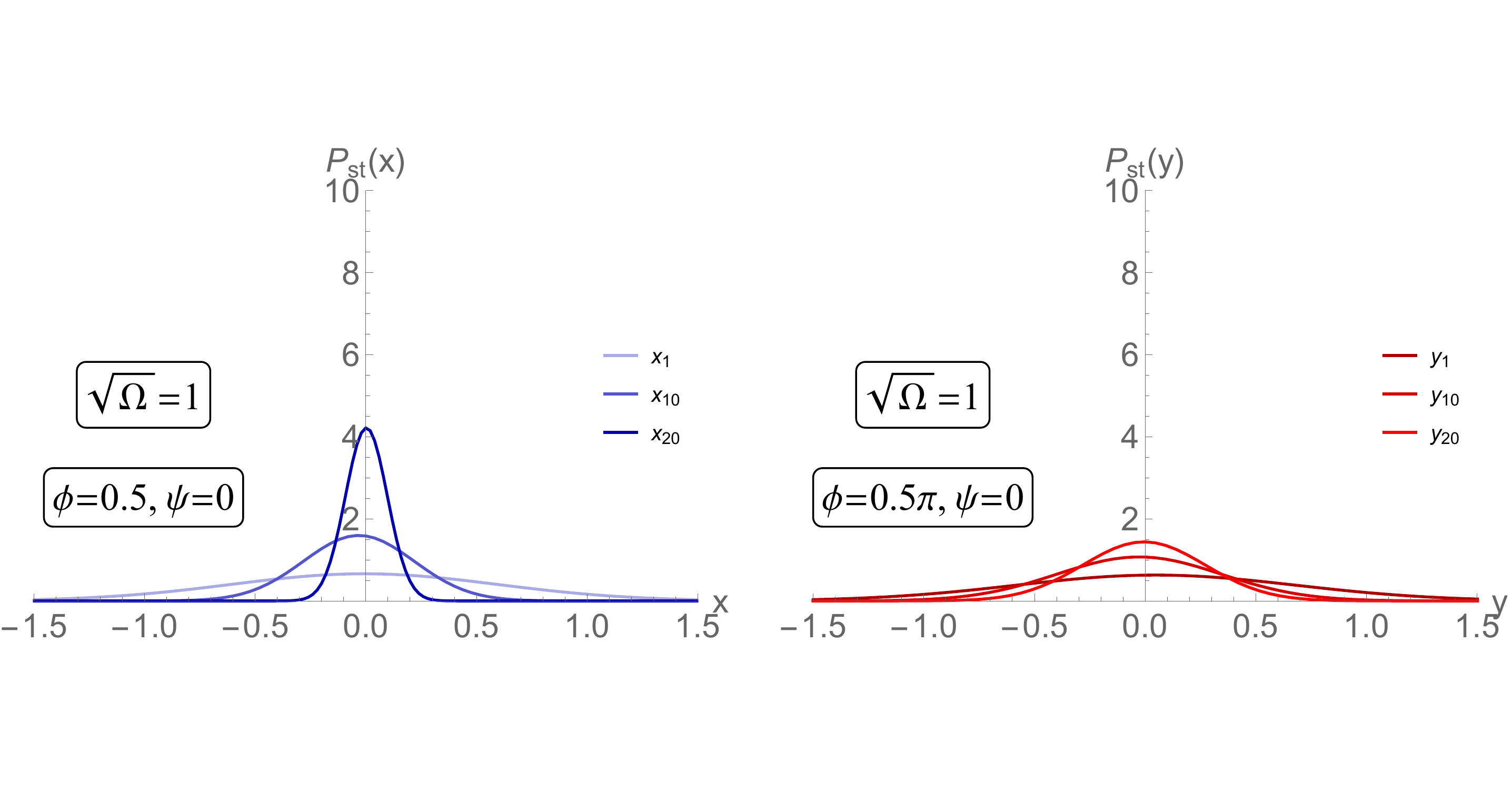}
\label{subfig:marginalnormalfigRED1}}
\caption{
Plots of Eq.(\ref{marginalnormalall}) for parameters $\phi=0.5 \pi$, $\psi=0$ and $\sqrt{\Omega}=0.2$ and $1$ for the left and right plots respectively,
with Blue population normal modes left column and Red population normal modes right column.
}
\label{fig:marginalnormalfigBLUERED1}
\end{figure}

Thus for ${\cal K}>0$ we anticipate significantly noisier $\alpha$ around a constant value as well as noisy normal modes $x_r$ and $y_r$ and order parameters, also around
constant values. As noise strength $\Omega$ increases, these fluctuations will be larger with no sign of periodicity but evidence of fracturing as the tail of these
densities extend beyond the basin of attraction permitting non-linearities to temporarily kick in. 
We plot the densities for the normal modes in Fig.\ref{fig:marginalnormalfigBLUERED1} and see precisely the Gaussian shape of these.
In particular, we observe that the distributions for Red modes are much more spread out than those of Blue for stronger noise, with
increased localisation for higher Laplacian eigenmodes. The value of $\sigma_R \lambda_r^{(R)}$ for the lowest mode
in the equivalent of Gaussian Eq.(\ref{normalmodejoint}) for Red connot explain this. In fact it is a consequence of the combination
of the Gaussian with $\hat{\calP}_{st}(\alpha)$ where the projections $d^{(BR)}_r$ and  $d^{(RB)}_r$ appear. We recall from the main body that,
due to the partitionability of Blue (reflected in the Fiedler eigenvector structure), the former is significantly smaller than the latter.
Here, through the integration in Eq.(\ref{marginalnormalall}), the small value effectively leads to a narrower Gaussian for the Blue modes - a smoothed 
version of the sharp bound in Eq.\ref{eq:modes-bounds} for the noisy zero mode case. This means that, at the parameter values used
in our illustrative examples in this paper, the fracturing of Red should be evident first as $\Omega$ is increased.

For ${\cal K}<0$ we should find periodic $\alpha$ with variation in the period (as for the noisy zero mode case) but now with distinctly noisy
fluctuations on the normal modes and order parameters - with nevetheless a strong signal of periodicity. With stronger noise $\alpha$ should
simply drift according to the sign of the tilted ratchet potential. Normal modes and order parameters will be simply noisy. By similar arguments
using the comparative values of  $d^{(BR)}_r$ and  $d^{(RB)}_r$ we should observe more signs of fracturing in Red than Blue.

\subsection*{D.2 Numerical simulations}
\begin{figure}
\centering
\subfloat[][Order Parameters for Red and Blue]{\includegraphics[width=.9\linewidth]{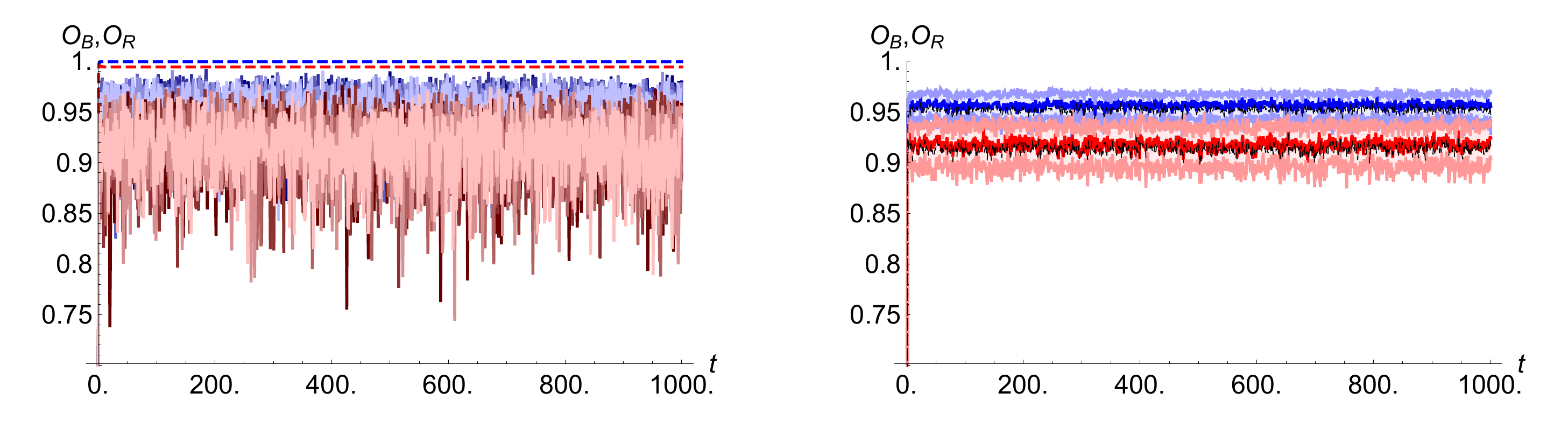}
\label{subfig:All-Order12}}

\subfloat[][Deterministic and Langevin simulations of $\alpha$]{\includegraphics[width=.9\linewidth]{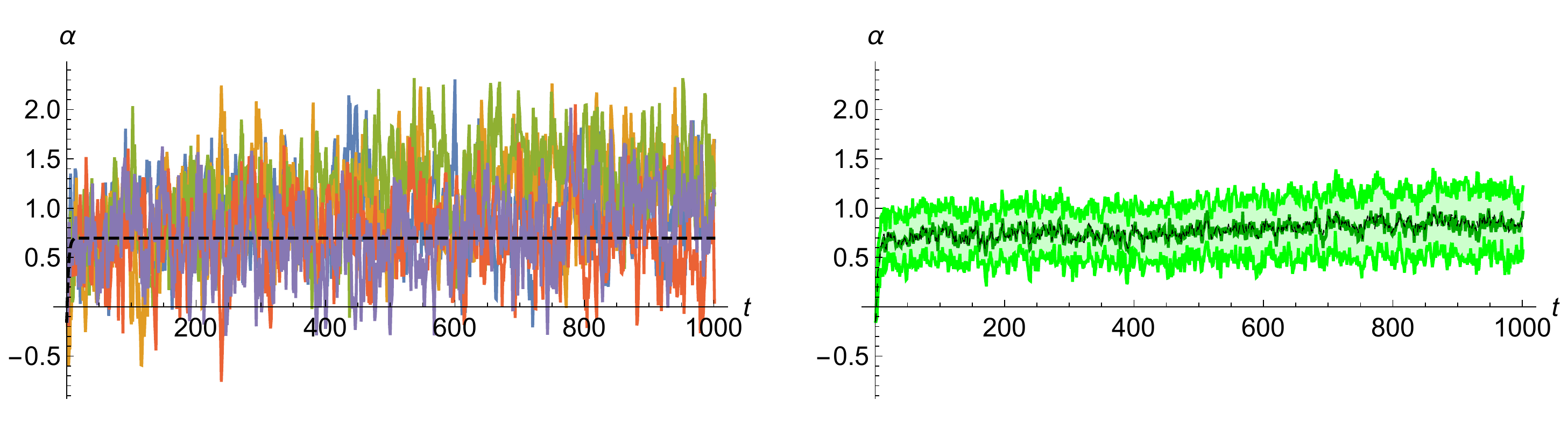}
\label{subfig:All-Alpha12}}

\subfloat[][Simulations of the $r=1,10,20$ modes]{\includegraphics[width=.9\linewidth]{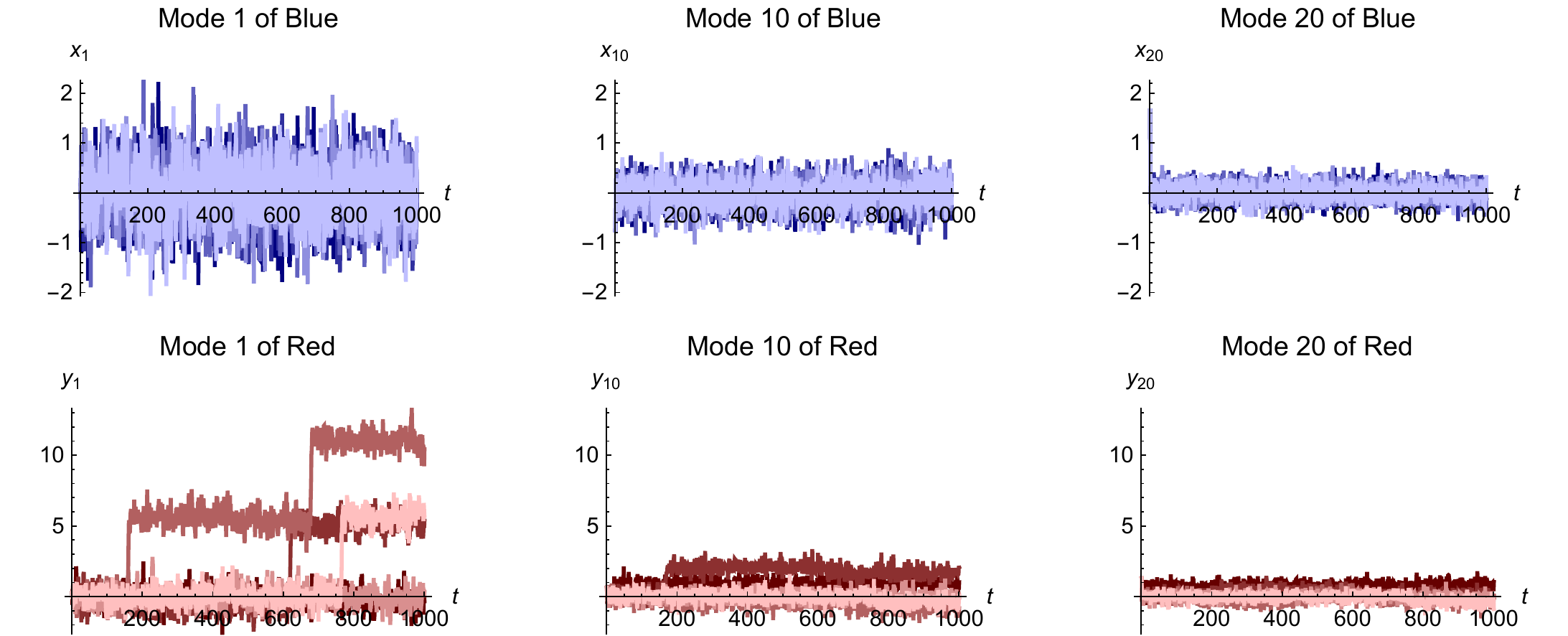}
\label{subfig:All-Modes12}}

\caption{Simulations of Eq.\eqref{BR-eq} with and without noise, with $\sqrt{\Omega}=1$ on all modes and parameters $\sigma_B=8,\,\sigma_R=0.5,\,\zeta_{BR}=\zeta_{RB}=0.4,\,\phi=0.5\pi,\,\psi=0$: Fig.\protect\subref{subfig:All-Order12} (Left) displays the deterministic local synchronisation order parameter $O_B$ (dashed blue) and $O_R$ (dashed red) and five paths of the Langevin local synchronisation order parameter $O_B$ and $O_R$, respectively, (Right) displays the median (blue/red), upper and lower quartile (light blue/red) and mean (dashed black) of 50 simulation of  $O_B$ and $O_R$, respectively. Fig.\protect\subref{subfig:All-Alpha12} (Left) displays deterministic $\alpha$ (Eq.\eqref{fixpoint-alph}) and 5 paths of the Langevin simulations for $\alpha$, (Right) displays the median (green), upper and lower quartile (light green) and mean (dashed black) of 50 simulation of $\alpha$. Fig.\protect\subref{subfig:All-Modes12} contains plots of the $r=1,10,20$ modes for 5 paths of the Langevin simulations. The top row contains the Blue modes and the bottom row contains the Red modes.}
\label{fig:All-Frac05-StdDev-1}
\end{figure}

\begin{figure}
\centering
\subfloat[][Order Parameters for Red and Blue]{\includegraphics[width=.9\linewidth]{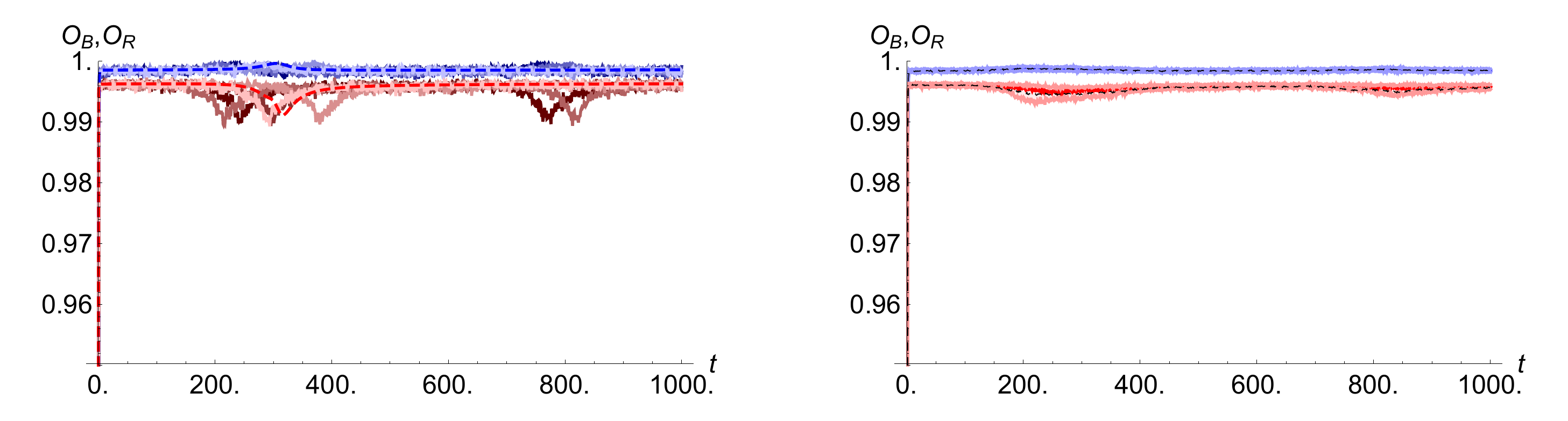}
\label{subfig:All-Order13}}

\subfloat[][Deterministic and Langevin simulations of $\alpha$]{\includegraphics[width=.9\linewidth]{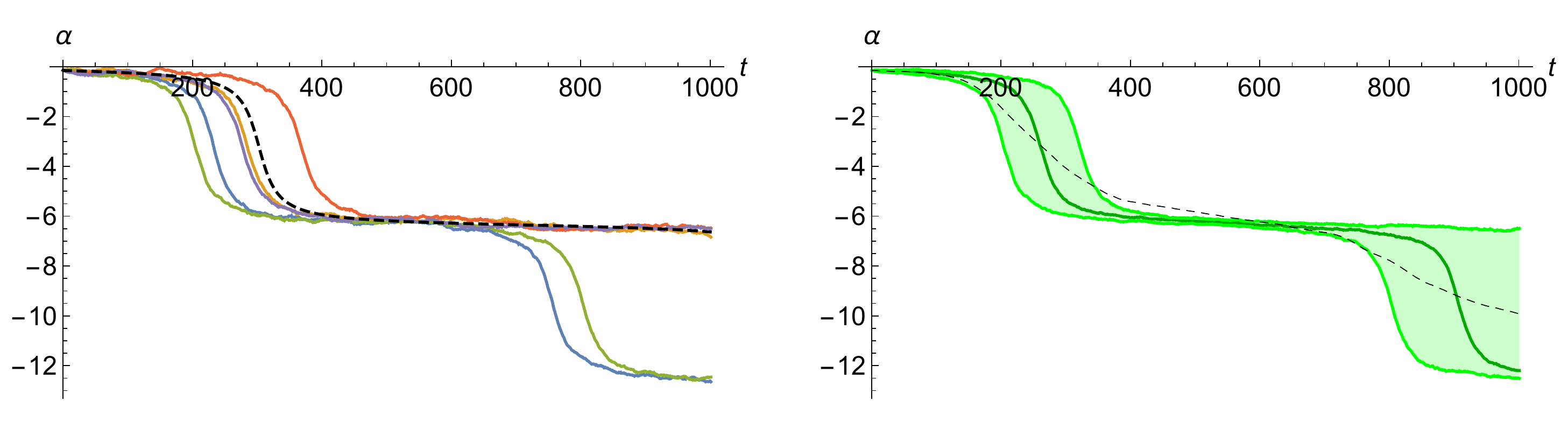}
\label{subfig:All-Alpha13}}

\subfloat[][Simulations of the $r=1,10,20$ modes]{\includegraphics[width=.9\linewidth]{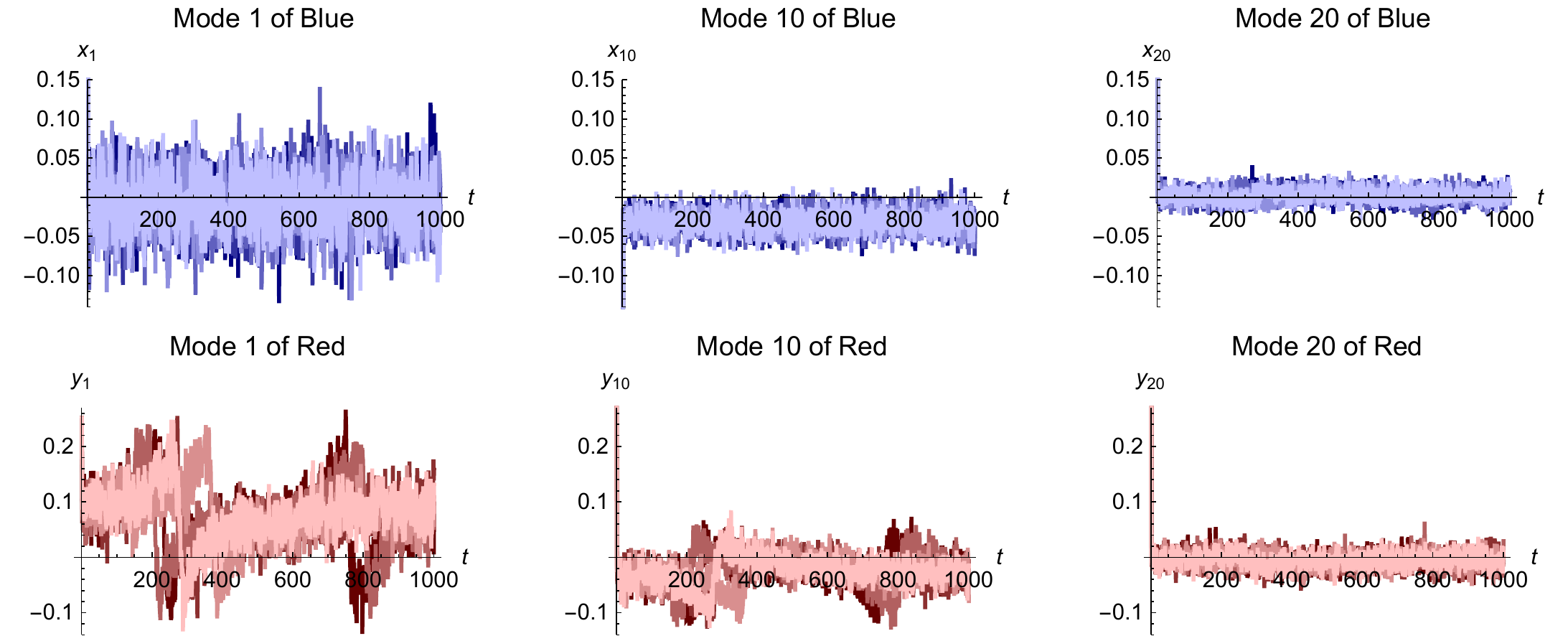}
\label{subfig:All-Modes13}}

\caption{Simulations of Eq.\eqref{BR-eq} with and without noise, with $\sqrt{\Omega}=0.05$ on all modes and parameters $\sigma_B=8,\,\sigma_R=0.5,\,\zeta_{BR}=\zeta_{RB}=0.4,\,\phi=0.95\pi,\,\psi=0$: Fig.\protect\subref{subfig:All-Order13} (Left) displays the deterministic local synchronisation order parameter $O_B$ (dashed blue) and $O_R$ (dashed red) and five paths of the Langevin local synchronisation order parameter $O_B$ and $O_R$, respectively, (Right) displays the median (blue/red), upper and lower quartile (light blue/red) and mean (dashed black) of 50 simulation of  $O_B$ and $O_R$, respectively. Fig.\protect\subref{subfig:All-Alpha13} (Left) displays deterministic $\alpha$ (Eq.\eqref{fixpoint-alph}) and 5 paths of the Langevin simulations for $\alpha$, (Right) displays the median (green), upper and lower quartile (light green) and mean (dashed black) of 50 simulation of $\alpha$. Fig.\protect\subref{subfig:All-Modes13} contains plots of the $r=1,10,20$ modes for 5 paths of the Langevin simulations. The top row contains the Blue modes and the bottom row contains the Red modes.}
\label{fig:All-Frac095-StdDev-005}
\end{figure}

\begin{figure}
\centering
\subfloat[][Order Parameters for Red and Blue]{\includegraphics[width=.9\linewidth]{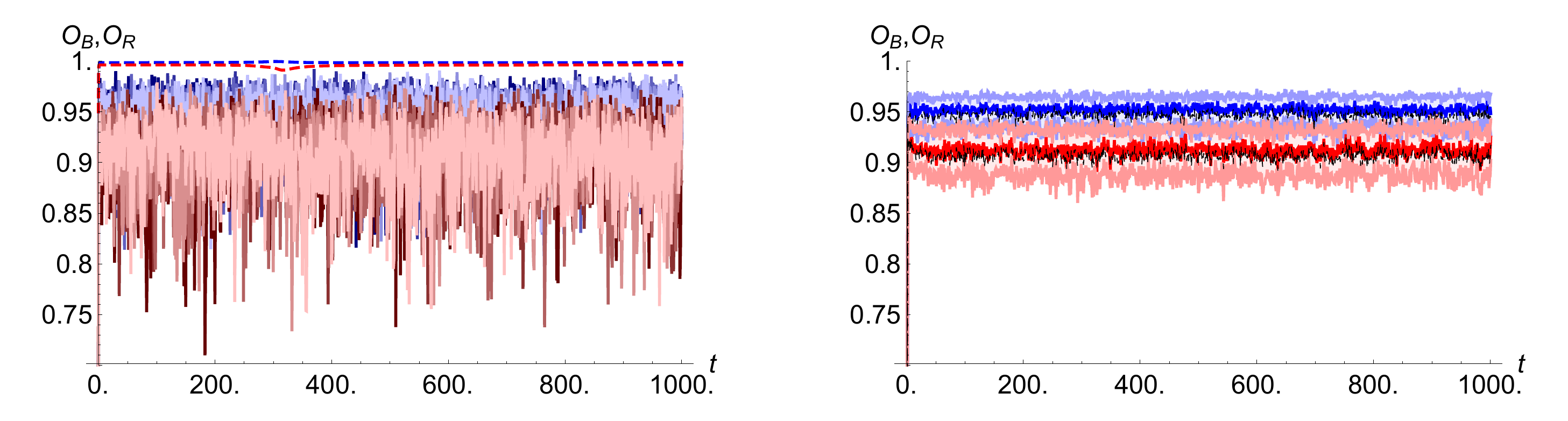}
\label{subfig:All-Order14}}

\subfloat[][Deterministic and Langevin simulations of $\alpha$]{\includegraphics[width=.9\linewidth]{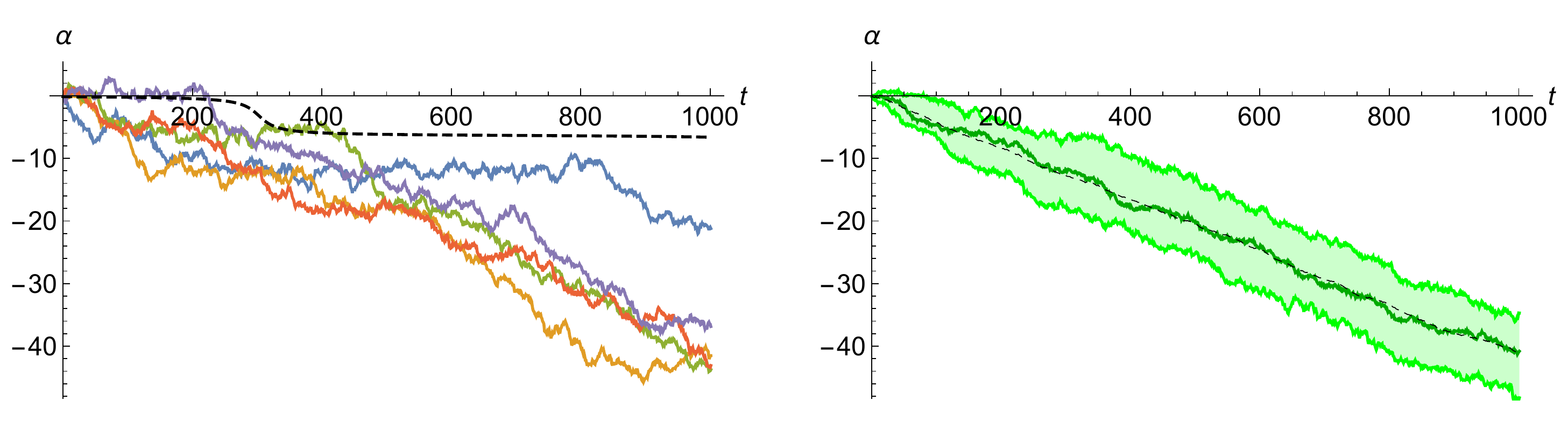}
\label{subfig:All-Alpha14}}

\subfloat[][Simulations of the $r=1,10,20$ modes]{\includegraphics[width=.9\linewidth]{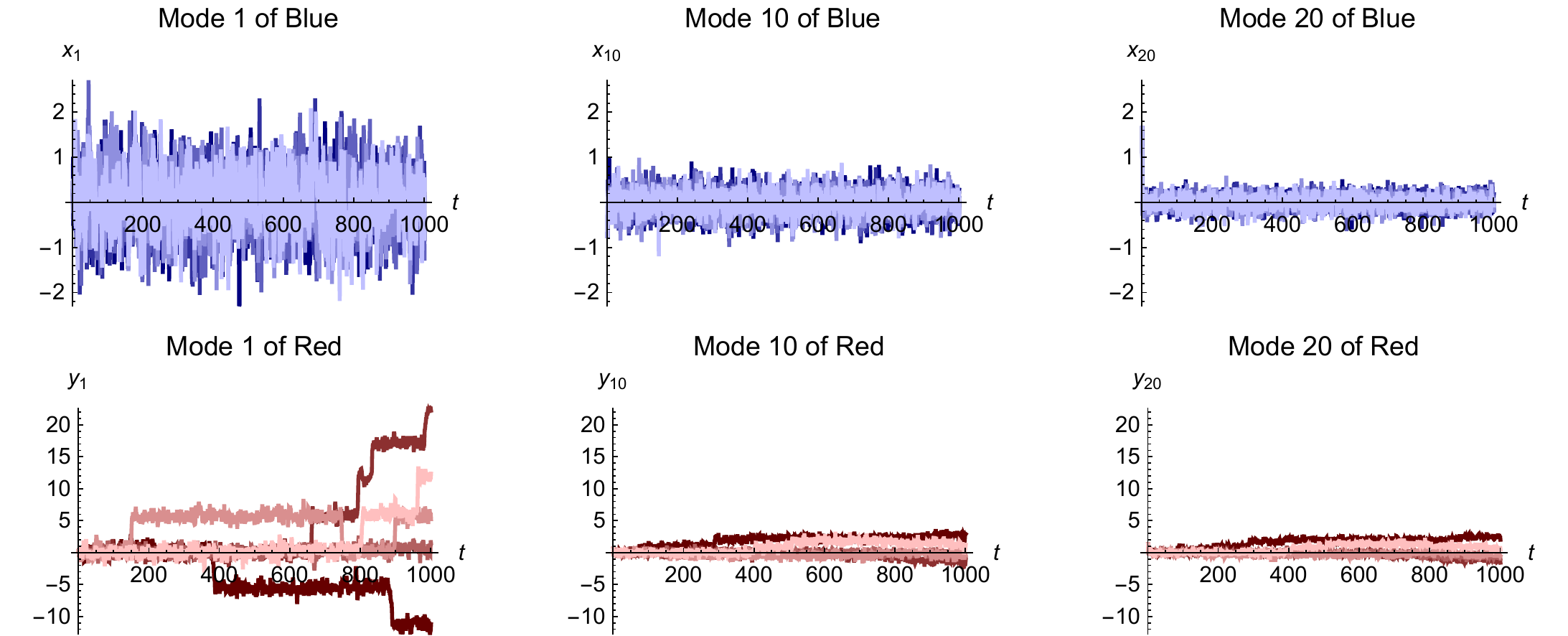}
\label{subfig:All-Modes14}}

\caption{Simulations of Eq.\eqref{BR-eq} with and without noise, with $\sqrt{\Omega}=1$ on all modes and parameters $\sigma_B=8,\,\sigma_R=0.5,\,\zeta_{BR}=\zeta_{RB}=0.4,\,\phi=0.95\pi,\,\psi=0$: Fig.\protect\subref{subfig:All-Order14} (Left) displays the deterministic local synchronisation order parameter $O_B$ (dashed blue) and $O_R$ (dashed red) and five paths of the Langevin local synchronisation order parameter $O_B$ and $O_R$, respectively, (Right) displays the median (blue/red), upper and lower quartile (light blue/red) and mean (dashed black) of 50 simulation of  $O_B$ and $O_R$, respectively. Fig.\protect\subref{subfig:All-Alpha14} (Left) displays deterministic $\alpha$ (Eq.\eqref{fixpoint-alph}) and 5 paths of the Langevin simulations for $\alpha$, (Right) displays the median (green), upper and lower quartile (light green) and mean (dashed black) of 50 simulation of $\alpha$. Fig.\protect\subref{subfig:All-Modes14} contains plots of the $r=1,10,20$ modes for 5 paths of the Langevin simulations. The top row contains the Blue modes and the bottom row contains the Red modes.}
\label{fig:All-Frac095-StdDev-1}
\end{figure}

Having dissected so many cases we may be brief in the results here. Fig.\ref{fig:All-Frac05-StdDev-1} 
gives the usual sequence of plots from
numerical simulations for $\phi=0.5\pi$, $\sqrt\Omega=1$. We see precisely what analytical results predicted:
noisy fluctuations on everything - order parameters, $\alpha$ and normal modes, with $\alpha$ showing no drift, consistent
with ${\cal K}>0$ here. Moreover, we observe the predicted fragmentation in the lowest Red mode $y_1$.
Fig.\ref{fig:All-Frac095-StdDev-005} and \ref{fig:All-Frac095-StdDev-1} show the results for $\phi=0.95\pi$ where the time-periodicity
of $\alpha$ is evident for low $\Omega$ going to general negative drift (and a hint of periodicity in the median over paths) for higher $\Omega$.
Again, this is consistent with ${\cal K}<0$.
However, the key observation across all of these results is that the key differences between behaviours arises from the noise on zero modes -
according to the sign of ${\cal K}$ in the deterministic behaviour - overlaid with Gaussian noise on normal modes. Both of these effects fold into
the behaviour of the order parameters.
\FloatBarrier

\FloatBarrier

\section*{Appendix E: Formalism for Red fragmentation}
In the scenario where the dynamics of the Red population show fragmentation into two sub-populations we define three centroids:
\begin{eqnarray*}
B=\frac{1}{N} \sum_{i \in {\cal B}} \beta_i,\;\; P_1=\frac{1}{M_1} \sum_{i \in {\cal R}_1} \rho_i,\;\; P_2=\frac{1}{M_2} \sum_{i \in {\cal R}_2} \rho_i.
\end{eqnarray*}
The projection of cross-Laplacians on fluctuations - terms that are deemed small in our approximations - now take the form:
\begin{eqnarray*}
{\cal L}_i &=& \left\{  \begin{array}{ll}
\zeta_{BR} \cos(\alpha_{B R_1}-\phi) \sum_{j \in {\cal B} \cup {\cal R}_1}L^{(B R_1)}_{ij}v_j  & i \in {\cal B}\\
\zeta_{RB} \cos(\alpha_{B R_1} + \psi) \sum_{j \in {\cal B} \cup {\cal R}_1}L^{(BR_1)}_{ij}v_j \\
+ \sigma_R \cos (\alpha_{R_1 R_2}) \sum_{j \in {\cal R}_1 \cup {\cal R}_2}L^{(R_1 R_2)}_{ij}v_j  & i \in {\cal R}_1 \\
\sigma_R \cos (\alpha_{R_1 R_2}) \sum_{j \in {\cal R}_1 \cup {\cal R}_2}L^{(R_1 R_2)}_{ij}v_j & i \in {\cal R}_2
\end{array} \right.
\end{eqnarray*}
and $v_i \in  \{ b_i, p^{(1)}_i, p^{(2)}_i \}$ for $i \in \{ {\cal B}, {\cal R}_1, {\cal R}_2  \}$ respectively.

As with the two cluster regime, in order to gain analytical insights into the linearised 3 cluster system we need to decide which of the Laplacians we wish to diagonalise, and assume the remainder are negligible. As with the two cluster regime, we chose to diagonalise the intra-Laplacians $L^{(B)}$, $L^{(R_1)}$ and $L^{(R_2)}$, and assumed that the interaction Laplacians in ${\cal L}$ offer a negligible contribution to the overall dynamics. The obvious pitfall of this assumption is that we lose all but the coarsest of information of the inter-network structure, namely the total number of connections per node $d^{(B R_1)}_i,  d^{(R_1 B)}_i,  d^{(R_1, R_2)}_i$ and $ d^{(R_2 R_1)}_i$. However, as shown in \cite{KallZup2015}, if the interaction networks are not too complicated, this assumption does quite well in detecting system fixed points and the onset of dynamical behaviour. Thus for analytical considerations we made the approximation
\begin{eqnarray}\label{supelLapapprox}
{\cal L} \approx 0.
\end{eqnarray}

Using the Laplacians in the three clusters for ${\cal B}$, ${\cal R}_1$ and ${\cal R}_2$, the corresponding eigenvalue-eigenvectors are
\begin{eqnarray*}
\sum_{j \in {\cal R}_a} L^{(R_a)}_{ij} e^{(R_a,r)}_j = \lambda^{(R_a)}_r e^{(R_a,r)}_i, \;\; \{i,r\} \in \{{\cal R}_a, {\cal R}^E_a \}, \;\; a \in \{1,2\}.
\end{eqnarray*}
As usual, each set of Laplacian eigenvalues has at least one zero eigenvalue, with the remaining being real, positive semi-definite \cite{Boll98}.

Taking advantage of the non-zero eigenvectors, we again expanded the fluctuations $b_i$ and $p^{(a)}_j$ in the following normal-modes:
\begin{eqnarray}
b_i = \sum_{r \in {\cal B}^E / \{0\}} x_r e^{(B,r)}_i \;\; i \in {\cal B}, \;\; p^{(a)}_j = \sum_{r \in {\cal R}^E_a / \{0\}} y^{(a)}_r e^{(R_a,r)}_j \;\; j \in {\cal R}_a.
\label{normalex2}
\end{eqnarray}



\section*{Acknowledgements}
The authors gratefully acknowledge discussions with 
Richard Taylor, Tony Dekker, Iain Macleod, Dale Roberts and Markus Brede.
ACK is supported through a Chief Defence Scientist Fellowship.

\end{document}